\documentclass[11pt,a4paper]{article}
\pdfoutput=1
\usepackage{jheppub, hyperref, cleveref}
\usepackage{amsfonts, amsmath, amssymb, latexsym}
\usepackage{graphicx, caption, subcaption}
\usepackage[normalem]{ulem}
\usepackage[dvipsnames]{xcolor}

\newcommand{\lp}{\left (}
\newcommand{\rp}{\right )}
\newcommand{\llp}{\left [}
\newcommand{\rrp}{\right ]}

\renewcommand{\d}{{\rm d}}

\newcommand{\Ic}{\mathcal{I}_c}
\newcommand{\Is}{\mathcal{I}_s}

\newcommand{\GW}{\text{\tiny GW}}
\newcommand{\NL}{\text{\tiny NL}}

\def\PBH{\text{\tiny  PBH}}

\newcommand{\be}{\begin{equation}}
\newcommand{\ee}{\end{equation}}

\def\be{\begin{equation}}
\def\ee{\end{equation}}

\def\beq{\begin{equation}}
\def\eeq{\end{equation}}

\newcommand{\bv}{{\bf{v}}}

\newcommand{\bn}{{\bf{n}}}

\def\bea{\begin{eqnarray}}
\def\eea{\end{eqnarray}}

\def\d{{\rm d}}
\def\dd{{\rm d}}

\def\d{{\rm d}}

\def\0{{\boldsymbol 0}}

\def\lsim{\mathrel{\rlap{\lower3pt\hbox{\hskip0pt$\sim$}}
   \raise1pt\hbox{$<$}}}         
\def\gsim{\mathrel{\rlap{\lower4pt\hbox{\hskip1pt$\sim$}}
   \raise1pt\hbox{$>$}}}         

 \newcommand{\sfootnote}[1]{}

\newcommand{\sapienza}{Dipartimento di Fisica, Sapienza Universit\`a
	di Roma, Piazzale Aldo Moro 5, 00185, Roma, Italy}
\newcommand{\infnroma}{INFN, Sezione di Roma, Piazzale Aldo Moro 2, 00185, Roma, Italy}

\title{Probing Anisotropies of the Stochastic Gravitational Wave Background with LISA}

\author[a, b, c]{ Nicola Bartolo}
\author[a, b, c]{\!\!, Daniele Bertacca}
\author[d]{\!\!, Robert Caldwell}
\author[e]{\!\!, Carlo R. Contaldi}
\author[f]{\!\!, Giulia Cusin}
\author[f]{\!\!, Valerio De Luca}
\author[g,h]{\!\!, Emanuela Dimastrogiovanni}
\author[i,j]{\!\!, Matteo Fasiello}
\author[l]{\!\!, Daniel G. Figueroa}
\author[m,n]{\!\!, Gabriele Franciolini}
\author[o, p]{\!\!, Alexander C. Jenkins}
\author[a, b]{\!\!, Marco Peloso}
\author[e]{\!\!, Mauro Pieroni}
\author[q, r]{\!\!, Arianna Renzini}
\author[a, b]{\!\!, Angelo Ricciardone\footnote{Project coordinator and corresponding author: angelo.ricciardone@pd.infn.it}}
\author[f]{\!\!, Antonio Riotto}
\author[p]{\!\!, Mairi Sakellariadou}
\author[s]{\!\!, Lorenzo Sorbo}
\author[t]{\!\!, Gianmassimo Tasinato}
\author[u]{\!\!, Jes\'us Torrado}
\author[v]{\!\!, Sebastien Clesse}
\author[i]{\!\!, Sachiko Kuroyanagi}

\author[]{\\ \centering \texttt{(For the LISA Cosmology Working Group)}}

\affiliation[a]{Dipartimento di Fisica e Astronomia ``G. Galilei'', Universit\`a degli Studi di Padova, via Marzolo 8,
I-35131, Padova, Italy}
\affiliation[b]{INFN, Sezione di Padova, via Marzolo 8, I-35131, Padova, Italy}
\affiliation[c]{INAF - Osservatorio Astronomico di Padova, Vicolo dell'Osservatorio 5, I-35122 Padova, Italy.}
\affiliation[d]{HB6127 Wilder Lab, Department of Physics \& Astronomy, Dartmouth College, Hanover, New Hampshire 03755 USA}
\affiliation[e]{Blackett Laboratory, Imperial College London, South Kensington Campus, London, SW7 2AZ, UK}
\affiliation[f]{D\'epartement de Physique Th\'eorique and Centre for Astroparticle Physics (CAP), Universit\'e de Gen\`eve, 24 quai E. Ansermet, CH-1211 Geneva, Switzerland}
\affiliation[g]{Van Swinderen Institute for Particle Physics and Gravity, University of Groningen, Nijenborgh 4, 9747 AG Groningen, The Netherlands}
\affiliation[h]{School of Physics, The University
of New South Wales, Sydney NSW 2052, Australia}
\affiliation[i]{Instituto de Fisica Téorica UAM-CSIC, C\slash $\,$ Nicolas Cabrera 13-15, Cantoblanco, 28049, Madrid, Spain}
\affiliation[j]{Institute of Cosmology \& Gravitation, University of Portsmouth, PO1 3FX, UK}
\affiliation[\ell]{Instituto de F\'isica Corpuscular (IFIC), CSIC-Universitat de Val\`{e}ncia, E-46980, Valencia, Spain.}
\affiliation[m]{\sapienza}
\affiliation[n]{\infnroma}
\affiliation[o]{Department of Physics \& Astronomy, University College London, \\Gower Street, London WC1E 6BT, United Kingdom}
\affiliation[p]{Theoretical Particle Physics and Cosmology Group,  Physics Department, \\ King's College London, University of London, Strand, London WC2R 2LS, United Kingdom}
\affiliation[q]{LIGO Laboratory,  California  Institute  of  Technology,  Pasadena,  California  91125,  USA}
\affiliation[r]{Department of Physics, California Institute of Technology, Pasadena, California 91125, USA}
\affiliation[s]{Amherst Center for Fundamental Interactions, Department of Physics,\\ University of Massachusetts, Amherst, MA 01003, U.S.A.}
\affiliation[t]{Physics Department, Swansea University, SA28PP, UK}
\affiliation[u]{Institute for Theoretical Particle Physics and Cosmology (TTK), RWTH Aachen University, D-52056 Aachen, Germany}
\affiliation[v]{Service de Physique Th\'eorique, Universit\'e Libre de Bruxelles, Boulevard du Triomphe, CP225, 1050 Brussels, Belgium. }

\abstract{We investigate the sensitivity of the Laser Interferometer Space Antenna (LISA) to the anisotropies of the Stochastic Gravitational Wave Background (SGWB). We first discuss
the main astrophysical and cosmological sources of SGWB which are characterized by anisotropies in the GW energy density, and we build a Signal-to-Noise estimator to quantify the sensitivity of LISA to different multipoles. We then perform a Fisher matrix analysis of the prospects of detectability of anisotropic features with LISA for individual multipoles, focusing on a SGWB with a power-law frequency profile. We compute the noise angular spectrum taking into account the specific scan strategy of the LISA detector. We analyze the case of the kinematic dipole and quadrupole generated by Doppler boosting an isotropic SGWB. We find that $\beta\, \Omega_{\rm GW}\sim 2\times 10^{-11}$ is required to observe a dipolar signal with LISA. The detector response to the quadrupole has a factor $\sim 10^3 \,\beta$ relative to that of the dipole. The characterization of the anisotropies, both from a theoretical perspective and from a map-making point of view, allows us to extract information that can be used to understand the origin of the SGWB, and to discriminate among distinct superimposed SGWB sources.}

\begin{document}
\begin{figure}
\begin{flushright}
\href{https://lisa.pages.in2p3.fr/consortium-userguide/wg_cosmo.html}{\includegraphics[width = 0.2 \textwidth]{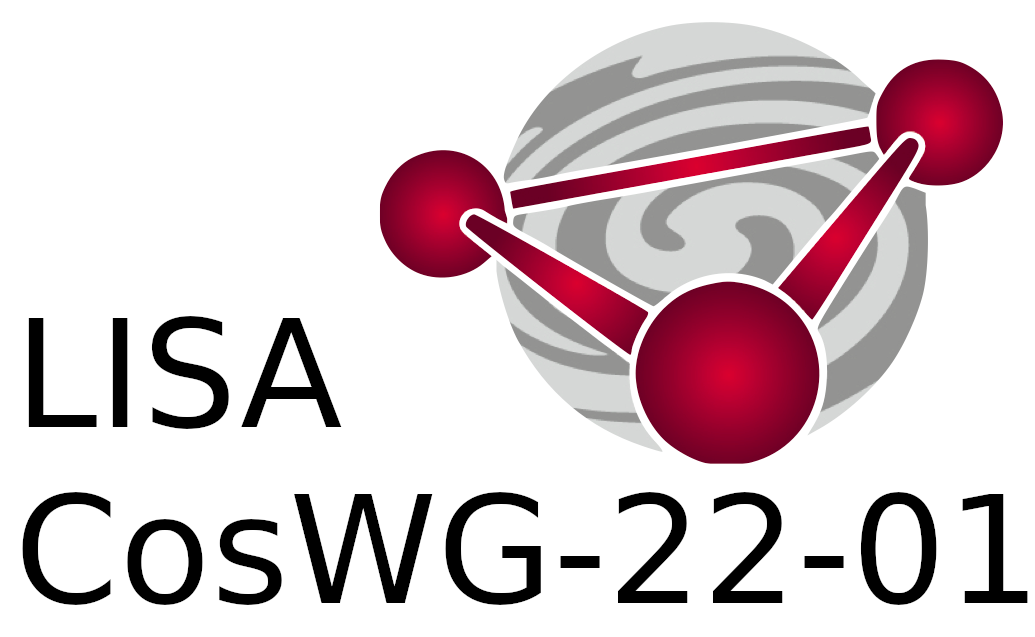}}\\[5mm]
\end{flushright}
\end{figure}

\maketitle

\section{\sc Introduction}
\label{sec:intro}

One of the main targets of the LISA gravitational wave (GW) detector is the detection of a stochastic gravitational wave background (SGWB), which can shed light on the physics of the early universe and on astrophysical population properties not accessible with resolved sources. There are many possible astrophysical and cosmological sources which contribute to the stochastic background (see e.g., \cite{Regimbau:2011rp,Maggiore:2018sht,Caprini:2018mtu} for recent reviews), and up to now we have only upper bounds on its amplitude in \cite{TheLIGOScientific:2016dpb}, and on parameters characterising its directional properties \cite{TheLIGOScientific:2016xzw,KAGRA:2021mth}, by the LIGO/Virgo collaboration. On the other hand we have a recent claim of a possible detection of a SGWB signal in the nano-Hertz regime by the NANOGrav collaboration \cite{Arzoumanian:2020vkk}. Typically it is expected that each source is characterized by a specific spectral shape \cite{Caprini:2019pxz, Flauger:2020qyi}, however, given the plethora of sources (both resolved and unresolved) which are present in the LISA band (i.e., milli-Hertz regime), it is important to study other features which can allow for a better characterization and detection of this signal. 

Interesting properties can be extracted by measuring the anisotropies in the SGWB. The first attempts for  extracting information on the anisotropies of the SGWB have been first done in \cite{Allen:1996gp} for the case of ground-based
interferometers,  in \cite{Cornish:2001hg} for space-based interferometers, and in \cite{Mingarelli:2013dsa,Taylor:2013esa} for pulsar timing arrays.

The aim of this work, developed within the LISA Cosmology Working Group, is to analyze the capabilities of LISA \cite{LISA:2017pwj} to detect anisotropies of the SGWB in the milli-Hertz band, making use of current instrument specifications, as well as of the latest theoretical
characterizations of sources of SGWB anisotropies. The work is  developed in two main parts:
The first part is more theoretical in nature, and reviews our present understanding of  cosmological and  astrophysical sources for the SGWB and the properties of its  anisotropies; The second part contains new results on the  characterisation of  {\it{angular}} response functions for LISA,  accompanied by    forecasts of the detectability 
of an anisotropic SGWB with LISA.

The theory part of our work starts with a review of a Boltzmann equation approach for analyzing anisotropies of the SGWB~\cite{Contaldi:2016koz,Bartolo:2019oiq,Bartolo:2019yeu,Cusin:2018avf, Pitrou:2019rjz}, similarly to what is commonly done for the Cosmic Microwave Background (CMB).
This method is  convenient for distinguishing effects on anisotropies  sourced at the moment of GW production, from anisotropies developed as GWs propagate through our inhomogeneous universe. We then discuss
 early universe sources of the SGWB, and we describe SGWB anisotropies produced from inflationary mechanisms and from the formation of primordial black holes (PBH). Similarly to  CMB photons, gravitons are also affected by the Sachs-Wolfe and Integrated Sachs-Wolfe effects, both related to the propagation of  GW through a perturbed universe. Besides these contributions, we discuss  the intrinsic SGWB anisotropy generated at the moment of production, whose   frequency-dependence  represents a peculiar signature of GW. We then discuss a case where GW anisotropies are induced by primordial non-Gaussianity, showing  that in certain scenarios  such a contribution can be relatively large. Then we study the anisotropies expected in some post-inflationary mechanisms, like preheating in a scale invariant model~\cite{Bethke:2013aba,Bethke:2013vca} --  even if the GW background in this model is typically peaked at larger frequencies \cite{Figueroa:2017vfa}, beyond the LISA frequency band. We finally review another two main GW sources that are characterized by anisotropies: phase transitions and topological defects. For phase transitions, if only cosmological adiabatic perturbations are considered, the fluctuations in any causally produced GW background will follow those in the CMB, and hence they are expected to be  small~\cite{Geller:2018mwu,Kumar:2021ffi}.
 
    For GW sourced by topological defects, anisotropies induced by a network of Nambu-Goto cosmic string loops have been computed in \cite{Jenkins:2018lvb,Kuroyanagi:2016ugi,Olmez:2011cg}.
It has been shown that while the angular power spectrum $C_\ell$ -- the quantity characterising the multipole decomposition of the SGWB spectrum --  depends on the model of the loop network,
the anisotropies are driven by local Poisson fluctuations in the number of loops, and the resulting angular power spectrum is spectrally white (i.e., $C_\ell=\text{constant}$ with respect to $\ell$), regardless of the particular loop distribution \cite{Jenkins:2018lvb}.

 We then present the case of anisotropies generated from astrophysical sources of GWs.  LISA will be sensitive to several astrophysical sources such as Super Massive Binary Black Holes (SMBBHs) with masses $ \sim 10^4-10^7 M_\odot$, stellar origin Binary Black Holes (SOBBHs), Extreme Mass Ratio Inspirals (EMRIs) and Galactic white dwarf Binaries (GBs). Beyond these resolvable sources, measurements by LISA will also be affected by a huge number of unresolvable events which will sum up incoherently, forming a SGWB~\cite{Farmer:2003pa,Regimbau:2009rk,Regimbau:2011rp}. At least two SGWB components are guaranteed to be present in the LISA band: a contribution due mostly due to GB inspirals in the low-frequency band of (up to $\sim 10^{-3} $~Hz), and a contribution from extra-galactic BBH mergers expected at slightly higher frequencies ($\sim 10^{-3} - 10^{-2}$~Hz). 
The analytic derivation of the energy density anisotropies for an SGWB has been well studied in the literature \cite{Cusin:2017fwz, Contaldi:2016koz, Cusin:2017mjm, Cusin:2018avf, Pitrou:2019rjz, Bertacca:2019fnt}. Predictions for the energy density angular power spectrum have been presented in \cite{Cusin:2018rsq,  Jenkins:2018uac, Jenkins:2018kxc, Cusin:2019jpv, Cusin:2019jhg, Bertacca:2019fnt} in the Hz band and in \cite{Cusin:2019jhg} in the mHz band (see~\cite{Bellomo:2021mer} for a recent numerical code to estimate the angular spectrum of the anisotropies of the astrophysical GWB). 
Anisotropies show a range of variability depending on the underlying astrophysical model for star formation, mass distribution and collapse, and on the considered cosmological perturbation effects. Due to its stochastic nature, we characterise the anisotropies in terms of their angular power spectrum taking into account all the cosmological and astrophysical dependencies.

We then move to the second  part of this paper containing  original results on
prospects  of detection of anisotropies of the  SGWB with LISA, given the current instrument specifications. 
The characterisation of the angular resolution of space-based detectors as 
 LISA has been pioneered  in  \cite{Peterseim:1997ic,Cutler:1997ta,Moore:1999zw},
 and previous studies on LISA capabilities in detecting and characterising SGWB anisotropies include
  \cite{Ungarelli:2001xu,Seto:2004np,Kudoh:2004he,Taruya:2005yf,Taruya:2006kqa}. 
We start our analysis 
 computing the {\it angular} response functions of LISA to the different multipoles for a statistically isotropic SGWB. We work in the A, E, T Time-Delay-Interferometry basis (see \cite{Tinto:2001ii,Tinto:2002de,Hogan:2001jn,Tinto:2004wu,Christensen:1992wi,Adams:2010vc}, as well as the comprehensive review \cite{Romano:2016dpx}) and we compute the angular response as a function of frequency for the auto-correlation channels (AA, EE, and TT) and cross-correlated ones (i.e.  AE, AT). We also give their analytic expression in the low frequency limit. We develop an estimator for the angular power spectrum $C_{\ell}$, giving a simple analytic tool to estimate the total sensitivity of LISA to an anisotropic signal. With these tools we estimate the minimal amplitude  of GW energy density needed for  detecting higher multipoles. As a concrete example, we analyse the case of the kinematic dipole and quadrupole generated by boosting with a factor $\beta\equiv v/c$ an isotropic SGWB. We find that for one year of observation, $\beta\, \Omega_{\rm GW}\sim 2\times 10^{-11}$ is required to observe a dipolar signal with LISA. We also find that the detector response to the quadrupole has a factor $\sim 10^3 \,\beta$ relative to that of the dipole.
 
We then  perform  a Fisher matrix analysis aimed at forecasting the amplitude required on the lowest multipoles of the SGWB angular power spectrum for being detectable with LISA, given the current information on LISA strain and angular resolution sensitivity. We  consider a power-law SGWB spectrum peaking at some multiple $\ell$ characterised by a fiducial amplitude and spectral tilt. 

The peak in sensitivity for $\ell =1$ occurs at higher frequencies than that for $\ell=0,2$. Therefore, if we choose the pivot scale of the power-law signal to coincide with the peak sensitivity frequency of the $\ell=0,2$ multipoles, so that their detectability is only weakly sensitive to the spectral index, we then find a greater sensitivity for $\ell =1$ in the case of a positive spectral tilt. 

Finally we  apply the maximum likelihood map-making method for stochastic backgrounds proposed in~\cite{Contaldi:2020rht} to the LISA detector, in order to provide estimates for the noise angular power spectrum $\cal N_\ell$. We simulate and map the noise directly in the sky domain, and we take into account the specific scan strategy of LISA, which describes how the sky signal is sampled as a function of time. 

\medskip

The structure of this paper is as follows: in Section~\ref{sec:cosmo} and~\ref{sec:astro} we review the main cosmological and astrophysical GW sources and their predicted angular power spectra; in Section~\ref{sec:response} we present the LISA angular response function to different multipoles and the Signal-to-Noise (SNR) estimator for anisotropic signals. In Section~\ref{sec:fisher} we perform a Fisher matrix analysis for the amplitude and spectral tilt of a SGWB signal characterized by a power-law behaviour. Finally in Section~\ref{sec:mapmaking} we compute the noise angular power spectrum of LISA for different multipoles using a map-making approach. A conclusion and some technical appendices
conclude the work.

\section{\sc Cosmological Sources of Anisotropies}
\label{sec:cosmo}

\subsection{Theoretical framework}
\label{theoretical_framework}
The SGWB energy is controlled by the energy density 
spectrum $\Omega_{\rm GW}$ defined as
\begin{equation}
\label{def1omGW}
\Omega_{\rm GW}\, \equiv \,\frac{\rm {d} \rho_{\rm GW}}{\rho_{{\rm c},0}\,\rm{d} \ln q} \;, 
\end{equation}
with $d \rho_{\rm GW}$ being the energy density in GW contained in the comoving momentum interval $q$ to $q+ d q$, and $\rho_{{\rm c},0}$ corresponding to the critical energy density of the present-day universe. As we are going to discuss, we expect that the quantity $\Omega_{\rm GW}$  is characterized by an averaged isotropic component plus a direction-dependent component. Both the isotropic and the anisotropic contributions are 
 two key observables that can be targeted by the GW LISA detector. 
 Several cosmological sources can produce a monopole GW energy density within the reach of the LISA sensitivity:  inflationary models beyond vanilla single-field scenarios, where the inflaton is coupled with extra (gauge) fields~\cite{Barnaby:2010vf,Cook:2011hg,Sorbo:2011rz, Barnaby:2011qe,Dimastrogiovanni:2016fuu,Peloso:2016gqs, Domcke:2016bkh} to models with features in the scalar power spectrum~\cite{Flauger:2009ab,Braglia:2020eai,Fumagalli:2020nvq}, or models where space-time symmetries are broken during inflation  \cite{Endlich:2013jia,Koh:2013msa,Cannone:2014uqa,Cannone:2015rra,Bartolo:2015qvr,Ricciardone:2016lym,Bartolo:2015qvr,Cannone:2014uqa,Akhshik:2014gja,Akhshik:2014bla}, or scenarios where non-attractor phases characterize the Universe evolution, \cite{Leach:2001zf, Namjoo:2012aa, Mylova:2018yap}, or second-order scalar induced GWs which are also responsible for PBH formation~\cite{Acquaviva:2002ud, Mollerach:2003nq, Carbone:2004iv, Ananda:2006af, Baumann:2007zm, Saito:2009jt, Garcia-Bellido:2016dkw, Cai:2018dig, Bartolo:2018rku, Bartolo:2018evs, Unal:2018yaa, Wang:2019kaf, Cai:2019elf, DeLuca:2019ufz,Inomata:2019yww, Yuan:2019fwv,Ozsoy:2019lyy, Pi:2020otn, Yuan:2020iwf,Tasinato:2020vdk}. Also post-inflationary mechanisms can generate GW signals within the reach of the LISA detector: expected signals come from first order phase transitions beyond the Standard Model of particle physics, and from the subsequent generation of topological defects, including the irreducible SGWB from any network of cosmic defects. Forecasts about  the detection of the isotropic monopole contribution  have been  performed in previous publications of the LISA Cosmology Working group:
for inflationary scenarios in \cite{Bartolo:2016ami},  for phase transitions in \cite{Caprini:2015zlo,Caprini:2019egz}, and for cosmic strings in \cite{Auclair:2019wcv}. 
 
 All such backgrounds are also expected to display anisotropies (direction dependence) in the GW energy density $\Omega_{\rm GW} (f, \hat{n})$, which can be generated either at the time of their production ~\cite{Bethke:2013aba, Bethke:2013vca,Ricciardone:2017kre, Geller:2018mwu, Bartolo:2019zvb, Adshead:2020bji, Malhotra:2020ket} or during their propagation in our perturbed universe~\cite{Contaldi:2016koz,Bartolo:2019oiq,Bartolo:2019yeu, Domcke:2020xmn}. For this reason anisotropies in the SGWB energy density can  be considered  as a new tool to characterize and distinguish various generation mechanisms of primordial GW. At the same time, they can be considered as a probe of the evolution of cosmological perturbations.\\
 As shown in \cite{ Alba:2015cms,Contaldi:2016koz, Bartolo:2019oiq, Bartolo:2019yeu}, SGWB anisotropies show strong analogies with those of the Cosmic Microwave Background (CMB), at least in the geometrical optics limit~\cite{Dodelson:2003ft,Bartolo:2006cu,Bartolo:2006fj,DallArmi:2020dar}. For this reason they can be treated using the Boltzmann equation approach, i.e.,  computing and evolving the equation for the gravitons distribution function $f$ in a perturbed FLRW background, analogously to what is done for CMB photons. At zeroth order in the perturbations, the isotropy and homogeneity of the background imply that the graviton distribution depends only on time and on their frequency. The gravitons propagate freely, and their physical momentum redshifts during the propagation, as CMB photons. There is however a marked difference between the graviton and the photon distribution, namely the initial population of gravitons is not expected to be thermal, as we have in mind specific production mechanisms, such as inflation \cite{Barnaby:2010vf,Cook:2011hg}, phase transitions \cite{Geller:2018mwu}, or enhanced density perturbations leading to primordial black holes (PBH) \cite{Bartolo:2018evs,Bartolo:2018rku,Bartolo:2019zvb}, occurring at energy densities much smaller to what be required for the thermalization of the produced gravitons. This induces a sort of `memory' of the initial state in the distribution. 
 
 The production mechanism could occur inhomogeneously in the observed universe, in a way that correlates to the large scale perturbations. This would result in an anisotropic signal arriving on Earth.  Besides this initial condition, an additional anisotropic contribution is induced by the GW propagation in our perturbed universe. Working at the linearized level in a regime of a large hierarchy $q \gg k$ between the GW (comoving) momentum $q$ and the (comoving) momentum $k$ of the large scale perturbations, the graviton propagation  is affected by a Sachs-Wolfe (SW) effect, which is dominating on large scales, and by an Integrated Sachs-Wolfe (ISW), similarly to CMB photons. An important difference with respect to the CMB photons is associated with the `decoupling' time of the two species: while the CMB temperature anisotropies are generated only at the last scattering surface, or afterward, the universe is instead transparent to GWs at all energies below the Planck scale. For this reason, the SGWB provides a snapshot of the universe right after inflation, and its anisotropies retain precious information about the primordial cosmological evolution.
 
 The Boltzmann equation for the graviton distribution function $f ( x^{\mu}, p^{\mu})$, with  $x^{\mu}$ the graviton position  and $p^{\mu} = d x^{\mu}/d \lambda$ its momentum, is given by 
\begin{equation}
\label{Boltzgeneric}
\mathcal{L}[f] = \mathcal{C}[f(\lambda)] + \mathcal{I}[f(\lambda)]\,,
\end{equation}  
where  $\mathcal{L}\equiv d/ d\lambda$ is the Liouville operator, while $\mathcal{C}$ and $ \mathcal{I}$ account, respectively, for the collision of GWs along their path, and for their emissivity from cosmological and astrophysical sources~\cite{Contaldi:2016koz}.  In the case of a cosmological SGWB, the emissivity term can be treated as an initial condition on the GW distribution, while, as we will see in section \ref{sec:astro}, in the case of an astrophysical background it is related to the astrophysical process that generate the GW signal at various redshifts, such as the black hole merging. On the other hand, we disregard the GW collision term since  it affects  the distribution  at higher orders  in an expansion series in the gravitational strength $1/M_{\rm Pl}$, where $M_{\rm Pl}$ is the Planck mass. We assume that our universe is well described by a perturbed FLRW metric
\begin{equation}
\label{metric}
ds^2=a^2(\eta)\left[
-e^{2\Phi} d\eta^2+(e^{-2\Psi}\delta_{ij}+ h_{ij}) 
dx^i dx^j\right]\, ,
\end{equation}
where $a(\eta)$ is the scale factor as a function of the conformal time  $\eta$, $\Phi$ and $\Psi$
scalar fluctuations, and $h_{ij}$ the transverse-traceless 
tensor fluctuations. We can then solve the Boltzmann equation \eqref{Boltzgeneric}, at background and linear levels. The background  Boltzmann equation simply reads $\partial \bar{f}/ \partial \eta=0$, and it is solved by any distribution that is function only of the comoving momentum $q$, namely $f = {\bar f} \left( q \right)$.
This implies that the physical momentum of the individual gravitons redshifts proportionally to $1/a$. \\
At linearized  level, the evolution equation for $f$ becomes \cite{Contaldi:2016koz, Bartolo:2019oiq, Bartolo:2019yeu}
\begin{equation}
\frac{\partial f}{\partial \eta}+
n^i \, \frac{\partial f}{\partial x^i} +
\left[  \frac{\partial \Psi}{\partial \eta} - {\hat n}^i \, \frac{\partial \Phi}{\partial x^i} + \frac{1}{2}  {\hat n}^i {\hat n}^j \frac{\partial  h_{ij} }{\partial \eta} \right] q \,   \frac{\partial f}{\partial q} = 0 \,, 
\label{Dfc2}
\end{equation}
where ${\hat n}^i = {\hat q}^i$ is the direction of motion of the gravitons.  
The distribution function $f$ is related to the GW energy density by 
\begin{eqnarray}
\rho_{\rm GW} \left( \eta_0 ,\, \vec{x} \right) &=& \frac{1}{a_0^4}  \int d^3 q \, q \, f \left( \eta_0 ,\, \vec{x} ,\, q ,\, {\hat n} \right) 
\equiv \rho_{{\rm c},0} \, \int d \ln q \; \Omega_{\rm GW} \left( \eta_0, \vec{x} ,\, q  \right) \,,
\label{rho-GW}
\end{eqnarray}
where we  use the spectral energy density  $\Omega_{\rm GW}$ introduced in Eq. \eqref{def1omGW}, which  depends also on the  position $\vec{x}$ where the energy density is evaluated.
 The suffix $0$ indicates a quantity evaluated today. We can account for a possibly anisotropic dependence  by defining the quantity $\omega_{\rm GW}$ through 
 \begin{equation}
 \label{deflom}
 \Omega_{\rm GW}( \eta_0 ,\, \vec{x} ,\, q)\, = \,\int d^2 {\hat n} \, \omega_{\rm GW} ( \eta_0 ,\, \vec{x} ,\, q ,\, {\hat n} )/4\pi\,,
 \end{equation}
 and then the bar quantity $\bar \Omega_{\rm GW}( \eta_0, \, q)$
  is defined as spatial average (over the evaluation point $\vec{x}$) of the above quantity $\Omega_{\rm GW} \left( \eta_0 ,\, \vec{x} ,\, q \right)$. With these ingredients we can introduce
 the density contrast 
\begin{equation}
\label{gwdensitycontrast}
\delta_{\rm GW} \left( \eta_0 ,\, \vec{x} ,\, q ,\, {\hat n} \right) 
\equiv  \frac{\delta \omega_{\rm GW} ( \eta_0 ,\, \vec{x} ,\, q ,\, {\hat n} )}{\bar \Omega_{\rm GW} ( \eta_0 ,\, q ) } \equiv \frac{\omega_{\rm GW} ( \eta_0 ,\, \vec{x} ,\, q ,\, {\hat n} ) -\bar{\Omega}_{\rm GW}(\eta_0,q)}{\bar{\Omega}_{\rm GW}(\eta_0,q)}\, ,
\end{equation}
where the homogeneous and isotropic fractional energy density is obtained from the zeroth order distributions function ${\bar f}$. 

We decompose, as for the CMB, the density contrast in spherical harmonics, 
\begin{equation}
\delta_{\rm GW} \left( \eta_0 ,\, \vec{x} ,\, q ,\, {\hat n} \right)  = \sum_\ell \sum_{m=-\ell}^\ell \delta_{\rm GW,\ell m} \left( \eta_0 ,\, \vec{x} ,\, q \right) \, Y_{\ell m} ( {\hat n} )\,,
\label{dec}
\end{equation}
and, under the assumption of statistical isotropy, we define the multipole coefficients through 
\begin{equation}
\left\langle \delta_{\rm GW,\ell m} \delta_{\rm GW,\ell' m'}^* \right\rangle = 
C_\ell^{\rm GW} \left( \eta_0 ,\, q \right) \delta_{\ell \ell'} \, \delta_{m m'} \,. 
\label{Cell-def}
\end{equation} 

As shown in \cite{Contaldi:2016koz, Bartolo:2019oiq, Bartolo:2019yeu}, it is useful to re-define the graviton distribution function as $\delta f \equiv  - q \, \frac{\partial {\bar f}}{\partial q} \, \Gamma \left( \eta ,\, \vec{x} ,\, q ,\, {\hat n} \right), $ to simplify the first order Boltzmann equation, that now in Fourier space reads\footnote{In the CMB case $\Gamma_{\rm CMB} = \delta\,T /T$. } 
\begin{equation}
\Gamma'+ i \, k \, \mu\, \Gamma =  \Psi' - i k \, \mu \, \Phi -  \frac{1}{2}n^i n^j  \, h_{ij}' \, ,
\label{Boltfirstgamma1}
\end{equation}
where the terms on the right hand side (rhs)  define the so-called source function $S (\eta, \vec{k}, {\hat n})$, prime denotes a derivative with respect to conformal time, and $\mu$ is the cosine of the angle between $\vec{k}$ and ${\hat n}$. The GW density contrast is related to the $\Gamma$ and to the background energy density fractional contribution $\bar{\Omega}_{\rm GW}$~\cite{Bartolo:2019oiq, Bartolo:2019yeu},
\begin{equation}
\delta_{\rm GW}  = \left[ 4 -  \frac{\partial \ln \, {\bar \Omega}_{\rm GW}\left( \eta_0 ,\, q \right)}{ \partial \ln \, q }\right] \, \Gamma \left( \eta_0 ,\, \vec{k} ,\, q ,\, {\hat n} \right)\,,
\label{delta-Gama}
\end{equation}
where we recall that $\vec{q} = q {\hat n}$ is the graviton comoving momentum. Many of the cosmological GW scenarios mentioned above have a GW spectrum well described by a simple power law in frequency (i.e., $\bar{\Omega}_{\rm GW}\propto q^{n_T}$). In these cases the previous relation reduces  to $\delta_{\rm GW}=(4-n_T)\Gamma$, where $n_T$ is the tensor spectral index.\\
  The solution of the Eq. \eqref{Boltfirstgamma1} can be decomposed as 
  \begin{equation} 
  \label{dec}
\Gamma \left( \eta ,\, \vec{k} ,\, q ,\, {\hat n} \right) =  \Gamma_I \left( \eta ,\, \vec{k} ,\, q ,\, {\hat n} \right) +  \Gamma_S \left( \eta ,\, \vec{k} ,\,  {\hat n} \right) +  \Gamma_T \left( \eta ,\, \vec{k} ,\,  {\hat n} \right) \;, 
\end{equation}
where $I$, $S$, and $T$ stand for {\it Initial}, {\it Scalar} and {\it Tensor} sourced terms respectively. The scalar and tensor terms correspond to the induced anisotropies arising from the propagation of GWs in a background with large-scale perturbations, and they are therefore ubiquitous for all the cosmological (and astrophysical) sources. On the contrary, the {\it initial} term is related to the initial anisotropy contribution, and it is therefore dependent on the specific mechanism for the GW production (as we review in the next sections, it can for instance arise from large scalar-tensor-tensor or tensor-tensor-tensor primordial non-Gaussianity, or in the case of preheating). 

Inserting the three terms of (\ref{dec}) into (\ref{delta-Gama}), and expanding in spherical harmonics, one obtains the Initial, Scalar, and Tensor contributions to the correlators 
\begin{equation} 
C_{\ell}^{\rm GW} = C_{\ell,I}^{\rm GW} \left( q \right) + C_{\ell,S}^{\rm GW} + C_{\ell,T}^{\rm GW} \;, 
\label{Cell}
\end{equation} 
which evaluate to~\cite{Bartolo:2019oiq, Bartolo:2019yeu}  
\begin{eqnarray}
C_{\ell,I}^{\rm GW} \left( q \right)  &=&  4 \pi \, \left( 4 -  \frac{\partial \ln \, {\bar \Omega}_{\rm GW}}{\partial \ln \, q } \right)^2 \, \int \frac{d k}{k} \,  \left[  j_\ell \left(k \left( \eta_0 - \eta_{\rm in} \right) \right) \right]^2 
\,  P_{I} \left( q ,\, k \right) \,, \nonumber\\ 
C_{\ell,S}^{\rm GW}  &=&  4 \pi \, \left( 4 -  \frac{\partial \ln \, {\bar \Omega}_{\rm GW}}{\partial \ln \, q } \right)^2 \, \int \frac{dk}{k} \,  {\cal T}_\ell^{\left( S \right) \,2} \left( k ,\, \eta_0 ,\, \eta_{\rm in} \right) 
\, P_{\zeta} \left( k \right)  \;, \nonumber\\ 
C_{\ell,T}^{\rm GW}  &=&  4 \pi \, \left( 4 -  \frac{\partial \ln \, {\bar \Omega}_{\rm GW}}{\partial \ln \, q } \right)^2  \, \int \frac{d k}{k} \, {\cal T}_\ell^{\left( T \right) \,2}   \left( k ,\, \eta_0 ,\, \eta_{\rm in} \right) \sum_{\lambda=\pm2}  P_\lambda \left( k \right)   \;, 
\label{Cell-res}
\end{eqnarray}
where $P_I$, $P_\zeta$, and $P_\lambda$ are, respectively, the power spectrum of the initial condition term, of the scalar primordial density perturbations, and of the tensor priomordial modes with helicity $\lambda$~\cite{Bartolo:2019oiq, Bartolo:2019yeu}. Moreover, $j_{\ell}$ are spherical Bessel functions, while the expressions for the scalar and tensor transfer functions are
\begin{eqnarray}
{\cal T}_\ell^S \left( k ,\, \eta_0 ,\, \eta_{\rm in} \right) &\equiv& T_\Phi \left( \eta_{\rm in} ,\, k \right) \, j_\ell \left( k \left( \eta_0 - \eta_{\rm in} \right) \right) + \int_{\eta_{\rm in}}^{\eta_0} d \eta' \, \frac{\partial \left[ T_\Psi \left( \eta ,\, k \right) +  T_\Phi \left( \eta ,\, k \right) \right] }{\partial \eta} \, j_\ell \left( k \left( \eta - \eta_{\rm in} \right) \right) \,,\nonumber\\ 
{\cal T}_\ell^T \left( k ,\, \eta_0 ,\, \eta_{\rm in} \right) &\equiv&  \sqrt{ \frac{\left(\ell +2 \right)!}{\left( \ell - 2 \right)!}} \, \frac{1}{4} \int_{\eta_{\rm in}}^{\eta_0} d \eta \, \frac{ \partial \chi \left( \eta ,\, k \right)}{\partial \eta} \,  \frac{j_\ell \left( k \left( \eta_0 - \eta \right) \right)}{k^2 \left( \eta_0 - \eta \right)^2}\,, 
\label{transf}
\end{eqnarray} 
where $T_\Phi$ and $T_\Psi$ encode the evolution of the scalar perturbations in Eq. (\ref{metric}) in terms of the primordial variable $\zeta$, namely $\Phi \left( \eta ,\, \vec{k} \right) \equiv T_\Phi \left( \eta ,\, k \right) \zeta \left( \vec{k} \right)$, and $\Psi \left( \eta ,\, \vec{k} \right) \equiv T_\Psi \left( \eta ,\, k \right) \zeta \left( \vec{k} \right)$. Analogously, the mode function $h \left( \eta ,\, k \right)$ encodes the time dependence of the tensor perturbations~\cite{Bartolo:2019oiq, Bartolo:2019yeu}. As we discuss below, the spherical harmonic coefficients also have a non-vanishing three point correlation function, that can be related to the primordial bispectrum of the initial condition term and of the primordial scalar and tensor modes~\cite{Bartolo:2019oiq, Bartolo:2019yeu}.

\subsection{Production mechanisms}
\subsubsection{Inflation}

Inflation, a period of accelerated expansion in the very early universe, stands as one of the main pillars of our  understanding of the universe origin and evolution. Primordial quantum fluctuations, magnified by the expansion, provide the seeds for structure formation. The minimal (and observationally viable) implementation of the inflationary mechanism, comprises of a single scalar field slowly rolling down its potential, at an energy scale $E\sim \sqrt{M_P H}$, where $M_p$ and $H$, denote, respectively, the Planck mass and the energy scale during inflation. It is generally assumed that general relativity is the theory of gravity at this energy scale. Upon considering perturbations around a homogeneous and isotropic solution, it becomes clear that tensor fluctuations in the gravity sector, i.e. gravitational waves, are a universal prediction of inflation.\\ The existence of a cosmological stochastic gravitational wave background (SGWB) can be tested across a wide range of scales, from its effects on the CMB B-mode polarisation, all the way to direct detection via laser interferometers. In what follows, we shall focus on the latter possibility and clarify how anisotropies in the GW energy density, imprinted at the epoch of the SGWB generation, may directly probe inflationary dynamics.

These anisotropies, encoded in the first contribution $\Gamma_I (\eta_{\rm in},k,q)$ in Eq.~(\ref{dec}),  carry the imprints from the initial conditions because the Universe is essentially transparent to GWs.~This is to be compared to CMB photons for which  anisotropies at the initial epoch ``$\eta_{\rm in}$'' are erased by the multiple collisions photons suffer prior to the recombination epoch. We stress that, interestingly, anisotropies due to initial condition are  strongly model dependent and thus provide the opportunity to test and distinguish among different inflationary models. To give one example, in the case of single-field adiabatic initial conditions (and for scale-invariant primordial gravitational waves) one would get, in the language of Eq.~(\ref{dec}):
\begin{eqnarray}
\label{da2p8}
\Gamma \left( \eta_{\rm in} ,\, k \right)= - \frac{1}{2} \Phi(\eta_{\rm in}, k)\; ,
\end{eqnarray}
where $\Phi(\eta_{\rm in}, k)$ is the gravitational potential perturbation (in Poisson gauge), see~\cite{Bartolo:2019oiq,Bartolo:2019yeu,Alba:2015cms,Ricciardone:2021kel}.

Anisotropies provide a precious handle on the particle content of the very early Universe. We would like now to single out the two necessary conditions underlying the effectiveness of anisotropies specifically as a probe of inflationary interactions: (i) naturally, a primordial GW spectrum amplitude at small scales that is well-above the sensitivity curve of laser interferometers such as LISA; (iia) a sufficiently sizeable long-short mode coupling (i.e. squeezed primordial non-Gaussianity)~\cite{Dai:2013kra,Bartolo:2019oiq,Bartolo:2019yeu,Bartolo:2019zvb,Dimastrogiovanni:2021mfs}, or (iib) an anisotropic background tout court~\cite{Bartolo:2019oiq,Bartolo:2019yeu}.  

Each of the property in (i) and (ii) are unlikely to characterize single-field slow-roll (SFSR) models  of inflation. Indeed, the typical frequency profile of SFSR realisations is that of a slightly red-tilted GW spectrum, with a signal below the LISA sensitivity threshold\footnote{Noteworthy exceptions include models where an attractor phase is preceded by non-attractor evolution, see e.g. \cite{Ozsoy:2019slf}.}. Non-Gaussianities associated to the same SFSR paradigm are also small. Remarkably, there is a growing literature on multi-field inflationary realisations that comply with both requirements. Interesting examples of anisotropies induced by primordial non-Gaussianities include those occurring in models with light spin-2 field(s) during inflation \cite{Iacconi:2019vgc,Iacconi:2020yxn} and set-ups with a non-standard symmetry breaking patterns (see e.g.  \cite{Endlich:2012pz,Endlich:2013jia,Celoria:2021cxq}). For examples of anisotropies engendered by an anisotropic background we refer the reader to \cite{Bartolo:2019oiq,Bartolo:2019yeu}, where the case of GWs sourced by gauge fields in axion inflation is discussed. This set-up leads to anisotropies with a significant frequency dependence, in contradistinction to what happens for CMB photons.

A general treatment of anisotropies from initial (i.e. inflationary) conditions is made possible by the Boltzmann equation and the theoretical framework expounded in section \ref{theoretical_framework}. In the remainder of this subsection, we shall describe and highlight the importance of anisotropies as a probe of primordial non-Gaussianities in the sense of (iia) defined above. We will put aside (iib) as well as assume, and later quantify, a sufficiently large primordial bispectrum so as to render the anisotropy via long-short mode coupling the leading contribution. It is convenient, before elaborating on the explicit form of non-Gaussianities-induced anisotropies, to make contact with the form they take in the context of the Boltzmann treatment. The effect of a squeezed scalar-tensor-tensor (STT) primordial bispectrum on GW anisotropies is captured by the ``initial conditions'' term $\Gamma_I$ in Eq.~(\ref{dec}) via:
\begin{eqnarray}
\label{dictionary}
\Big[4-\frac{\partial\, {\rm ln}\, \bar{\Omega}_{\rm GW}(q)}{\partial\, {\rm ln}\,q}  \Big] \Gamma_{I}(\eta_{\rm in}, {\bf k}_L,q,\hat{n})= F_{\rm NL}({\bf k}_L,{q})\zeta({\bf k}_L) \;, 
\end{eqnarray}
where ${\bf k}_L$ underscores the specific bispectrum configuration (squeezed) under scrutiny and  $F_{\rm NL}$ is a placeholder for primordial non-Gaussianity of the STT type. An analogous relation exists for anisotropies induced by TTT-type correlators, i.e. GW non-Gaussianities. 

The anisotropies of the GW energy density induced by, respectively, squeezed STT and TTT non-Gaussianity, have the following form \cite{Jeong:2012df,Dimastrogiovanni:2019bfl,Adshead:2020bji}:
\begin{eqnarray}
\label{ani-inf-stt}
  \delta_{\rm GW}^{\rm STT}(q,\hat{n}) = \int_{k_L\ll q}\frac{d^3k_L}{\left(2\pi\right)^3}\,e^{-i  d\, \hat{n}\cdot{\bf k}_L}  F^{\text{STT},\,sq}_{\rm NL}({\bf k}_L,{\bf q})  \zeta({\bf k}_L)\;,
\end{eqnarray}
and
\begin{eqnarray}
\label{ani-inf-ttt}
  \delta_{\rm GW}^{\rm TTT}(q,\hat{n}) = \int_{k_L\ll q}\frac{d^3k_L}{\left(2\pi\right)^3}\,e^{-i  d\, \hat{n}\cdot{\bf k}_L}  F^{\text{TTT},\,sq}_{\rm NL}({\bf k}_L,{\bf q})  \sum_{s}\gamma^s({\bf k}_L)\epsilon_{ij}^{s}(\hat{k}_L)\hat{n}^i \hat{n}^j\;,
\end{eqnarray}
where $d=\eta_{\rm 0}-\eta_{\rm in}$ is the elapsed from horizon re-entry to the present for the mode $q$, and the non-linearity parameters have been defined as

\begin{eqnarray}
\label{ani-inf-stt_fnl}
F^{\text{STT},\,sq}_{\rm NL}({\bf k}_L,{\bf q})  \equiv \frac{B_
   {\rm STT}^{sq}({\bf k}_L,{\bf q}-{\bf k}_L /2, -{\bf q}-{\bf k}_L /2)}{P_{\zeta}(k_L)P_{\gamma}(q)},
\end{eqnarray}

\begin{eqnarray}
\label{ani-inf-ttt_fnl}
  F^{\text{TTT},\,sq}_{\rm NL}({\bf k}_L,{\bf q})  \equiv \frac{B_
   {\rm TTT}^{sq}({\bf k}_L,{\bf q}-{\bf k}_L /2, -{\bf q}-{\bf k}_L /2)}{P_{\gamma}(k_L)P_{\gamma}(q)},
\end{eqnarray}
and the bispectra $B^{sq}$ are understood as defined in standard fashion from the squeezed limit of the three-point function in Fourier space.

\noindent The bispectrum component that appears in Eqs. (\ref{ani-inf-stt})-(\ref{ani-inf-ttt}) is the leading physical contribution to the three-point functions. It is often the case that those bispectrum diagrams that include interactions mediated by additional (w.r.t. the single-field slow-roll case) fields  give the largest contribution in terms of non-Gaussianities, squeezed or otherwise.

In order to identify the regime where non-Gaussianities provide the leading contribution to anisotropies, it suffices to report here that, schematically:
\begin{eqnarray}
\delta^{\rm STT}_{\rm GW} \sim F^{\text{STT},\,sq}_{\rm NL}\times  \sqrt{A_{S}} \; , \qquad \qquad \delta^{\rm TTT}_{\rm GW} \sim F^{\text{TTT},\,sq}_{\rm NL}\times \sqrt{ r\,A_{S}}\; ,
\end{eqnarray}
where $A_S$ is the amplitude of the primordial scalar power spectrum and $r$ is the tensor-to-scalar ratio. The regimes of interest are then, respectively, those where the conditions $F^{\text{STT},\,sq}_{\rm NL}\gg 1$ and $F^{\text{TTT},\,sq}_{\rm NL}\sqrt{r} \gg1$ hold true. It is instructive to recall, for illustrative purposes, the analytical approximation to the angular power spectrum of STT-induced anisotropies:
\begin{eqnarray}
\label{auto_aniso_prim}
C_\ell^{GW,{\rm STT}} = \left( F^{\text{STT},\,sq}_{\rm NL} \right)^2 \frac{2\pi A_S}{\ell (\ell+1)}\; ,
\end{eqnarray}
which has been obtained under the simplifying assumptions of a direction-independent, scale-invariant, $F^{\text{STT},\,sq}_{\rm NL}$ as well as a scale-invariant $P_{\zeta}$. Note that, in the large $F^{\,sq}_{\rm NL}$ limit, due diligence requires that one implements the constraints on the same quantities available at CMB scales.

The dependence of certain contributions to anisotropies on primordial scalar modes (as e.g. Eqs.~(\ref{da2p8}) and (\ref{dictionary}) indicate), provide the intriguing opportunity of cross-correlation with CMB temperature anisotropies. Naturally the latter are also dependent on scalar perturbations, as e.g. the following expression, obtained in the Sachs-Wolfe limit, indicates \cite{Dodelson:2003ft}:  
\begin{eqnarray}
\delta^{\rm SW}_{\ell m}=\frac{4\pi}{5}i^{\ell}\int \frac{d^3p}{(2\pi)^3}Y^*_{\ell m}(\hat{p}) j_{\ell}(p\,r_{\rm lss})\,\zeta({\bf p}) \;. 
\end{eqnarray}
We refer the interested reader to the literature in \cite{Adshead:2020bji,Malhotra:2020ket,Ricciardone:2021kel,Braglia:2021fxn,Dimastrogiovanni:2021mfs} for a thorough treatment of the topic. We find it worthwhile to briefly mention the following notion. In the case of primordial non-Gaussianity, the effectiveness of cross-correlations as a tool to constrain the non-linearity parameter hinges on two independent aspects: the amplitude and the angular dependence of the bispectrum. For example, a quadrupolar angular dependence cross-correlated with temperature anisotropies may well be suppressed with respect to the case of a monopolar $\delta_{\rm GW}$.
 
\subsubsection{Preheating and phase transitions}

In standard preheating scenarios, a daughter or 'preheat' field $\chi$ is coupled to an inflaton $\phi$ via some interaction term involving the two fields. If the inflaton potential exhibits a monomial shape at the stages following inflation, the inflaton oscillates around the minimum of its potential after inflation, inducing a non-adiabatic time evolution in the interactive mass of the preheat field. This leads to an efficient resonant production of the daughter species~\cite{Traschen:1990sw,Kofman:1994rk,Shtanov:1994ce,Kofman:1997yn,Greene:1997fu}, the efficiency of which depends on the inflaton-daughter coupling, as well as on the details of the inflaton potential (see e.g.~\cite{Amin:2014eta,Figueroa:2016wxr} for more recent analysis). This particle production mechanism is known as {\it parametric resonance}, and it corresponds to a non-perturbative, non-linear, and out-of-equilibrium effect. We speak about {\it broad} resonance when the choice of interaction and inflationary model leads to an excitation of the $\chi$ field modes within broad band(s) of momenta. In this case, a significant production of gravitational waves (GWs) takes place~\cite{Khlebnikov:1997di,Easther:2006gt,GarciaBellido:2007af,Dufaux:2007pt,Figueroa:2017vfa}.

In large field inflationary models, the daughter field is typically 'heavy' during inflation, as the inflaton field takes super-Planckian amplitudes. It is possible however, to find some coupling values for which the daughter field is light during most of the inflationary era, but becomes heavy only towards the last $e$-foldings of inflation (when the inflaton rolls down its potential towards smaller values). In this case, after inflation ends, $\chi$ displays amplified perturbations on super-horizon scales, just as the inflaton. At the onset of preheating, sub-horizon vacuum fluctuations serve as an initial condition for parametric resonance, but these are super-imposed over almost homogeneous values $\chi_{\rm i}$ of the daughter field\footnote{Such initial values are actually constant over regions that extend beyond the Hubble radius, as they are generated by super-Hubble fluctuations. The super-Hubble scale at which $\chi_{\rm i}$ varies spatially depends on the modelling, and it is determined essentially by the number of $e$-folds during which $\chi$ remains light.}. This is precisely the crucial ingredient for the development of anisotropies in the GW background. The value of $\chi_{\rm i}$ changes at super-horizon scales according to a variance $\sigma_\chi^2 \sim {H_{\rm inf}^2\over 4\pi^2}\Delta N$, where $\Delta N$ is the number of $e$-folds for which $\chi$ is a light degree of freedom, and $H_{\rm inf}$ is the inflationary Hubble scale. Initial quantum fluctuations of the daughter field $\chi$ at sub-horizon scales are exponentially stimulated via parametric resonance. When non linearities become relevant in the system, i.e.~when $\chi$ back-reacts on the inflaton $\phi$, the dynamics of the sub-horizon modes $\chi_k$ are influenced by the value of $\chi_{\rm i}$ within each given patch. The spatial distribution of the field $\chi$, and hence of the source of the GWs, will be then different at causally disconnected regions. As a result, a different amount of GWs is produced at each super-horizon region, in correspondence with the different values of $\chi_{\rm i}$. 

The anisotropies in the GW energy density spectrum from preheating have been studied in detail in the scale invariant model $V (\phi) = \frac{1}{4} \lambda \phi^4 + {1\over 2}\phi^2\chi^2$~\cite{Bethke:2013aba,Bethke:2013vca}, chosen because of its computational convenience. GW anisotropies should be however a relatively common phenomenon arising in other preheating scenarios, as long as the appropriate conditions are met. In the mentioned scenario, the lightness of $\chi$ before the last $e$-folds of inflation is guaranteed if the coupling constant is taken to be $g^{2}/\lambda \sim \mathcal{O}(1)$. The dynamics of preheating proceeds as usual, but the initial conditions at the onset of parametric resonance are such that at each super-horizon volume there are different values $\chi_{\rm i}$, drawn from a Gaussian distribution with variance $\sigma_{\chi_{\rm i}}^2 \sim {H_{\rm inf}^2\over 4\pi^2}\Delta N$. In practice one just needs to run simulations with free values of $\chi_{\rm i}$, simply restricted to $\chi_{\rm i} > H_{\rm inf}/2\pi$. 

\begin{figure}[t!]
	\centering
	\includegraphics[width=0.45\columnwidth]{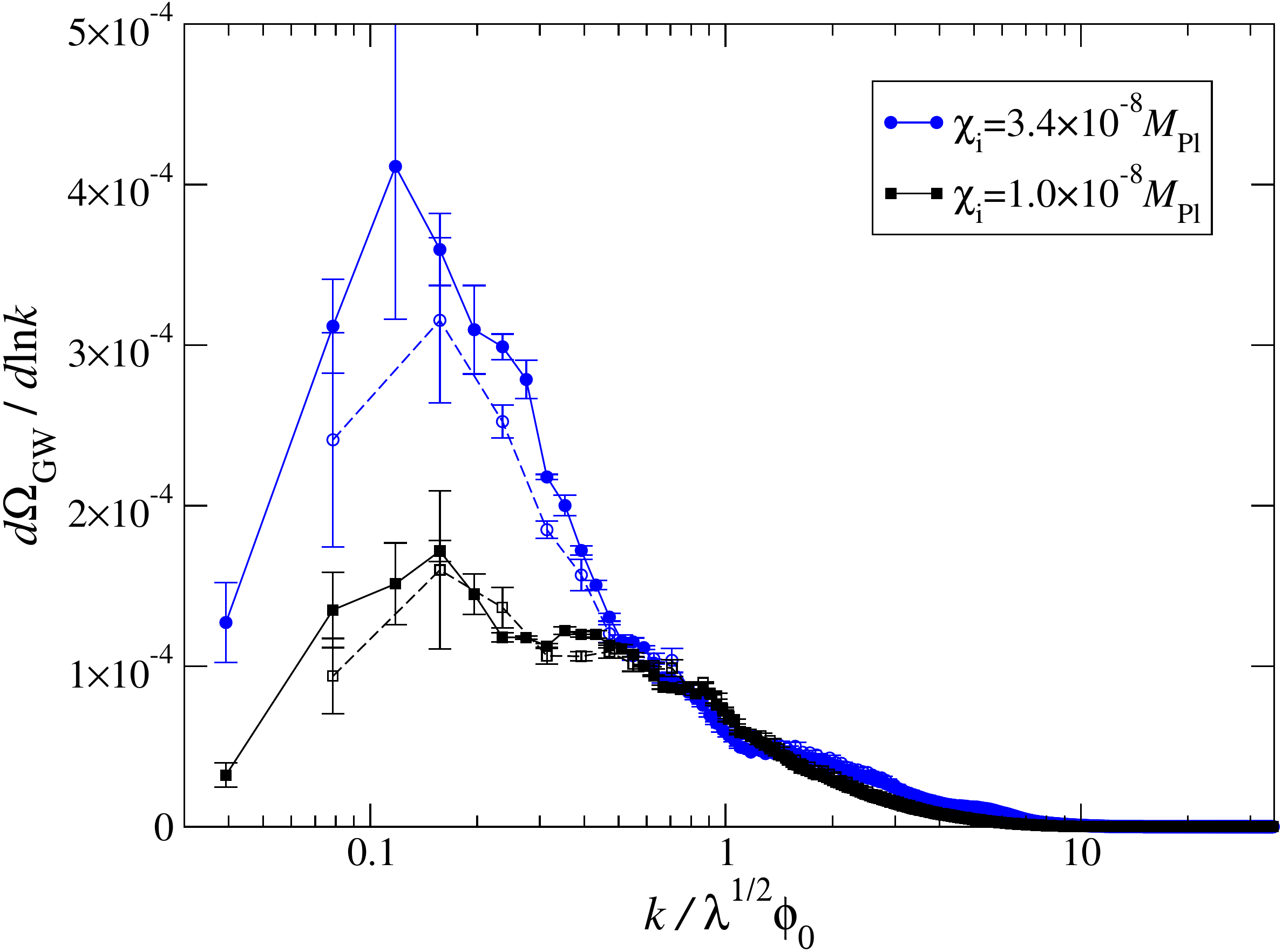}
	\hspace{.2 cm} 
	\includegraphics[width=0.45\columnwidth]{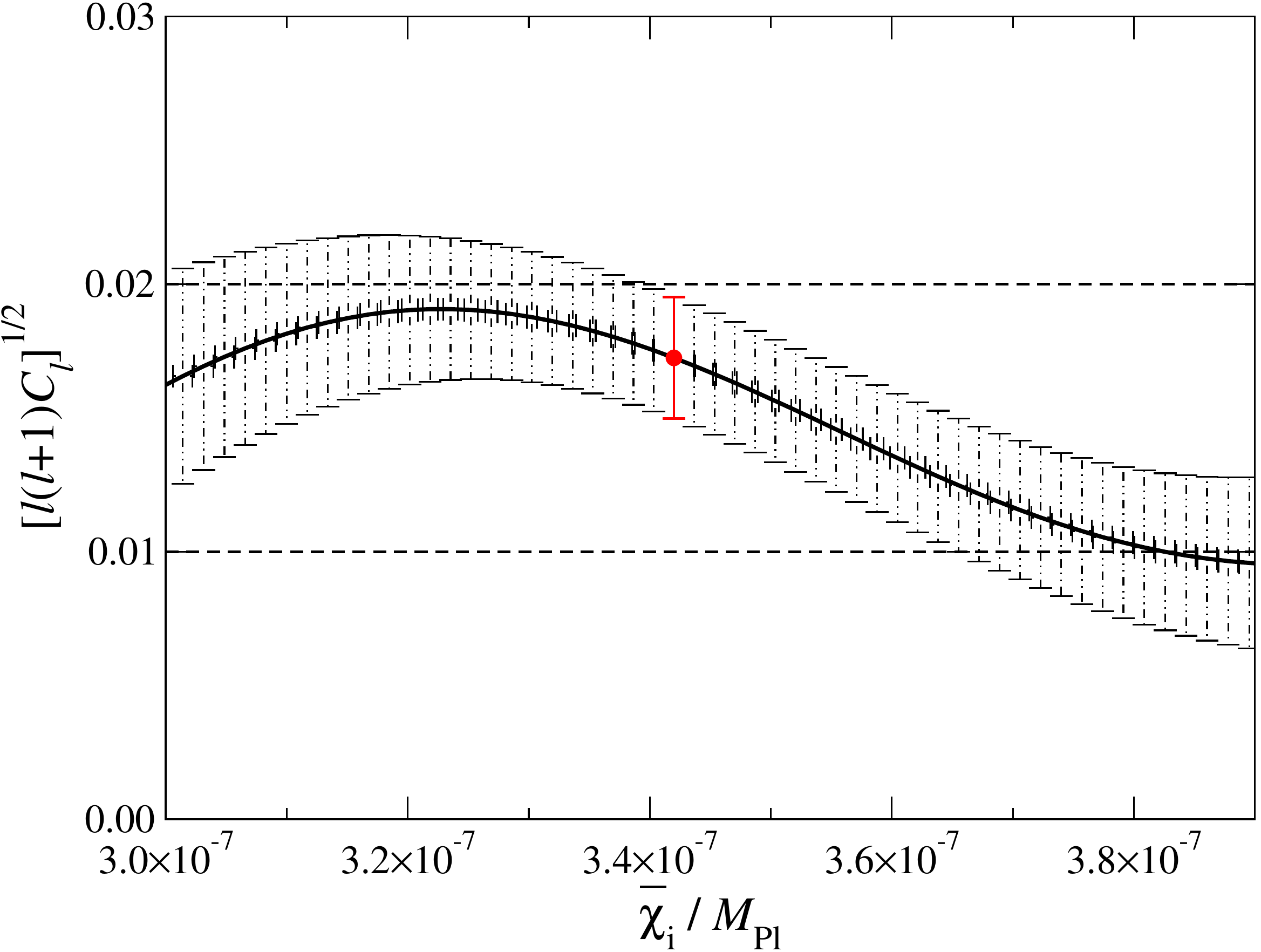}
	\caption{{\it Left: Energy density spectrum of GW background from preheating for $\chi_{\rm i} = 3.4\times 10^{-8}M_{\rm Pl}$ (upper, blue curves) and  $\chi_{\rm i}=1.0\times 10^{-8}M_{\rm Pl}$  (lower, black curves), averaged over five random realizations of the initial sub-horizon fluctuations of $\chi$ (dashed and solid lines are simulations with $N = 512$ and $N = 1024$ points per dimension, respectively). Right: Relative amplitude of angular power spectrum of the GW background from preheating as a function of the average field value $\overline{\chi}_i$. The red dot shows the amplitude for the reference value $\overline{\chi}_{\rm i}=3.42\times 10^{-7}M_{\rm Pl}$. Both plots are taken from Ref.~\cite{Bethke:2013vca}.}}
	\label{fig:anisotropiesPreH}
\end{figure}

Employing the `separate Universe' approach, Refs.~\cite{Bethke:2013aba,Bethke:2013vca} compared the peaks of the GW energy density spectrum from simulations with different initial values of $\chi_{\rm i}$, run for the choice $g^{2}/\lambda =2$. While the GW backgrounds were always peaked at the same frequency, as expected, the peak amplitudes of the GW spectra differed significantly. For example, in the left panel of Fig.~\ref{fig:anisotropiesPreH} we show two GW spectra obtained for slightly different values of $\chi_{\rm i}$, and it is clearly appreciated that one amplitude is larger than the other by a factor $\sim 2-3$. In other words, the actual value of $\chi_{\rm i}$ influences the evolution of the sub-horizon gradients of $\chi$, and hence the production of GWs. To be concrete, $\Omega_{\rm GW}$ was observed to vary up to a factor $\sim 5$ between slightly different values of $\chi_{\rm i}$ (the non-linear dynamics is actually chaotic~\cite{Bond:2009xx}, so small variations of $\chi_{\rm i}$ can lead to a large variation of sub-horizon dynamics of the modes $\chi_k$). The level of anisotropy produced in the energy density of the resulting GW background is characterized by the angular power spectrum $C_\ell^{GW}$ of the relative GW spectral energy-density fluctuation [c.f.~Eq.~(\ref{Cell-def})], which can be written as a function of the $\chi_{\rm i}$ values. A general formula applicable to all scenarios characterized by a light spectator field during inflation is~\cite{Bethke:2013aba,Bethke:2013vca}~\footnote{From Eqs.~(31), (33) and (34) of \cite{Bethke:2013vca} one can verify that the quantities $C_\ell^{\rm GW}$ entering in this relation coincide with those defined here in Eq.~(\ref{Cell-def}).} 
\begin{equation}
\ell\left(\ell+1\right)C_\ell^{\rm GW}=\frac{H_{\rm inf}^{2}}{8\pi}\frac{\langle \delta \chi_{\rm i}\,\Omega_{\rm GW}(\chi_{\rm i})\rangle^{2}}{\sigma_{\chi_{\rm i}}^{4}\langle\Omega_{\rm GW}\rangle^{2}}\,,
\end{equation}
where $\delta \chi_i \equiv \chi_i - \overline{\chi}_i$, with $\overline{\chi}_i$ the mean value over the
currently observable universe. This implies that the angular power spectrum of the GW energy density anisotropy is scale invariant, i.e.~characterised by a {\it plateau} at small multi-poles, $\ell\left(\ell+1\right)C_\ell^{GW} \propto const.$, analogous to the large angular scale Sachs-Wolfe {\it plateau} for the temperature anisotropies in the CMB. In the analysed preheating scenario, the relative amplitude of the GW energy density spectrum, for a reference value of $\overline{\chi}_i = 3.42 \cdot 10^{-7} M_{\rm Pl}$ (here $M_{\rm Pl} \simeq 1.22\times10^{19}$ GeV is the Planck mass), was found to have spatial fluctuations as $\sqrt{l\left(l+1\right)C_\ell^{\rm GW}} = 0.017 \pm 0.003$. For other values of $\overline{\chi}_i$ the anisotropy amplitude is also similar, always at the $\mathcal{O}(1)\,\%$ level, see right panel of Fig.~\ref{fig:anisotropiesPreH}. For comparison, recall that the relative amplitude of the CMB temperature fluctuations is of the order of $\mathcal{O}(0.001)\,\%$. The GW anisotropies obtained in this model are therefore very large. 

The details of the GW anisotropy, if ever observed, could provide a powerful way to differentiate between different inflationary and preheating sectors. The GW background from scale invariant preheating just discussed, is however peaked at very large frequencies~\cite{Figueroa:2017vfa}, way above the observational frequency window accessible to LISA (or to any other ground- or space-based planned detector for this matter). So the example mentioned only serves as a proof of principle, at least for what can be detected in the foreseeable future. 

The mechanism just described corresponds to the imprint of intrinsic anisotropies in the energy density of the GW background from preheating. However, in general, other effects causing anisotropy can be also present. As a matter of fact, any causally sourced GW background will exhibit, in general, anisotropies in the spatial distribution of its energy density at large scales. This is simply due to Doppler, Sachs-Wolfe, and Integrated Sachs-Wolfe effects~\cite{Contaldi:2016koz}, similarly as the anisotropies arising in the photons of the CMB. This type of anisotropies concern actually not only preheating, but also GW backgrounds from cosmological phase transitions, and in general from any causally driven mechanism creating GWs after inflation (as well as inflationary GWs themselves). For cosmological adiabatic perturbations, the fluctuations in any causally produced GW background will simply follow those in the CMB, and hence they are expected to be very small~\cite{Geller:2018mwu,Kumar:2021ffi}, of the order of $\sim 10^{-5}$. If primordial fluctuations carry however an isocurvature component, this need no longer be true. Ref.~\cite{Kumar:2021ffi} has recently shown that in non-minimal inflationary and reheating settings leading to large non-Gaussian perturbations, a non-Gaussian GW background is also expected, even when the rest of the cosmological
fluids inherit predominantly Gaussian fluctuations. Primordial isocurvature perturbations can survive in the GW background say from a cosmological phase transition, exhibiting significant non-Gaussianity, while
obeying observational bounds from the CMB or Large-Scale Structure surveys. Probing such inherited non-Gaussianity in causally generated GW backgrounds seems to be however a marginal
possibility at LISA~\cite{Kumar:2021ffi}, and rather more futuristic proposals such as the detectors DECIGO or BBO are needed.

\subsubsection{Topological defects}

Gravitational wave sources with an inhomogeneous spatial distribution would lead to anisotropies in the SGWB, in addition to the anisotropies induced by the nature of spacetime along the line of propagation of the GWs. 
An inhomogeneous distribution of cosmic strings, 
formed generically \cite{Jeannerot:2003qv} in the early Universe as a result of a phase transition, followed by a spontaneous symmetry breaking characterized by a vacuum manifold with non-contractible closed curves, will lead to anisotropies in the SGWB.

Several studies \cite{Jenkins:2018lvb,Kuroyanagi:2016ugi,Olmez:2011cg} in the literature have calculated the anisotropies induced by a network of Nambu-Goto cosmic string loops, addressing the question of whether the model for the loop distribution will affect the angular power spectrum.
It has been shown that while the amplitude of the resulting power spectrum $C_\ell$ depends on the model of the loop network,
 the anisotropies are driven by local Poisson fluctuations in the number of loops, and the resulting angular power spectrum is spectrally white (i.e., $C_\ell=\text{constant}$ with respect to $\ell$), regardless of the particular loop distribution \cite{Jenkins:2018lvb}.

\begin{figure}
    \centering
    \includegraphics[width=0.7\textwidth]{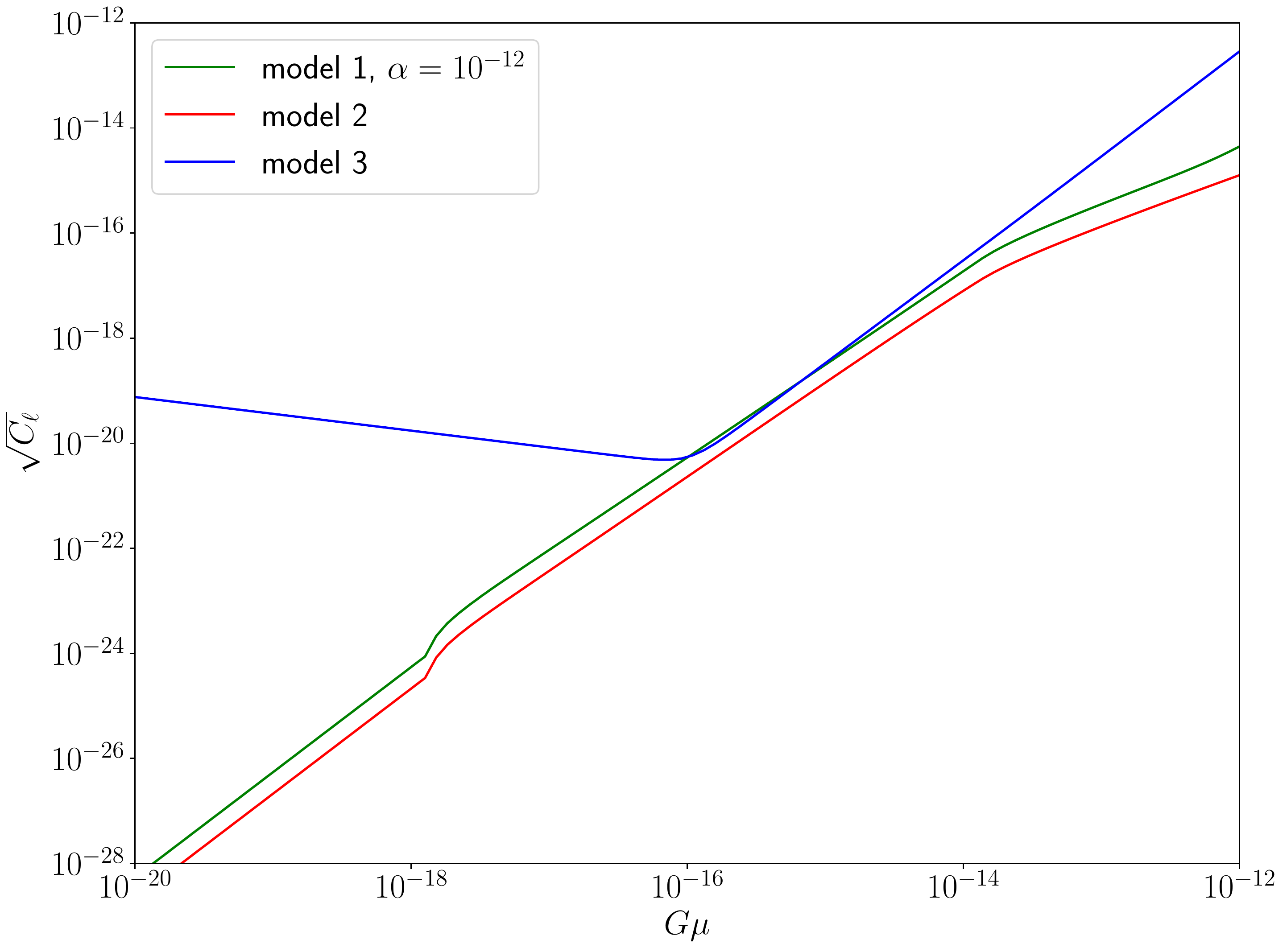}
    \caption{\it Amplitude of the SGWB anisotropies for different cosmic string network models, as a function of the string tension. We use a representative LISA-band GW frequency of $1~\mathrm{mHz}$. Note that the spectra here are not normalised with respect to the monopole, so $\sqrt{C_\ell}$ is proportional to $\Omega_\mathrm{\rm GW}$. As discussed in the text, the spectra are $\ell-$independent.}
    \label{fig:C_ell-cosmic-strings}
\end{figure}

We show in Fig.~\ref{fig:C_ell-cosmic-strings} the amplitude of the SGWB angular power spectrum as a function of the string tension $G\mu$ for three cosmic string loop distributions, dubbed ``Model 1, 2, 3".
The first, Model 1, is the original one-scale model where all loops have the same size set by a free parameter $\alpha$, chosen here to be $\alpha=10^{-12}$.
While this model is rather obsolete, we illustrate it here since 
it has been shown that it leads to significant anisotropies in the PTA frequency band~\cite{Kuroyanagi:2016ugi}.
Models 2~\cite{BlancoPillado:2011dq,Blanco-Pillado:2013qja} and 3~\cite{Ringeval:2005kr,Lorenz:2010sm} are based on different computer simulations and they differ on the way they model the production and cascade of loops from the super-horizon cosmic string network.

We find that, regardless of the adopted cosmic string loop model and the considered string tension, the predicted angular power spectrum $C_\ell$ is too small to be detected with LISA.
Note that in both models 2 and 3, the monopole should be detectable by LISA for $G\mu\gtrsim10^{-17}$~\cite{Auclair:2019wcv} (though the presence of astrophysical foregrounds reduces the sensitivity somewhat to $G\mu\gtrsim10^{-16}$~\cite{Boileau:2021gbr}).

\begin{figure}
    \centering
    \includegraphics[width=0.7\textwidth]{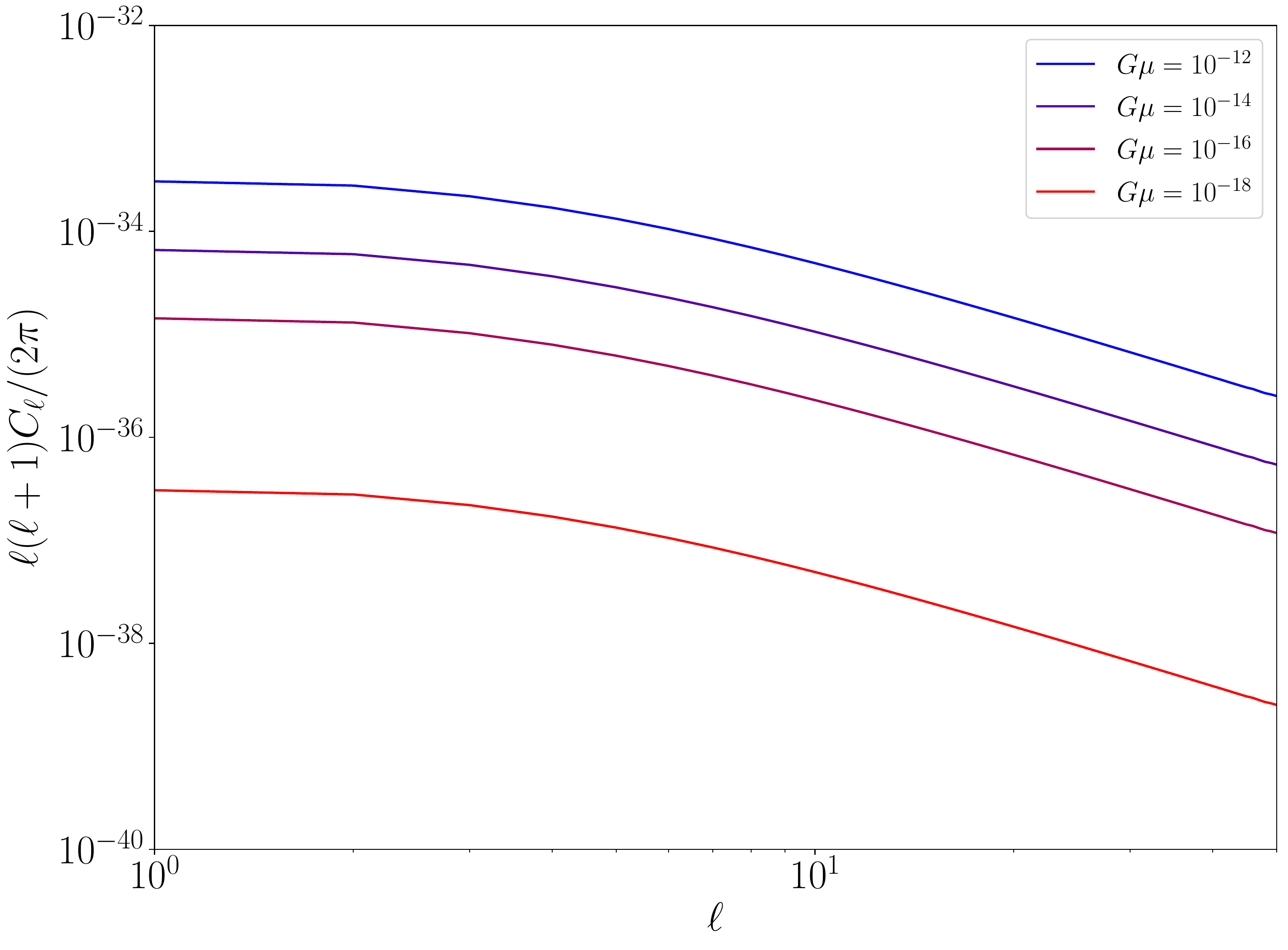}
    \caption{\it Angular power spectrum from a hypothetical population of cosmic string loops in the Milky Way halo.}
    \label{fig:C_ell-cosmic-strings-milky-way}
\end{figure}

Aside from the extra-galactic population of cosmic string loops discussed above, several authors have studied the possibility of a population of loops being captured in the halo of the Milky Way~\cite{Chernoff:2009tp,Khakhaleva-Li:2020wbr,Jain:2020dct}.
These loops would then give rise to an anisotropic SGWB signal which would trace the density profile of the galactic halo. 
Using the results from Ref.~\cite{Jain:2020dct} we calculate here the corresponding $C_\ell$ spectrum, which is shown in Fig.~\ref{fig:C_ell-cosmic-strings-milky-way}.
Again, this signal is too weak to be detected by LISA.

\subsubsection{Primordial Black Holes}
In this section we review the amount of  angular anisotropies inherited by the induced 
SGWB from primordial scalar perturbations in the scenario associated to the production of Primordial Black Holes (PBHs), see Ref.~\cite{Bartolo:2019zvb} for details. The standard formation scenario of PBHs requires an enhancement of curvature perturbations at small scales (denoted $\lambda_\PBH \approx 1/k_\PBH$ in this section) producing the collapse of large overdense region in the early (radiation-dominated) universe. This also predicts a copious amount of GWs induced at second order by the same scalar perturbations leading to a potential GW signature of the PBH production~\cite{Acquaviva:2002ud, Mollerach:2003nq, Ananda:2006af, Baumann:2007zm, Saito:2009jt, Garcia-Bellido:2016dkw, Cai:2018dig, Bartolo:2018rku, Bartolo:2018evs, Unal:2018yaa, Wang:2019kaf, Cai:2019elf, DeLuca:2019ufz, Inomata:2019yww, Yuan:2019fwv, Pi:2020otn, Yuan:2020iwf}.
Since the GW emission in this mechanism mostly occurs when the corresponding perturbation scales cross the horizon, one can relate the GWs frequency to the PBHs mass $M_\PBH$ as (see for example \cite{Saito:2009jt})
\begin{equation}
	f \simeq 6  \, \text{mHz} \sqrt{\gamma} \lp \frac{M_\PBH}{ 10^{-12}M_\odot}\rp ^{-1/2},
\end{equation}
where $\gamma$ is an efficiency factor relating the horizon scale and the PBH mass at formation epoch.
Therefore, the SGWB peak frequencies fall within the LISA sensitivity band for PBH masses between around  $M_\PBH \sim  10^{-15 } M_\odot$ and $M_\PBH \sim  10^{-8}  M_\odot$ \cite{Saito:2008jc,Garcia-Bellido:2017aan,Bartolo:2018rku,Cai:2018dig,Unal:2018yaa}.

Following subsection \ref{theoretical_framework}, we adopt the following definition of the line element
\begin{equation}
	\d s^2 = a^2 \left \{ - \lp1+ 2 \Psi  \rp \d \eta ^2 + \llp \lp1 - 2 \Psi   \rp \delta _{ij} +\frac{1}{2} h_{ij} \rrp  \d x^i \d x^j \right \} , 
\end{equation}
in terms of the scalar $\Psi$ and tensor $h_{ij}$ perturbation in the Newtonian gauge, assuming no anisotropic stress.
From the Einstein equation one can write down the equation of motion for the GWs as
\begin{equation}
h_{ij}''+2\mathcal H h_{ij}'-\nabla^2 h_{ij}=-16 \mathcal T_{ij}{}^{ \ell m}
\llp 
\Psi\partial_\ell \partial_m\Psi+2\partial_\ell\Psi\partial_m\Psi-\partial_\ell\left(\frac{\Psi'}{\mathcal H}+\Psi\right)\partial_m\left(\frac{\Psi'}{\mathcal H}+\Psi\right)
\rrp,
\label{eq: eom GW1}
\end{equation}
where the source term on the r.h.s is evaluated assuming a radiation dominated era. The prime denotes derivative with respect to conformal time $\eta$, and $\mathcal H \equiv a'/a$ is the conformal Hubble parameter. 

Using the equations of motion at first order in perturbation theory one can express the scalar perturbation as a function of the gauge invariant comoving curvature perturbation \cite{Lyth:1998xn}.  Employing the standard decomposition of the tensor perturbation in terms of the polarization tensors $e_{ij}^{\lambda}$ and  helicity modes $h_\lambda$, one finds~\cite{Espinosa:2018eve}
\begin{equation} 
h_\lambda ( \eta ,\, \vec{k} ) = \frac{4}{9 k^3 \eta} \int \frac{d^3 p}{\left( 2 \pi \right)^3}  
\, {\rm e}_\lambda^* ( \vec{k} ,\,\vec{p} ) \zeta ( \vec{p} )  \zeta ( \vec{k} - \vec{p} ) 
\left[ {\cal I}_c ( x, y )  \cos \left( k \eta \right) 
+  {\cal I}_s (x, y) 
\sin \left( k \eta \right) \right],
\label{h-sourced}
\end{equation} 
where we have introduced the dimensionless variables $x=p/k$ and $y=|\vec{k}-\vec{p}|/k$, the contracted polarization tensors ${\rm e}_\lambda ( \vec{k} ,\,\vec{p} ) \equiv {\rm e}_{ij,\lambda} ({\hat k} ) \vec{p}_i \vec{p}_j$, and the two oscillating functions ${\cal I}_{c,s}$ \cite{Espinosa:2018eve,Kohri:2018awv}
\begin{align}
\label{eq: Ic, Is tau0=0}
\Ic(x,y) &= -36\pi\frac{(s^2+d^2-2)^2}{(s^2-d^2)^3}\theta(s-1)\ ,\\
\Is(x,y) &= -36\frac{(s^2+d^2-2)}{(s^2-d^2)^2}\left[\frac{(s^2+d^2-2)}{(s^2-d^2)}\log\frac{(1-d^2)}{|s^2-1|}+2\right], 
\end{align}
in terms of $d \equiv |x-y|/\sqrt{3}$, $s \equiv (x+y)/\sqrt{3} $ with $(d,s) \in [0,1/\sqrt{3}]\times[1/\sqrt{3},+\infty)$.

The energy density associated to the gravitational modes is given by \cite{Misner:1974qy,Maggiore:1999vm,Flanagan:2005yc}
\begin{equation}
    \rho_\GW = \frac{M_p^2}{4} \langle \dot{h}_{ab} \left( t ,\, \vec{x} \right)  \dot{h}_{ab} \left( t ,\, \vec{x} \right) \rangle_T,
\end{equation} 
where the angular brackets denotes a time average on a timescale $T$, much smaller than the cosmological timescale ($T H \ll 1$) but much larger than the GW phase oscillations ($T k_i \gg 1$).
Adopting the standard assumption of a Gaussian scalar curvature perturbation $\zeta$, one finds the fractional GW energy density 
\begin{align} 
&\left\langle \rho_\GW \left( \eta ,\, \vec{x} \right) \right\rangle
\equiv \rho_{c,0} ( \eta) \, \int d \ln k \; \Omega_\GW \left( \eta ,\, k \right) \nonumber\\
&=  
\frac{2 \pi^4 M_p^2}{81  \eta^2 a^2}  \, \int \frac{d^3 k_1 d^3 p_1}{\left( 2 \pi \right)^{6} } 
\frac{1}{k_1^4}\, 
\frac{\left[ p_1^2 -  ( \vec{k}_1 \cdot \vec{p}_1)^2/k_1^2 \right]^2}{p_1^3 \, \left\vert \vec{k_1} - \vec{p}_1 \right\vert^3} \, 
{\cal P}_\zeta ( p_1) 
{\cal P}_\zeta( |\vec{k_1} - \vec{p}_1|)   \llp{\cal I}_c^2( \vec{k}_1 ,\, \vec{p}_1) + {\cal I}_s^2( \vec{k}_1 ,\, \vec{p}_1) \rrp,
\label{rho1-par}
\end{align} 
in terms of the critical energy density of a spatially flat universe $\rho_c = 3 H^2 M_p^2$ and the curvature perturbation power spectrum $ {\cal P}_\zeta$.

The predicted amount of angular anisotropies can be determined by the two-point correlation function of the density field $\rho_\GW$ in different angular directions. For a Gaussian curvature perturbation one expects those to be undetectable, given the capability of the LISA experiment to measure anisotropies between spatial points separated by  non-negligible fractions of the present horizon \cite{Bartolo:2019zvb}. Indeed, according to the Equivalence Principle, the anisotropies will be highly suppressed by a factor  $(k_\PBH \vert \vec{x} - \vec{y} \vert )^{-2} \ll 1$,  since the characteristic scales of the scalar perturbations are much smaller than those spatial distances, $k_\PBH \, \vert \vec{x} - \vec{y} \vert \gg 1$, and the emission takes place near horizon crossing.

This conclusion does not hold in the presence of primordial non-Gaussianity correlating short ($\zeta_\PBH$) and long scales ($\zeta_L$). Indeed, large scale modulation of the power spectrum may lead to anisotropies at large-scales imprinted at formation \cite{Bartolo:2019zvb}.
Assuming a local, scale-invariant, shape of non-Gaussianity $\zeta = \zeta_g + \frac{3}{5} f_\NL \, \zeta_g^2$ and keeping into account propagation effects (see Sec.~\ref{sec: propagation effects} for details), one can compute the two-point correlation function of the GW energy density contrast in spherical harmonics $\delta_{\GW,\ell m}$ as in Eq. (\ref{Cell-def}), obtaining 
\begin{equation}
\sqrt{\frac{\ell \left( \ell+1 \right)}{2 \pi} \, C_\ell^{\rm GW} \left( k \right)}  \simeq \frac{3}{5} 
\left\vert 1 + {\tilde f}_{\NL} \left( k \right) \right\vert \, \left\vert  4-\frac{\partial \ln {\Omega}_{\GW} (\eta ,\, k ) }{\partial \ln k} \right\vert \,  {\cal P}_{\zeta_L}^{1/2},
\end{equation}
in terms of the power spectrum at large scales ${\cal P}_{\zeta_L}$ and  the momentum dependent non-Gaussian parameter 
\begin{equation}
{\tilde f}_\NL \left( k \right) \equiv  \frac{8 \, f_\NL}{4-\frac{\partial \ln {\Omega}_{\GW} (\eta ,\, k ) }{\partial \ln k}}.
\label{GammaI-time}
\end{equation}

Non-Gaussianity in the curvature perturbation is constrained to fall in the range $- 11.1 \leq f_\NL \leq 9.3$ at $95\% \, {\rm C.L.}$ \cite{Akrami:2019izv} by the Planck collaboration. It is important to stress that its presence would also generate a significant variation of the PBH abundance on large scales given the impact of long modes inducing a modulation of the power on small scales.
As isocurvature modes in the DM density fluid are strongly constrained by CMB observations \cite{Akrami:2018odb}, 
one can put an upper bound on the  fraction of the Dark Matter (DM) in our universe composed by PBHs formed in the presence of non-Gaussianities \cite{Tada:2015noa,Young:2015kda}.

\begin{figure}[t!]
	\centering
	\includegraphics[width=0.45\columnwidth]{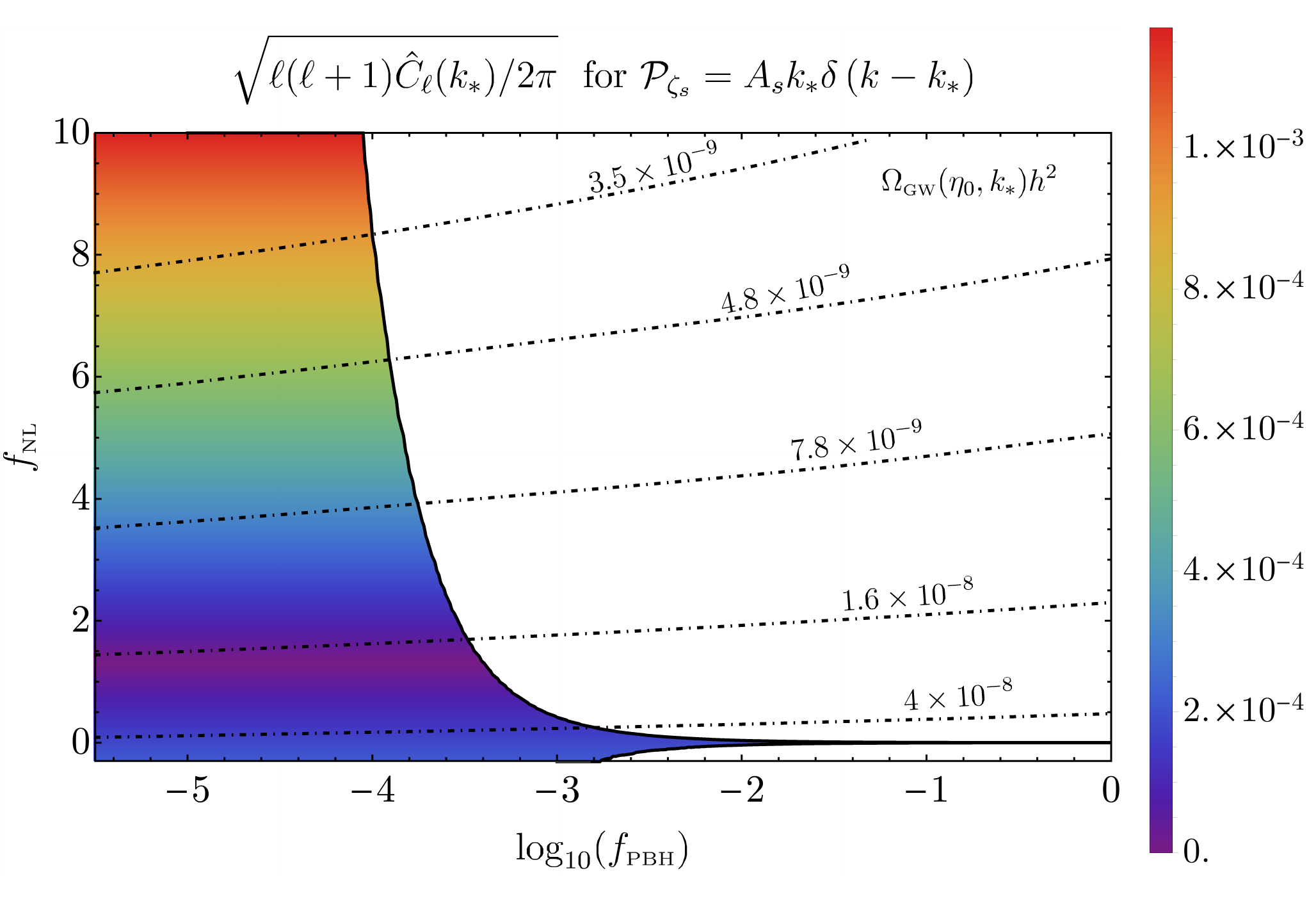}
	\hspace{.2 cm} 
	\includegraphics[width=0.45\columnwidth]{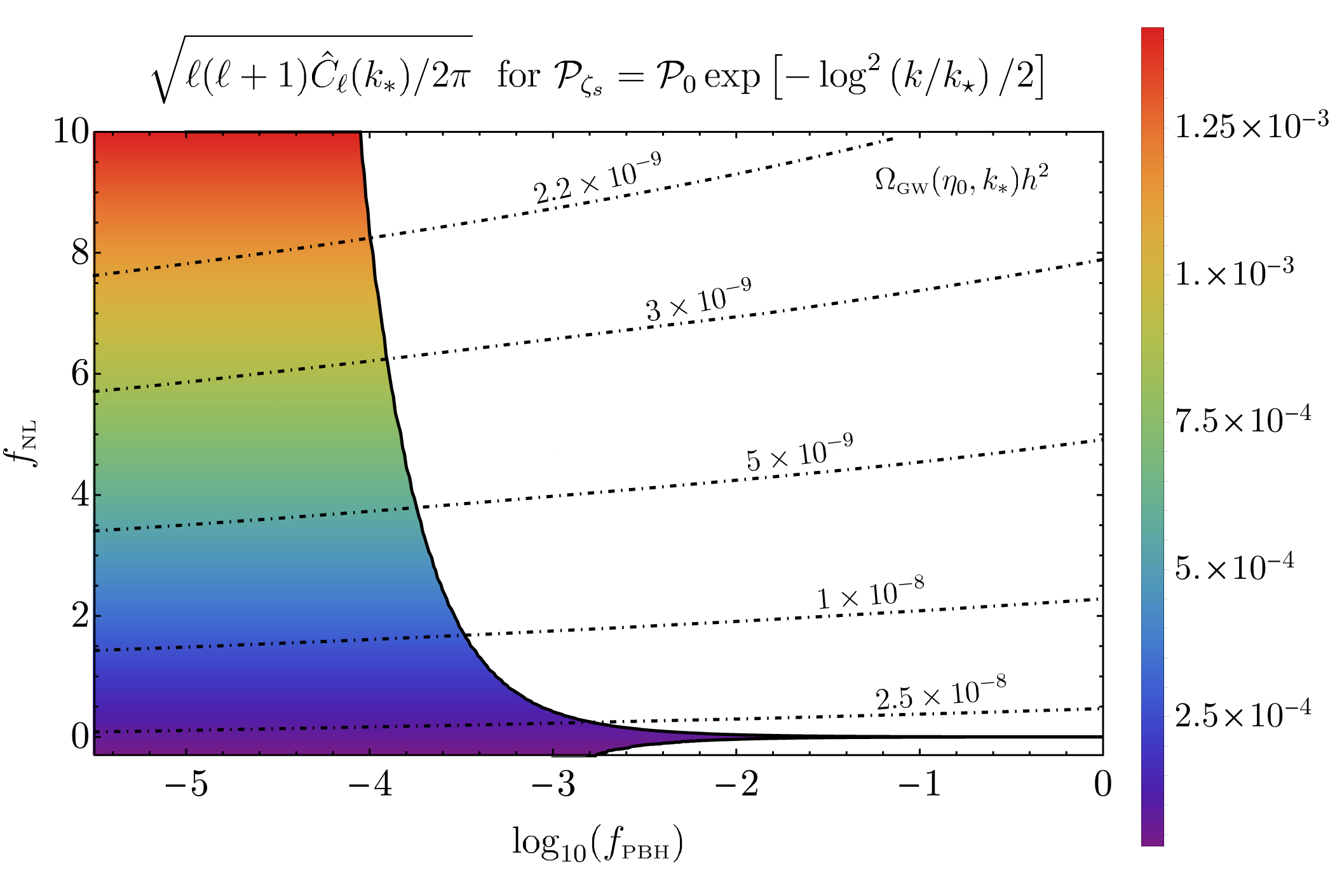}
	\caption{\it Contour plot of the GW anisotropy in the parameter space of $f_\PBH$ and $f_\NL$ allowed by the Planck constraints for a monochromatic (left) and lognormal (right) power spectrum at small scales. 
	We fixed the SGWB characteristic scale  around the maximum sensitivity of LISA. The dot-dashed lines identify the corresponding present day GWs abundance. Figure taken from Ref.~\cite{Bartolo:2019zvb} (with ${\hat C}_\ell \left( k_* \right)$ corresponding to our $C_\ell^{\rm GW}$).
	}
	\label{fig:anisotropiesPBH}
\end{figure}

For a monochromatic and lognormal power spectra of curvature perturbations at small scales, peaked at the LISA maximum sensitivity frequency, the GW anisotropy are plotted in fig.~\ref{fig:anisotropiesPBH}, where the coloured region identifies the range of parameters allowed by the Planck constraints and the dot-dashed lines identify the present epoch GWs abundance evaluated at the peak frequency. The  non-linear parameter has been assumed to be $f_\NL > -1/3$ to avoid the inconsistency of the perturbative approach in the PBH abundance computation happening at larger negative values, see the related discussion in \cite{Young:2013oia,Yoo:2019pma}. The main finding is that 
a large fraction of DM composed by PBHs would imply a highly isotropic and Gaussian SGWB, up to propagation effects. On the other hand, the detection of a sizeable amount of anisotropy in the signal related to the PBH scenario would imply that PBHs can account only for a small fraction of the DM in the universe \cite{Bartolo:2019zvb}.


\subsection{Propagation effects}
\label{sec: propagation effects}
Independently from the initial anisotropies in the SGWB of cosmological origin, we do  expect a minimal level of anisotropies in \emph{all} of the scenarios described above due to the propagation of GWs through (large-scale) cosmic inhomogeneities, while travelling from the generation surface till 
the observation point. Such anisotropies represent an unavoidable, irreducible contribution which indeed carries precious cosmological information, being sensitive to the evolution of cosmological perturbations and to the initial conditions from which cosmic structures formed. Employing the general formalism of Boltzmann equations explained above, the underlying cosmological perturbations leave specific imprints in the statistics of the SGWB anisotropies in terms of, e.g., angular power spectra. 

\begin{figure}[t!]
\centering 
\includegraphics[width=0.48\textwidth]{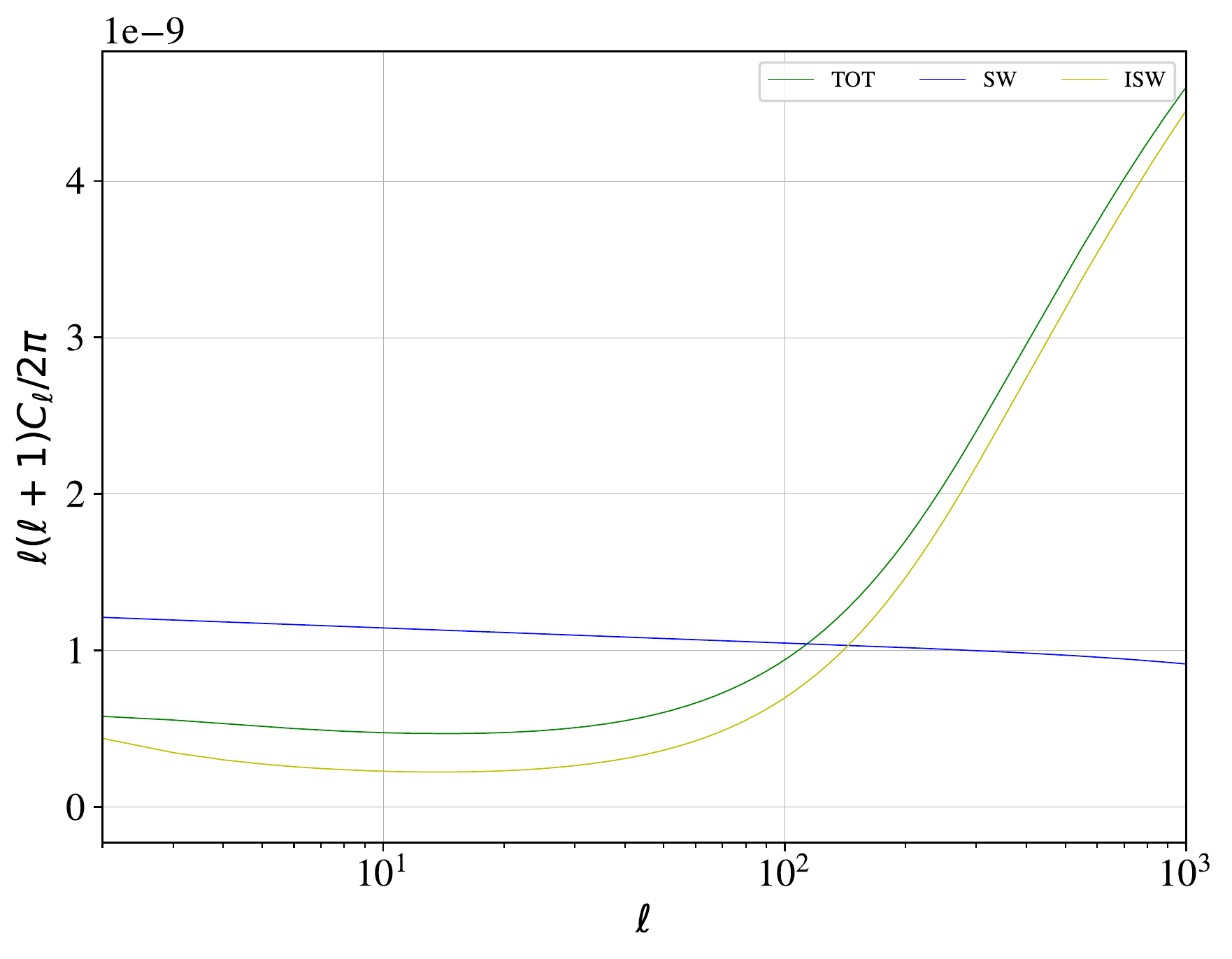}
 \includegraphics[width=0.48\textwidth]{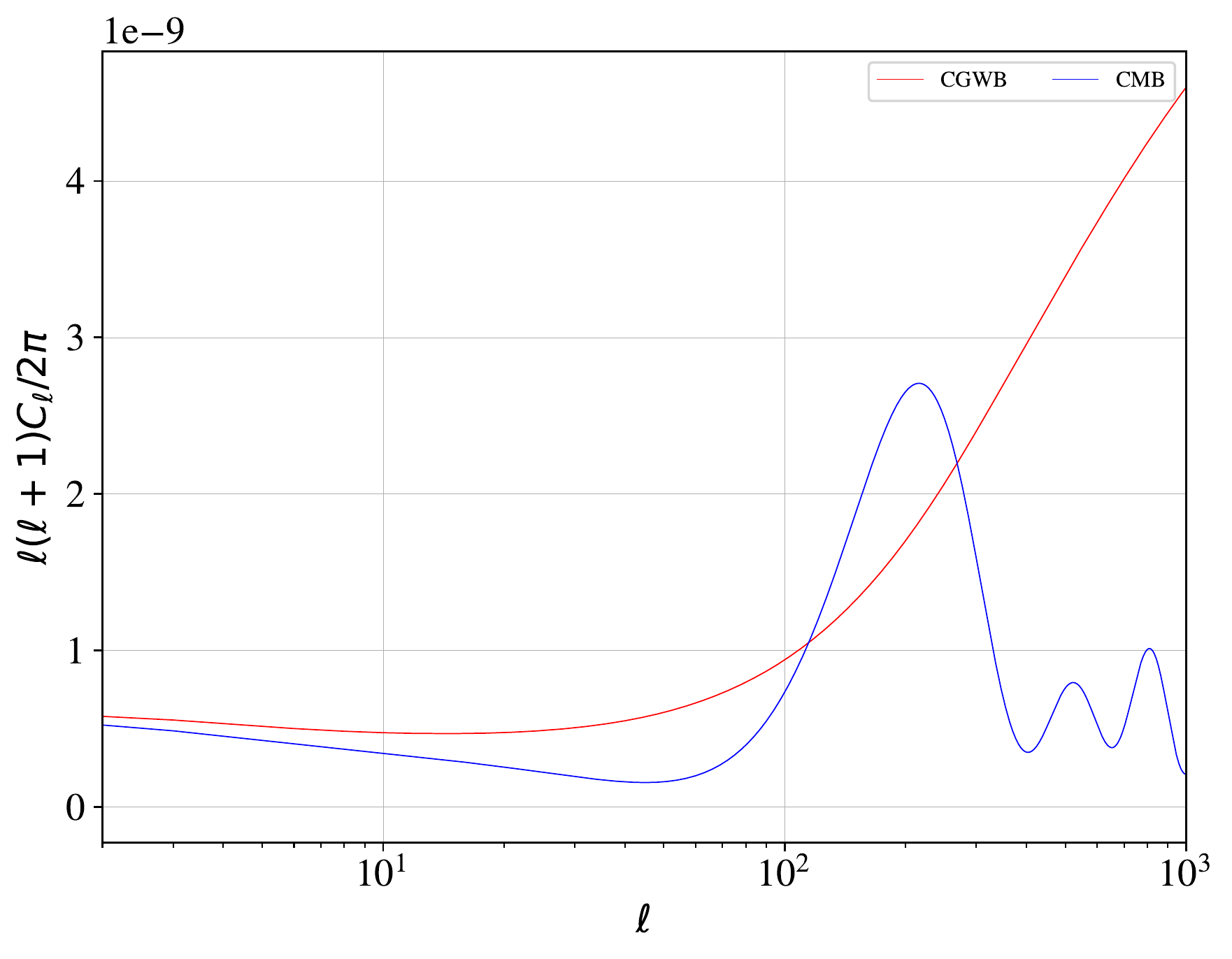}
\vskip -0.4cm
\caption{\it  Left plot: SW, ISW and total contribution to the angular power-spectrum of the cosmological SGWB. Right plot: comparison between the SGWB spectrum and the CMB one.}
\label{SW_figure}
\label{clcgwb}
\end{figure}

From Eqs. \eqref{transf}, we can infer some properties about the SGWB anisotropies { due to their propagation through cosmological perturbations}: similarly to CMB, gravitons are affected by the Sachs-Wolfe contribution, which represents the energy lost by a graviton which escapes from a potential well, and by the Integrated Sachs-Wolfe (ISW) effect, {\bf due to tensor and scalar perturbations, the latter producing} an anisotropy which is roughly proportional to the total variation of the potentials $\Delta \Phi+\Delta\Psi$.  An important point to stress here is the ``initial'' time $\eta_i$, which has an impact both on the SW and on the ISW contributions. The numerical evaluation of the angular power spectrum for the cosmological SGWB has been performed in~\cite{DallArmi:2020dar} (see also~\cite{Braglia:2021fxn}), modifying the publicly available code CLASS, usually employed for the computation of CMB anisotropies \cite{Lesgourgues:2011re} and adapting it to the SGWB.\\
In Fig.~\ref{clcgwb} we report the angular power spectrum of the cosmological SGWB { due to propagation effects sourced by scalar perturbations} and we compared it to the CMB one coming from temperature anisotropies. We can see that the SGWB spectrum shows a larger amplitude compared to the CMB. This can be explained considering the graviton ``decoupling'' time, which occurs earlier compared to CMB photons and so gravitons feel for longer time the propagation effects. In such a figure, we also report the contribution from the SW and the ISW separately, to show their behaviour at different angular scales.

From the left plot we can see that at large angular scales (i.e., low $\ell$),  the SW contribution is dominating while moving to smaller scales (i.e., $\ell  \gtrsim 100$),  the ISW contribution starts to be larger. On the other hand, from the right plot we can quantify the expected difference among the CMB and SGWB anisotropies.

Interestingly enough, by measuring or constraining angular anisotropies of the SGWB, it is also possible to probe the level of primordial non-Gaussianity possibly present both in the scalar and tensor cosmological perturbations through which the SGWB propagates. Indeed such primordial non-Gaussianity will left be imprinted into the GWs passing through the background large-scale underlying inhomogeneities, similarly to what happens  for CMB photons. This entails to go beyond the power spectra statistics and to compute higher-order correlation functions, such as the angular bispectra of the GW energy density fluctuations~\cite{Bartolo:2019oiq,Bartolo:2019yeu}
\begin{eqnarray} 
\left\langle \delta_{\rm GW,\ell_1 m_1}  \delta_{\rm GW,\ell_2 m_2}  \delta_{\rm GW,\ell_3 m_3} \right\rangle \equiv 
\left( \begin{array}{ccc} 
\ell_1 & \ell_2 & \ell_3 \\ 
m_1 & m+2 & m_3   
\end{array} \right) \, b_{\ell \ell' \ell''}^{\rm GW} \;, 
\label{Cell-Bell-def}
\end{eqnarray} 
where $b_{\ell \ell' \ell''}^{\rm GW}$ is the so-called reduced bispectrum (see e.g.~\cite{Komatsu:2001rj, Gangui:1993tt}).  
For example, as shown in~\cite{Bartolo:2019oiq,Bartolo:2019yeu}, in the case of primordial local non-Gaussianity in the curvature perturbations 
\begin{equation}
\zeta \left( \vec{x} \right) = \zeta_g \left( \vec{x} \right) + \frac{3}{5} \, f_{\rm NL} \, \zeta_g^2 \left( \vec{x} \right) \,,
\end{equation} 
$\zeta_g \left( \vec{x} \right)$ being the linear Gaussian part of the perturbation, one finds 
\begin{eqnarray}
b_{\ell_1 \ell_2 \ell_3,S}^{\rm GW} &\simeq&  \frac{2 \, f_{\rm NL}}{4 -  \frac{\partial \ln \, {\bar \Omega}_{\rm GW}}{\partial \ln \, q }} \,  
\left[ C_{\ell_1,S}^{\rm GW} \, C_{\ell_2,S}^{\rm GW} + C_{\ell_1,S}^{\rm GW} \, C_{\ell_3,S}^{\rm GW} + C_{\ell_2,S}^{\rm GW} \, C_{\ell_3,S}^{\rm GW} \right] \,. 
\label{reducedbis}
\end{eqnarray} 

It is important to stress that similar results follow in the case of primordial non-Gaussianity in the large-scale {\it tensor} perturbations. Therefore  
the 3-point correlation function of GW energy density anisotropies provides for the first time a way to probe at interferometers primordial non-Gaussianity of large-scale tensor modes through the imprints that the latter leave in the spatial distribution of GW energy density as described by 
the second equation in~(\ref{transf}). Indeed, for a sufficiently high SGWB, it might happen that primordial (scalar/tensor) non-Gaussianity can be measurable through the detection of the SGWB anisotropies at interferometers, even in cases where such primordial non-Gaussianity are not measurable at CMB scales.

As it is clear from the previous discussion, the seeds that give rise to anisotropies during the propagation, are the same for photons and gravitons. This naturally induces a cross-correlation among these two messengers. Recently, the cross-correlation between CMB and SGWB anisotropies induced during the propagation has been studied in \cite{Ricciardone:2021kel, Braglia:2021fxn}, and focusing on the initial anisotropy in \cite{Adshead:2020bji, Malhotra:2020ket}. All these studies have shown that such a cross correlation signal will be within the reach of the LISA detector, and it will be extremely useful to extract information on cosmological parameters, pre-recombination physics and the non-linear parameter $f_{NL}$ to measure primordial non-Gaussianity.

\section{\sc Astrophysical Sources of Anisotropies}
\label{sec:astro}
 
The astrophysical stochastic gravitational-wave background (AGWB) is generated by the incoherent superposition of signals emitted by a large number of resolved and unresolved astrophysical sources. In  different frequency bands, several astrophysical sources can contribute to the AGWB, as merging stellar-mass black hole (SOBHB) or binary neutron stars (BNS) \cite{TheLIGOScientific:2016wyq, Regimbau:2016ike, Mandic:2016lcn,Bavera:2021wmw, Dvorkin:2016okx, Nakazato:2016nkj, Dvorkin:2016wac, Evangelista:2014oba}, super-massive black hole binaries (SMBHB) \cite{Kelley:2017lek}, rotating neutron stars \cite{Surace:2015ppq, Talukder:2014eba, Lasky:2013jfa}, stellar core collapse \cite{Crocker:2017agi, Crocker:2015taa} and population III binaries \cite{Kowalska:2012ba} (see, e.g., \cite{Regimbau:2011rp} for a review). As the cosmological GW background, also the AGWB is characterized by an isotropic energy density contribution and through the spatial angular power spectrum encoding its anisotropy.

 Based on the recent observations of merging black holes and neutron star binaries by the  LIGO and Virgo detectors~\cite{TheVirgo:2014hva,TheLIGOScientific:2014jea,Abbott:2016blz,Abbott:2016nmj,TheLIGOScientific:2016pea,TheLIGOScientific:2017qsa}, 
 it is estimated that the stochastic background from unresolved stellar-mass compact binaries may be detected within a few years of operation of such a network \cite{Abbott:2021xxi}. 
 Its anisotropic component is constrained by LIGO/Virgo observations up to $\ell=4$ \citep{2019arXiv190308844T} which results in upper limits on the amplitude of the dimensionless energy density per units of logarithmic frequency in the range $\Omega_{\rm GW}(f=25 \text{Hz},\Theta)<0.64-2.47\times 10^{-8}$ sr$^{-1}$ for a population of merging binary compact objects, where $\Theta$ denotes the angular dependence. Methods to measure and map the AGWB in the LIGO and LISA frequency ranges are discussed in \cite{Allen:1996gp, Cornish:2001hg, Mitra:2007mc, Thrane:2009fp, Romano:2015uma, Romano:2016dpx, Renzini:2018vkx, TheLIGOScientific:2016xzw,  Contaldi:2020rht, Alonso:2020mva}. 

Traditionally, the energy density of the AGWB  has been modeled and parameterized under the assumption that both our universe and the distribution of sources are  homogeneous and isotropic (see e.g. Refs.~\cite{Dvorkin:2016okx,Regimbau:2011rp}). This is a rather crude approximation: GW sources are located in galaxies embedded in the cosmic web; moreover, once a GW signal is emitted, it is deflected by the presence of massive structures, such as galaxies and compact objects.  
It follows that the  energy flux from all astrophysical sources has a stochastic, directional dependence. 

The first prediction of the AGWB angular power spectrum was presented in ~\cite{Cusin:2018rsq, Jenkins:2018uac} following the framework introduced in Refs.~\cite{Cusin:2017mjm, Cusin:2017fwz}. This framework is flexible and splits the cosmological large-scale structure and sub-galactic scales so that it can be applied to any source contributions and any frequency band. The astrophysical dependence of the angular power spectrum on the detail of the underlying astrophysical model has been studied in \cite{Jenkins:2018kxc, Jenkins:2018lvb, Cusin:2019jpv,  Cusin:2019jhg, Jenkins:2019uzp, Jenkins:2019nks, Wang:2021djr,Capurri:2021zli} and different formal aspects of the derivation of anisotropies and their interpretation are discussed in \cite{Contaldi:2016koz, Cusin:2018avf, Bertacca:2019fnt, Pitrou:2019rjz, Alonso:2020mva}. 
The relative importance of cosmological and astrophysical effects depends on the frequency band chosen, hence offering the possibility to distinguish different astrophysical processes. Due to their stochastic nature, anisotropies can be statistically characterized in terms of their angular power spectrum  and they also correlate with other cosmological observables such as weak lensing, galaxy number counts and CMB anisotropies.

The study of the cross correlations with electromagnetic observables  provides complementary information and might improve the signal to noise of the anisotropic searches \cite{Cusin:2019jpv, Alonso:2020mva, Yang:2020usq, Mukherjee:2020jxa}. Moreover, cross-correlating the background that collects contribution from sources at all redshifts along the line of sight, with EM observables (such as galaxy number counts) at a given redshift, allows one to get a tomographic reconstruction of the redshift distribution of sources \cite{Cusin:2018rsq, Alonso:2020mva, Cusin:2019jpv,Yang:2020usq}.

\subsection{GW energy density for  astrophysical sources}

The total present-day GW energy density per logarithmic frequency  $f_{\rm o}$ (where $f_{\rm o}=q$, see the previous section) and solid angle~$\Omega_{\rm o}$ along the line-of-sight~$\hat s$ (note that $\hat s=-\hat n$)  of the AGWB is defined as~\cite{Allen:1997ad, Maggiore:1999vm}. 
\begin{equation}
\frac{\omega^{\rm tot}_{\rm GW} \left(f_{\rm o}, \hat s\right)}{4 \pi}= \frac{ f_{\rm o} }{\rho_{{\rm c},0}} \frac{ {\rm d} \rho^{\rm tot}_{\rm GW}}{ {\rm d} f_{\rm o} {\rm d} \Omega_{\rm o} }\left(f_{\rm o}, \hat s\right) \,,
\end{equation}
%
and it represents the fractional contribution of GWs to the critical energy density of the Universe today $\rho_{{\rm c},0} =3H_0^2/(8\pi G)$;  $\rm {d} \rho^{\rm tot}_{\rm GW}$ is the total energy density of GWs   in the frequency interval of today $\{f_{\rm o},f_{\rm o} +{\rm d} f_{\rm o}\}$. See  also the definition in Eq. 
\eqref{deflom}.
Such a quantity contains both a background (monopole) contribution in the observed frame, which is homogeneous and isotropic, i.e. $\bar{\omega}^{\rm tot}_{\rm GW} = \bar{\Omega}^{\rm tot}_{\rm GW}$, and a direction-dependent contribution $ \Delta \omega_{\rm GW}(f_{\rm o}, \hat{s})= \omega_{\rm GW}(f_{\rm o}, \hat{s}) - \bar \omega_{\rm GW}(f_{\rm o})$. 

As usual, we consider {\it the local wave zone} approximation at the source position: in other words, the observer ``at the emitted position" is defined in a region with a comoving distance to the source ``sufficiently large" such that the gravitational field is ``weak enough" but still ``local'', i.e., its wavelength is small w.r.t. the comoving distance from the observer $\chi$ (see also \cite{Bertacca:2017vod, Bertacca:2019fnt}).
Considering an observer that measures a GW signal in a fixed direction $\hat{n}$, one expects that the total gravitational energy density in such a direction is given by the sum of  all the (unresolved) astrophysical contributions along the line of sight contained in a given volume $dV_{\rm e}(\hat{s})$ and can be expressed as    
\begin{equation}
\frac{{\rm d} \rho^{\rm tot}_{\rm GW}}{{\rm d} f_{\rm o} {\rm d} \Omega_{\rm o} }= \frac{{\rm d} \mathcal{E}^{\rm tot}_{\rm GW}}{{\rm d} f_{\rm o} {\rm d} {\cal T}_{\rm o} {\rm d} A_{\rm o} {\rm d} \Omega_{\rm o}} \;,
\end{equation}
from which we can build the  total gravitational energy density $\Omega^{\rm tot}_{\rm GW}=\int {\rm d} \Omega_{\rm o} \, \omega^{\rm tot}_{\rm GW}/4\pi $ where
\begin{equation}\label{GWdensity}
\omega^{\rm tot}_{\rm GW} \left(f_{\rm o}, \hat s \right) =\frac{4\pi f_{\rm o} }{\rho_{{\rm c},0}}\sum_i 
 \int  n_{\rm h}^{[i]}(x_{\rm e}^\alpha, \vec{\theta} )
\frac{{\rm d}  {}{\mathcal{E}^{[i]}_{\rm GW}} [x^{\mu}_{\rm o}, \vec{\theta} (x^{\mu}_{\rm e})]} {{\rm d} f_{\rm o} {\rm d} {\cal T}_{\rm o}  {\rm d} A_{\rm o}}
 \left|\frac{{\rm d} V_{\rm e}}{{\rm d} \Omega_{\rm o} {\rm d} \chi}\right| {\rm d}\chi  \,  {\rm d}\vec{\theta} \;,
\end{equation}
and $[i]$ is related to the summation over all unresolved astrophysical sources that produce the SGWB. Here ${\rm d} A_{\rm o}$ is the unit surface element at observer~\cite{Cusin:2017mjm}.  The vector $\vec{\theta}= \{M_{\rm h}, M^*, \vec{m}, \vec{\theta}^*\}$ represents all the parameters which are: the halo mass $M_h$, the mass of stars that give origin to the sources $M^*$; $\vec{m}$ indicates the masses of the compact objects and $\mathbf{\theta}^*$ includes the astrophysical parameters related to the model (like spin, orbital parameters, star formation rate). In Eq. (\ref{GWdensity}), $n_{\rm h}^{[i]}$ is the (physical) number of halos at a given mass $M_{\rm h}$, within the physical volume ${\rm d} V_{\rm e}$, {\it weighted}  with the parameters $\vec{\theta}$ of the sources at $x_{\rm e}^\mu$. The letter ``${ \rm e}$'' stands for ``evaluated at the emission (source)'' while ``${ \rm o}$'' for ``evaluated at the observer''. Using the energy conservation we have
\begin{equation}
\label{Spectrum}
\frac{{\rm d}  {\mathcal{E}^{[i]}_{\rm GW}}_{\rm o} } {{\rm d}  f_{\rm o} {\rm d} {\cal T}_{\rm o}  {\rm d} A_{\rm o}} 
={1 \over (1+z)^3 {\mathcal D}_{\rm A}^2(z)}  
\frac{{\rm d}  {\mathcal{E}^{[i]}_{\rm GW}}_{\rm e} }{{\rm d}  f_{\rm e} {\rm d} {\cal T}_{\rm e}  {\rm d} \Omega_{\rm e}}\;,
\end{equation}
where we have redefined ${\mathcal{E}^{[i]}_{\rm GW}} [x^{\mu}_{\rm o}, \vec{\theta} (x^{\mu}_{\rm e})]={\mathcal{E}^{[i]}_{\rm GW}}_{\rm o}.$
Here ${\cal T}$ is the proper time of the observer and ${\mathcal D}_{\rm A}$ is  the angular diameter distance. Now, defining the {\it total} GW density as
\begin{equation}
n^{[i]}(x_{\rm e}^\alpha, \vec{\theta} )\equiv n_h^{[i]}(x_{\rm e}^\alpha) \frac{{\rm d} {\mathcal{E}^{[i]}_{\rm GW}}_{\rm e}(z, f_{\rm e}, x^{\mu}_{\rm e}, \vec{\theta} ) } {{\rm d}  f_{\rm e} {\rm d} {\cal T}_{\rm e}  {\rm d} \Omega_{\rm e}} \;,
\end{equation}
we can easily obtain the expression for  the energy density
\begin{equation}\label{rho_GW-2}
\frac{{\rm d} \rho^{\rm tot}_{\rm GW}}{{\rm d}  f_{\rm o} {\rm d} \Omega_{\rm o}} = \sum_{[i]} \int  a(x^0)^2 {n^{[i]}(x_{\rm e}^\alpha,  \vec{\theta} )  \over (1+z)^2} {\rm d} \chi  {\rm d}\vec{\theta} \;.
\end{equation}
Here,  $x^\mu (\chi)$ are the comoving coordinates in the {\it real frame} (the ``physical frame"), where $\chi$ is the comoving distance from the source to the detector.
The previous expression depends on the position, but 
we can also define a position independent, isotropic
version of it by integrating over a spatial volume: we
denote the corresponding quantity ${{\rm d} {\bar \rho}^{\rm tot}_{\rm GW}}/{{\rm d}  f_{\rm o} {\rm d} \Omega_{\rm o}} $ with a bar, as in  section \ref{theoretical_framework}.

We use the {\it observer} frame where we perform observations (also called ``cosmic GW laboratory'' in \cite{Bertacca:2017vod}). This is the correct frame where we want to reconstruct 3D maps/catalogs of  galaxies  by using both EM and GW signals. 
Let us point out that if we use unperturbed coordinates, instead of the observer coordinates, we are not able to interpret correctly the correlation between for istance the AGWB and EM sources from observed galaxies. This could induce a wrong estimate of our results~\cite{Bertacca:2017vod, Bertacca:2019fnt}. 
In particular, we consider coordinates that are flattened in our past gravitational wave cone so that the GW geodesic from the source can be defined with the following conformal space-time coordinates
\begin{equation}
\label{ObGWframe}
\bar{x}^\mu=(\bar \eta,\; \bar {\bf x})=(\eta_0-\bar \chi, \; \bar \chi \, \hat {\bf s}).
\end{equation}
Here, $\eta_0$ is the conformal time today, $\bar \chi(z)$ is the comoving distance to the observed redshift and $\hat {\bold s}$ is the observed direction of the GW, i.e. %
\begin{equation}
\hat s^i={{\bar x}^i \over \bar \chi}=\delta^{ij} {\partial {\bar \chi} \over \partial \bar x^j} \,.
\end{equation} 
Using $\bar \chi$ as an affine parameter in the observer's frame, the total derivative along the past GW-cone is 
\begin{equation}
{{\rm d} \over {\rm d} \bar \chi} = - {\partial \over \partial \bar \eta} + \hat s^i { \partial \over \partial \bar x^i} \,.
\end{equation}

Setting up a mapping between the observed frame and the ``physical frame"  in the following way $ x^\mu(\chi)=\bar{x}^\mu(\bar\chi)+\Delta x^{\mu} (\bar \chi)$, where $\Delta x^{\mu} (\bar \chi)$ is a suitable linear perturbation that shifts the comoving four-coordinates from the real-space frame to the observed frame.  Then using the decomposition of Eq. \eqref{GWdensity}, we obtain
%
\begin{equation} \label{back-rho_GW}
{ \bar \Omega^{\rm tot}_{\rm GW}} =\frac{ f_{\rm o}}{\rho_{{\rm c},0}} \frac{{\rm d} \bar \rho^{\rm tot}_{\rm GW}}{{\rm d}  f_{\rm o}} = \frac{ 4 \pi f_{\rm o}}{\rho_{{\rm c},0}}  \sum_{[i]} \int  {N^{[i]}(z,  f_{\rm e}, \vec{\theta} )\over  (1+z)}  \; {\rm d}  \bar \chi \,  {\rm d} \vec{\theta} \;,
\end{equation}
with $ N^{[i]}(z, f_{\rm e}, \vec{\theta})  = \bar n^{[i]}(z, f_{\rm e}, \vec{\theta}) /  (1+z)^3 $ the {\it total} comoving number density at a given redshift or $\bar \chi$. Notice that, by construction, the quantity $\bar \Omega^{\rm tot}_{\rm GW}$ is isotropic. 
 At linear order, we obtain the following AGWB energy density fluctuation
\begin{eqnarray}\label{DeltaOmega}
 {\Delta \omega^{\rm tot}_{\rm GW} \over 4 \pi} &=& \frac{ f_{\rm o}}{\rho_c}  \sum_{[i]} \int   {N^{[i]}(z, f_{\rm e}, \vec{\theta}\,) \over  (1+z)} \Bigg\{\delta^{[i]} +  \frac{{\rm d} \ln  N^{[i]}  }{{\rm d}  \ln \bar a} \, \Delta \ln a - \left(1 + {{\mathcal H}'\over {\mathcal H}^2} \right) \Delta \ln a +\delta  f \nonumber\\
 &&- \frac{1}{{\mathcal H}}{{\rm d} \Delta \ln a  \over {\rm d} \bar \chi}\Bigg\}  \; {\rm d}  \bar \chi  \,  {\rm d} \vec{\theta}  \;,
 \end{eqnarray}
where
\begin{equation}
\label{a}
\frac{a(\chi)}{\bar a(\bar \chi)} = 1 + \Delta \ln a=1+{\mathcal H}\Delta x^{0 },
\end{equation}
and $\delta  f$ is the linear perturbation of the frequency of the GW due to the anisotropies. 

Finally, let us mention that when the integration along the line of sight is performed, one should also consider the normalized selection window function $w(z)$, whose form depends, besides redshift,  on the sensitivity/characteristics of the GW detector (see \cite{Bertacca:2019fnt} and \cite{Bellomo:2021mer} for more details about the window function). So we finally have
\begin{equation} \label{back-rho_GW-FV}
{ \bar \omega^{\rm tot}_{\rm GW}\over 4\pi} = \frac{ f_{\rm o}}{\rho_{{\rm c},0}}  \sum_{[i]} \int  w(z){N^{[i]}(z,  f_{\rm e}, \vec{\theta} )\over  (1+z)}  \; {\rm d}  \bar \chi \,  {\rm d} \vec{\theta} \;,
\end{equation}
and 
\begin{eqnarray}\label{DeltaOmega-FV}
 \delta^{\rm tot}_{\rm GW} &\equiv&  \frac{\Delta \omega^{\rm tot}_{\rm GW}}{\bar \omega^{\rm tot}_{\rm GW}} 
 = \frac{4 \pi f_{\rm o}}{\bar \Omega^{\rm tot}_{\rm GW} \, \rho_{{\rm c},0}}  \sum_{[i]} \int  w(z) {N^{[i]}(z, f_{\rm e}, \vec{\theta}\,) \over  (1+z)} \Bigg\{\delta^{[i]} + \left[ \frac{{\rm d} \ln  N^{[i]}  }{{\rm d}  \ln \bar a}  - \left(1 + {{\mathcal H}'\over {\mathcal H}^2} \right) \right] \Delta \ln a  \nonumber\\
 && +\delta  f - \frac{1}{{\mathcal H}}{{\rm d} \Delta \ln a  \over {\rm d} \bar \chi}\Bigg\}  \; {\rm d}  \bar \chi  \,  {\rm d} \vec{\theta}  \;.
 \end{eqnarray}

\subsection*{Connection with the Halo and Stellar Mass Function and with the\\ Star Formation Rate}

In general, the isotropic component  of equation~\eqref{back-rho_GW} is given by~\cite{Phinney:2001di}
\begin{equation}
\bar{\Omega}^{\rm tot}_{\rm GW}(f_o) = \sum_{i} \bar{\Omega}^{[i]}_{\rm GW}(f_o) = \frac{4\pi f_{\rm o}}{\rho_{\rm c}}  \sum_i \int {\rm d}z {\rm d}{\bf \theta} ~ p^{[i]}({\vec \theta}\,) {R^{[i]} \over H(z)}  \frac{{\rm d} {\mathcal E}^{[i]}_{{\rm GW}, {\rm e}}}{{\rm d} f_{\rm e} {\rm d} \Omega_{\rm e}},
\label{sgw_background}
\end{equation}
where ${\rm d}  {\mathcal{E}^{[i]}_{\rm GW}}_{\rm e}  / {\rm d}  f_{\rm e} / {\rm d} \Omega_{\rm e} $ is the energy spectrum per unit solid angle, $p^{[i]}({\vec \theta}\,) $ is the probability distribution of the source parameters ${\vec \theta}$ and $R^{[i]}$ is the observed comoving merger rate density of $i$-th unresolved type of source. In particular, the event rate (per unit of redshift) can be derived from the cosmic star formation rate. For instance, assuming for simplicity that the gravitational emission occurs shortly after the birth of the progenitor, it turns out that
\begin{equation}
R^{[i]}=\frac{\lambda^{[i]}(z,\vec{\theta}\,)}{(1+z)} {{\rm d} \rho^{[i]}_* \over {\rm d} {\mathcal T}_{\rm e} }\,,
\end{equation}
 where the $(1+z)$ term corrects the cosmic star formation rate (SFR) by the time dilation due to the cosmic expansion and 
 ${\rm d}  \rho^{[i]}_*  / {\rm d} {\mathcal T}_{\rm e} $ is 
 the (density) cosmic SFR  in $M_\odot$, ${\rm Mpc}^3$ and ${\rm yr}^{-1}$. 
 Here $\lambda^{[i]}(z, \vec{\theta})$ is a generic function which depends on the initial mass function $M^*$ and, in general, on other parameters of the sources, as the halo mass $M_h$. So then we have
 \begin{equation}
 p^{[i]}({\vec \theta}\,) \lambda^{[i]}(z,\vec{\theta}\,){{\rm d} \rho^{[i]}_* \over {\rm d} {\mathcal T}_{\rm e} } =  w(z) N^{[i]}(z, f_{\rm e}, \vec{\theta}) =  w(z) \left[{\bar n_{\rm h}^{[i]}(z, f_{\rm e}, \vec{\theta}) \over (1+z)^3}\right]\frac{{\rm d} {\mathcal{E}^{[i]}_{\rm GW}}_{\rm e}(z, f_{\rm e}, x^{\mu}_{\rm e}, \vec{\theta} ) } {{\rm d}  f_{\rm e} {\rm d} {\cal T}_{\rm e}  {\rm d} \Omega_{\rm e}}\;.
 \end{equation}
 Now, let us consider events with short emission (i.e. {\it{burst sources}}), as merging binary sources (BH-BH, NS-NS and/or NS-BH). Then we have 
 \[ \frac{{\rm d}  {\mathcal{E}^{[i]}_{\rm GW}}_{\rm e} } {{\rm d}  f_{\rm e} {\rm d} {\mathcal T}_{\rm e}  {\rm d} \Omega_{\rm e}} 
 =  { {\rm d}  \over   {\rm d} {\mathcal T}_{\rm e}  }  \left({{\rm d}{ {\mathcal N}^{[i]}_{\rm GW}}_{\rm e} \over {\rm d} M^*}\right)  \frac{{\rm d}  {\mathcal{E}^{[i]}_{\rm GW}}_{\rm e} } {{\rm d}  f_{\rm e} {\rm d} \Omega_{\rm e}}\;,
 \]
 where $ {{\rm d} {{ \mathcal N}^{[i]}_{\rm GW}}_{\rm e}  /   {\rm d} {\mathcal T}_{\rm e}{\rm d} M^* }$ is the merging rate of events for each halo and at a given stellar mass $M^*$,  and the comoving density $\bar n_{\rm h}^{[i]}(z, f_{\rm e}, \vec{\theta}\,) / (1+z)^3$ of the halos can be rewritten as
 \begin{equation}
 {\bar n_{\rm h}^{[i]}(z, f_{\rm e}, \vec{\theta}) \over (1+z)^3 }= {{\rm d}{\bar N}^{[i]}_{\rm h} \over {\rm d} M_{\rm h}}\,,
 \end{equation}
 i.e. the comoving density at a given $M_{\rm h}$. In order to give a very simple example, let us assume that $\bar n_{\rm h}$ and ${\bar N}_{\rm h}$ are equal for all sources. In this case they do not depend only on $M_{\rm h}$ and we can remove the index $[i]$ and 
  $N_{\rm h}(M_{\rm h},z)$ can be related to the fraction of mass $F(M_{\rm h},z)$ that is bound at the epoch $z$ in halos of mass smaller than $M_{\rm h}$, i.e.
  \begin{equation}
 \frac{{\rm d} \bar N_{\rm h} (M_{\rm h},z) }{{\rm d}  M_{\rm h}} = {\bar \rho(z)\over M_{\rm h}}  \frac{{\rm d} F (M_{\rm h}) }{{\rm d}  M_{\rm h}}\;,
  \end{equation}
 where $\bar \rho(z)$ is the comoving background density (e.g., Press \& Schechter (1974) \cite{Press:1973iz},  Sheth \& Tormen  (1999) \cite{Sheth:1999mn} or Tinker (2008) \cite{Tinker:2008ff} mass fraction). Following \cite{Springel:2002ux, Hernquist:2002rg}, it is possible to express the mass function in terms of the multiplicity function of halos $g(M)$
   \begin{equation}
g(M_{\rm h})= \frac{{\rm d} F (M_{\rm h},z) }{{\rm d} \ln M_{\rm h}}\;.
  \end{equation}
Physically, this quantity gives the fraction of mass that is bound in halos per unit logarithmic interval in mass. Finally let introduce the (mean) SFR that it is connected with ${{\mathcal N}^{[i]}_{\rm GW}}_e$ in the following way
  \begin{equation}
{ {\rm d}  \over   {\rm d} {\mathcal T}_{\rm e}  }  \left({{\rm d}{ {\mathcal N}^{[i]}_{\rm GW}}_{\rm e} \over {\rm d} M^*}\right)=   
{{\rm d}{ {\mathcal N}^{[i]}_{\rm GW}}_{\rm e} \over {\rm d} M^*} \times {\rm SFR}\;.
 \end{equation}
Note that $s(M_{\rm h},z)$ defined in \cite{Springel:2002ux, Hernquist:2002rg} can be related to the ${\rm SFR} $ in the following way  $s(M_{\rm h},z)= (M^*/M_h)\times{\rm SFR}$.

\subsection{Projection/propagation effects}

As a first step, we compute the GW density fluctuation in the energy density in a spatially flat FLRW background in the Poisson Gauge 
\begin{equation}
 \label{Poiss-metric}
 {\rm d} s^2 = a(\eta)^2\left[-\left(1 + 2\Phi\right)  {\rm d}  \eta^2+ \delta_{ij} \left(1 -2\Phi \right)  {\rm d}   x^i  {\rm d}  x^j\right]\;. \\
\end{equation}
In this gauge, $v_\parallel=\hat s^i v_i=\hat{\mathbf{s}}\cdot\mathbf{v}$ (where $v^{i  }= \partial^i v$) and the GWs overdensity is written as
\begin{equation}
\delta^{[i] {\rm (P)}} =  \delta^{[i] {\rm (SC)}} - b_e {\mathcal H} v+ 3 {\mathcal H} v = b^{[i]} (\eta) \delta^{[i]}_{\rm m} - b_e {\mathcal H} v+ 3 {\mathcal H} v
\;,
\end{equation}
where we have used Synchronous Comoving gauge (SC) to define the bias.    
In the Poisson gauge we obtain (see~\cite{Bertacca:2019fnt, Bellomo:2021mer} for more details on the derivation)
%
\begin{eqnarray}
\label{DeltaOmegaPoisson}
 \delta^{\rm tot}_{\rm GW}  &=& \frac{4 \pi f_{\rm o}}{\bar \Omega^{\rm tot}_{\rm GW} \, \rho_{{\rm c},0}}  \sum_{[i]} \int w(z) {N^{[i]} [z, f_{\rm o} (1+z)] \over (1+z)}   \nonumber \\
 &\qquad& \times \Bigg\{ b^{[i]} \delta_{\rm m} \nonumber \\
 &\qquad&+ \left(b^{[i]}_{e} - 2 - \frac{\mathcal{H}'}{\mathcal{H}^2}\right) \hat{\mathbf{s}}\cdot\mathbf{v} - \frac{1}{\mathcal{H}} \hat{\mathbf{s}}\cdot { \bf \partial} (\hat{\mathbf{s}}\cdot\mathbf{v}) - (b^{[i]}_{e}-3) \mathcal{H} v + \nonumber \\
&\qquad& + \left(3 - b^{[i]}_{e} + \frac{\mathcal{H}'}{\mathcal{H}^2}\right) \Phi + \frac{1}{\mathcal{H}}\Phi' +2 \left(2 - b^{[i]}_\mathrm{e} + \frac{\mathcal{H}'}{\mathcal{H}^2}\right) \int_0^{\bar \chi} d\tilde \chi \Phi' + \nonumber\\
&\qquad& + \left(b^{[i]}_\mathrm{e} - 2 - \frac{\mathcal{H}'}{\mathcal{H}^2}\right) \left[-\mathcal{H}_0 \left(\int_{\bar \eta_{\rm in}}^{\bar \eta_0} d\tilde \eta {\Phi(\tilde \eta, {\bf 0} )\over (1+z(\tilde \eta ))}\right) + \Phi_{\rm o} -  \left(\hat{\mathbf{s}}\cdot\mathbf{v} \right)_{\rm o} \right]\Bigg\} \; {\rm d}  \bar \chi \;,\nonumber\\
\end{eqnarray}
where $\chi(z)$ is the comoving distance at redshift~$z$, $\eta_0$ is the conformal time today, $\mathcal{H}=a'/a$ is the Hubble expansion rate in conformal time. With ~``$\ '\ $'' we indicate derivatives with respect to the conformal time. We have also defined the evolution bias for each source
\begin{eqnarray}
b^{[i]}_{\rm e} = \frac{{\rm d} \ln N^{[i]}}{{\rm d} \ln \bar a}=-\frac{{\rm d} \ln N^{[i]}}{{\rm d} \ln (1+z)}.
\end{eqnarray}
Each of the four lines of Eq. \eqref{DeltaOmegaPoisson} is characterized by a specific  function: the gauge-invariant matter density fluctuation~$\delta$, the gauge invariant velocity~$\mathbf{v}$, and the Bardeen potential~$\Phi$. Notice that only the inclusion of all these terms allows to have a gauge-invariant observable. 
For a comparison and mapping between the various theoretical derivations of anisotropies presented in the literature and for a separate derivation
based on a Boltzmann approach, see Ref.\,\cite{Pitrou:2019rjz}.

\subsection{Angular power spectrum for astrophysical sources}
Similarly to CMB anisotropies a powerful observable to characterize the AGWB is the angular power spectrum that can be computed exploiting the spherical symmetry and working with spherical harmonics. In this section we expand the AGWB spectral energy density as
\begin{equation}
    \delta_{\rm GW}(\mathbf{s})=\sum_{\ell=0}^{\infty}\sum_{m=-\ell}^{+\ell}a_{\ell m}Y_{\ell m}(\mathbf{s})\,,
\end{equation}
where the coefficients $a_{\ell m}$ are given by
\begin{equation}
    a_{\ell m}=\int {\rm d}^2\mathbf{s} Y^{*}_{\ell m}(\mathbf{s})\delta_{\rm GW}(\mathbf{s})\,.
\end{equation}
The AGWB angular power spectrum then reads

\begin{equation}
    C_\ell^{\rm GW}
    \,=\,\sum_{m=-\ell}^{m=+\ell}\frac{\langle a_{\ell m} a_{\ell m}^*\rangle}{2\ell +1}  = \sum_{i,j;\alpha,\beta} C_\ell^{[ij]\alpha\beta} \,,
\end{equation}
where we have defined
\begin{eqnarray}
\label{Bertacca::90}
    C_\ell^{[ij]\alpha\beta}&\equiv&\sum_{m=-\ell}^{m=\ell}\frac{\langle a_{\ell m}^{[i]\alpha*}a_{\ell m}^{[j]\beta}\rangle}{2\ell +1} \,\nonumber\\
    &=&\int\frac{k^2 d{k}}{(2\pi)^3}\mathcal{S}_{\ell}^{[i]\alpha*}\mathcal{S}_\ell^{[i]\beta}P_m(k)\,
\end{eqnarray}
where $P_m(k)$ is the matter power spectrum today and $\mathcal{S}_{\ell}$ are the source functions which include all the effects described in Eq. \eqref{DeltaOmegaPoisson}. The index $[i]$ refers to the specific unresolved astrophysical source  while the greek index stands for the various contributions to the GW energy density anisotropies.\\
To have some physical insight into the information encoded in the anisotropies of the AGWB, we can use the Limber approximation. The general expression of the angular power spectrum reduces to \cite{Cusin:2017fwz}
\be\label{scaling-small-ell}
C^{\rm GW}_\ell(f) \simeq  \left(\ell+\tfrac{1}{2}\right)^{-1}\left(4\pi/\bar{\Omega}_{\text{GW}}\right)^2 \int \dd k P(k) \left|\partial_r \bar  \Omega (f,r)\right|^2\,,
\ee
where $\ell$ is the multipole in the spherical harmonic expansion, $P(k)$ is the galaxy power spectrum, $r$ is the (comoving) distance, related to the momentum $k$ via the Limber constraint $k r = \ell+1/2$. Each astrophysical model predicts a functional dependence of the astrophysical kernel $\partial_r \bar\Omega$, defined as
\be
\bar{\Omega}_{\rm GW}(f)=\int dr \partial_r \bar\Omega(f, r)\,.
\ee
It follows that the angular power spectrum depends on the astrophysical model chosen to describe sub-galactic physics. In particular, low $\ell$ are sensitive to the low-redshift value of the kernel $\partial_r \bar\Omega$. \\
The angular power spectrum of the anisotropies in the AGWB from merging stellar-mass binary BHs in the mHz band where LISA operates, has been computed in \cite{Cusin:2019jhg} using the astrophysical framework of \cite{PhysRevD.94.103011}.  It has been shown that AGWB anisotropies are very sensitive to sub-galactic astrophysical modeling. In particular, different descriptions of stellar evolution and black hole binary formation lead to fractional differences in the angular power spectrum of anisotropies up to $\sim 50\%$, independently on the global normalization (monopole) \cite{Cusin:2019jpv,  Cusin:2019jhg}.

Monopole and anisotropies contain complementary astrophysical information and studying the latter will allow one to break degeneracies between different astrophysical ingredients and potentially to constrain them separately.

\begin{figure}
    \centering
    \includegraphics[width=0.7\textwidth]{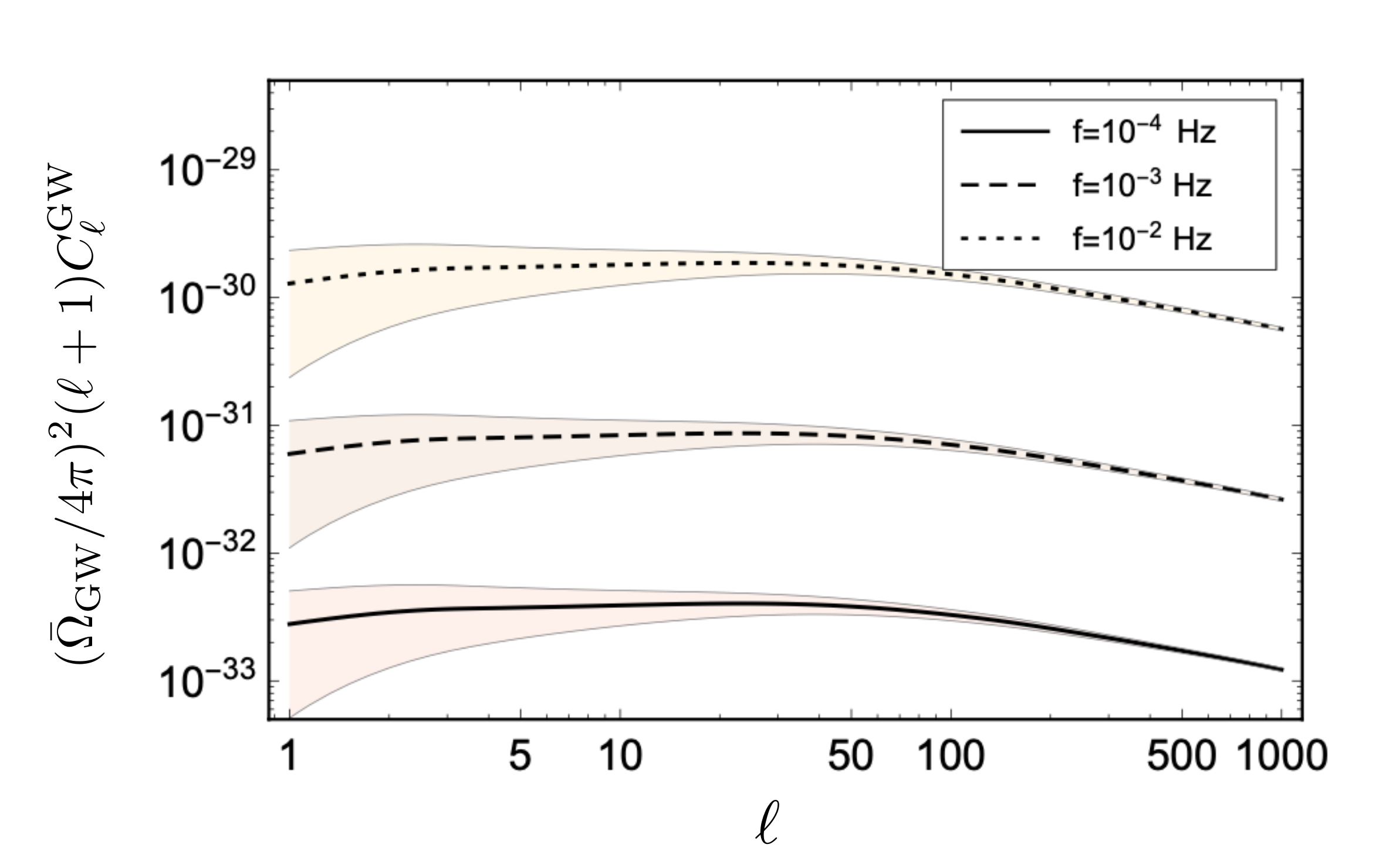}
    \caption{\it Angular power spectrum of anisotropies for three frequencies in the mHz band, for the reference astrophysical model of \cite{Cusin:2019jhg} and \cite{Cusin:2019jpv}. Multiplication by $(\ell+1)$ emphasizes the large-scale behaviour of Eq.\,(\ref{scaling-small-ell}), while we multiplied the spectrum by the monopole amplitude to show the frequency scaling of anisotropies. The shaded region corresponds to the cosmic variance limit. Adapted from \cite{Cusin:2019jhg}.  }
    \label{fig:C_ell-AGWB}
\end{figure}

\subsubsection{Systematic effects on the angular power spectrum} 

In estimating the anisotropies of the astrophysical gravitational-wave background, the finite number of the sources that contribute to the background at any given time and the  very short time each of them spends in the frequency of the interferometer, induce a white noise ($\ell$-independent) term, ${\cal W}$, in the angular power spectrum $C_\ell^{\rm GW}$: 
\begin{equation}
 C^{\rm GW}_\ell=C_\ell^{\rm LSS} +{\cal W}~,~{\cal W}\equiv{1\over r_H^3}\int{\rm d} r r^2{\cal V}(r)~,
 \end{equation}
 where ${\cal V}$ is some direction-independent (due to statistical isotropy) function describing the variance, $r_H \equiv 1/H_0$ is the Hubble radius and $C_\ell^{\rm LSS}$ stands for the angular power of the intrinsic, astrophysical anisotropy. 
The shot noise dominates over the true astrophysical power spectrum; the latter may be recovered with long enough observing runs and sufficient removal of a large number of foreground sources \cite{Jenkins:2019uzp}.

 To calculate the true, astrophysical angular power spectrum of a statistically-isotropic  gravitational-wave background, a novel method, based on combining
statistically-independent data segments, was proposed in \cite{Jenkins:2019nks}. The proposed estimator,  constructed from the cross-correlations between statistically-independent time intervals,  reads
\begin{equation}
    \hat {C_\ell}\equiv{2\over N_\tau(N_\tau-1)}
    \sum_{\mu=1}^{N_\tau}\sum_{\nu=\mu+1}^{N_\tau} {\hat C_\ell}^{\mu\nu}~,
    \label{est}
\end{equation}
where $N_\tau\equiv T/\tau$ (with $T$ the total observing time) denotes the number of segments and \begin{equation}
    {\hat C_\ell}^{\mu\nu}\equiv {1\over 2\ell+1}\sum_{m=-\ell}^{+\ell} \Omega_{\ell m}^\mu \Omega_{\ell m}^{\nu\star}~.
\end{equation}
The estimator \eqref{est} is unbiased since
\begin{equation}
    \langle {\hat C}_\ell
    \rangle_{S,\Omega} = C_\ell^{\rm GW}~,
\end{equation}
where the subscripts $S,\Omega$ stand for performing first the cosmological and the the shot noise average. In the limit of many data segments, $N_\tau\gg 1$, the estimator (\ref{est}) has the lowest-variance
\begin{equation}
    {\rm Var}[{\hat C}_\ell]_{S,\Omega}\simeq
    {2\over 2\ell +1}(C^{\rm GW}_\ell+{\cal W}_T)^2,
\end{equation}
and, in this sense, it is the most efficient one.
It is worth noting that the term ${\cal W}_T$ in the above equation is the same as the one appearing in the mean of the standard estimator $\langle C_\ell^{(\rm std)}\rangle_{S,\Omega}=C^{\rm GW}_\ell+{\cal W}_T$, hence (\ref{est}) is still affected by the shot noise. However, now the shot noise only adds to the variance of the estimator and it does not affect the angular power spectrum, as in the standard case.

Since the shot noise power may exceed the real astrophysical angular power spectrum by a factor as high as approximately $10^4$, according \cite{Jenkins:2019uzp}, the proposed method for estimating the true, astrophysical angular power spectrum of a statistically-isotropic astrophysical gravitational-wave background, is indeed a valuable tool. 

Another interesting method to alleviate this shot noise problem and extract information on the underlying GW population, is to make use of the cross-correlation of the AGWB background map with other cosmological observables such as galaxy distribution, see e.g. \cite{Alonso:2020mva}. Indeed, the shot noise level of the cross-spectrum
is primarily driven by the density of the much denser galaxy
survey (although the GW shot noise will still be a significant
contribution to the signal to noise of the cross-correlation).

\section{\sc LISA Angular Sensitivity}
\label{sec:response}

In this section we discuss the sensitivity of LISA to the anisotropies of the SGWB.

\subsection{LISA angular response functions}

We follow the notation of ref.~\cite{Flauger:2020qyi}, that we generalize to an anisotropic SGWB. We start from 
\begin{equation} 
h_{ab} \left( {\bf x} ,\, t \right) = \int_{-\infty}^{+\infty} df \int d \Omega_{\hat k} \, 
{\rm e}^{2 \pi i f \left( t - {\hat k} \cdot {\bf x} \right)} \, \sum_{\lambda} {\tilde h}_{\lambda} \left( f ,\, {\hat k} \right) 
e_{ab}^{\lambda} \left( {\hat k} \right) \;, 
\label{GW-deco}
\end{equation} 
where in the chiral basis ($\lambda = \pm 1$ denoting, respectively, the right and the left polarization), the polarization operators obey $e_{ab}^{\lambda*} \left( {\hat k} \right) = e_{ab}^{\lambda} \left( - {\hat k} \right) = e_{ab}^{-\lambda} \left( {\hat k} \right)$ and they are normalized according to $e_{ab}^{\lambda*} \left(  {\hat k} \right) e_{ab}^{\lambda'} \left(  {\hat k} \right) = \delta_{\lambda \lambda'}$, and where reality of the mode function is ensured by ${\tilde h}_\lambda^* \left( f ,\, {\hat k} \right) = {\tilde h}_{-\lambda} \left( -f ,\, {\hat k} \right)$.  An unpolarized and anisotropic SGWB is characterized by the intensity $I$, defined through 
\begin{equation}
\left\langle {\tilde h}_\lambda \left( f_1 ,\, {\hat k}_1 \right)  {\tilde h}_{\lambda'} \left( f_2 ,\,  {\hat k}_2 \right)  \right\rangle = 
\delta \left( f_1 + f_2 \right) \frac{\delta^{(2)} \left( {\hat k}_1 - {\hat k}_2 \right)}{4 \pi} \, \delta_{\lambda,- \lambda'} \, 
\sum_{\ell m} {\tilde I}_{\ell m} \left( \vert f_1 \vert \right) {\tilde Y}_{\ell m} \left( {\hat k}_1 \right) \;, 
\label{<hh>}
\end{equation} 
where ${\tilde Y}_{\ell m} \left( {\hat k} \right) \equiv \sqrt{4 \pi} \, Y_{\ell m} \left( {\hat k} \right)$, and $Y_{\ell m} \left( {\hat k} \right)$ are the standard spherical harmonics, with this normalization ${\tilde Y}_{00} \left( {\hat k} \right) = 1$. 

We want to relate the coefficients ${\tilde I}_{\ell m}$ to those of the fractional energy density in the decomposition (\ref{dec}). Starting from Eq. (\ref{GW-deco}) and from the intensity function defined in (\ref{<hh>}), we arrive to the following expression for the SGWB energy density over the critical energy density %
\begin{eqnarray}
&& \!\!\!\!\!\!\!\!
\frac{\rho_{\rm GW}}{\rho_{\rm crit}} = 
\frac{1}{32 \pi G} \, \left\langle \dot{h}_{ij} \, \dot{h}_{ij} \right\rangle \Bigg/ \frac{3 H_0^2}{8 \pi G} \nonumber\\ 
&&
= \frac{1}{12 H_0^2} 
\int_{-\infty}^{+\infty} df_1 \, df_2 \int d \Omega_{{\hat k}_1} \, 
d \Omega_{{\hat k}_2} 
\left( - 4 \pi^2 \, f_1 \, f_2 \right) 
{\rm e}^{2 \pi i f_1 \left( t - {\hat k}_1 \cdot {\bf x} \right)+2 \pi i f_2 \left( t - {\hat k}_2 \cdot {\bf x} \right)} \nonumber\\ 
&& \times  \sum_{\lambda_1,\lambda_2} 
e_{ij}^{\lambda_1} \left( {\hat k}_1 \right) e_{ij}^{\lambda_2} \left( {\hat k}_2 \right) 
\delta \left( f_1 + f_2 \right) 
\frac{\delta^{(2)} \left( {\hat k}_1 - {\hat k}_2 \right)}{4 \pi}
\delta_{\lambda_1,- \lambda_2} \, 
\sum_{\ell m} {\tilde I}_{\ell m} \left( \vert f_1 \vert \right) {\tilde Y}_{\ell m} \left( {\hat k}_1 \right) \;, \nonumber\\ 
\end{eqnarray} 
where $G$ is the Newton constant, while $H_0$ the present Hubble rate. Recalling the normalization of the polarization operators, we then find 
\begin{eqnarray}
\frac{\rho_{\rm GW}}{\rho_{\rm crit}} &=& \frac{\pi}{3 H_0^2}  
\int_0^{+\infty} d \ln f \, f^3 \,  \sum_{\ell m} {\tilde I}_{\ell m} \left( f \right) \, \int d \Omega_{\hat k} \, {\tilde Y}_{\ell m} \left( {\hat k} \right) \;. 
\end{eqnarray} 
Proceeding as in Section \ref{theoretical_framework}, we then arrive to 
\begin{equation}
{\tilde I}_{\ell m} \left( f  \right)  = \frac{1}{\sqrt{4 \pi}} \, \frac{3 H_0^2}{4 \pi^2} \, \frac{\Omega_{\rm GW} \left( f \right)}{f^3} \delta_{\rm GW,\ell m} \;. 
\label{I-to-delta}
\end{equation} 

Let us now discuss how to measure these coefficients. We consider two locations $\vec{x}_{1,2}$, at the unperturbed distance $L$ (by ``unperturbed'', we mean the quantity in absence of the SGWB), and a photon that, starting from $\vec{x}_2$ at the unperturbed time $t-L$, arrives at $\vec{x}_1$ at the unperturbed time $t$. The SGWB modifies the time of flight to $L + \Delta T_{12} \left( t \right)$, with 
\begin{equation}
\Delta T_{12} \left( t \right) = \frac{{\hat l}_{12}^a \, {\hat l}_{12}^b}{2} 
\int_0^L d s \, h_{ab} \left( t \left( s \right) ,\, \vec{x} \left( s \right) \right) \;, 
\label{deltaT}
\end{equation} 
where ${\hat l}_{12}$ is the unit vector going from $\vec{x}_1$ to $\vec{x}_2$. This time delay has an associated Doppler frequency shift~\cite{Flauger:2020qyi} 
\begin{equation}
\Delta F_{12} \left( t \right) \equiv \frac{\Delta \nu_{12} \left( t \right)}{\nu} = - \frac{d}{dt} \Delta T_{12} \left( t \right) \;. 
\end{equation} 
We denote by 
\begin{equation}
\Delta F_{1(2)} \left( t \right) \equiv \Delta F_{21} \left( t - L \right) +  \Delta F_{12} \left( t  \right) \;, 
\end{equation}
the frequency shift for the closed $\vec{x}_1 \to \vec{x}_2 \to \vec{x}_1$ path. Differences between closed path shifts originate the Time Delay Interferometry (TDI) 1.0 and 1.5 typically considered for LISA \cite{Flauger:2020qyi}. Specifically, the TDI 1.0 combination is given by the difference between the $\vec{x}_1 \to \vec{x}_2 \to \vec{x}_1$ and the $\vec{x}_1 \to \vec{x}_3 \to \vec{x}_1$ path: 
\begin{equation}
\Delta F_{1(23)}^{1.0} \left( t \right) \equiv \Delta F_{1(2)} \left( t \right) -  \Delta F_{1(3)} \left( t \right) \;, 
\label{TDI1.0}
\end{equation} 
while the TDI 1.5 combination is given by the difference between the $\vec{x}_1 \to \vec{x}_2 \to \vec{x}_1 \to \vec{x}_3 \to \vec{x}_1$ and the $\vec{x}_1 \to \vec{x}_3 \to \vec{x}_1 \to \vec{x}_2 \to \vec{x}_1$ path:  
\begin{eqnarray} 
\Delta F_{1(23)}^{1.5} \left( t \right) 
&\equiv&  \Delta F_{1(2)} \left( t - 2 L \right) + \Delta F_{1(3)} \left( t \right) -   
\Delta F_{1(3)} \left( t - 2 L \right) - \Delta F_{1(2)} \left( t \right) \nonumber\\ 
&=& \Delta F_{1(23)} \left( t - 2 L \right) + \Delta F_{1(32)} \left( t \right) \;. 
\label{TDI1.5}
\end{eqnarray} 

To simplify the notation, we denote~\footnote{Namely the index $4$ coincides with $1$, and the index $5$ coincides with $2$.} by $i \, {\rm mod} \, 3$ the i-th satellite of the LISA triangle, and define 
\begin{equation}
\Delta F_i \left( t \right) \equiv \Delta F_{i\left(i+1,i+2\right)} \;,\;\;\; i = 1 ,\, 2 ,\, 3 \,, 
\label{short-i}
\end{equation} 
%

We are interested in correlators between different measurements. The only statistical variable that participates non trivially in the correlator is the GW mode function, see Eq.~\eqref{<hh>}. Starting from the expression in Eq.~\eqref{TDI-integral} for the TDI measurement, we obtain 
\begin{eqnarray} 
\left\langle \Delta F_i \left( t \right) \Delta F_j \left( t \right) \right\rangle &=& 4 \sum_{\ell m} \int_0^\infty d f \, \left\vert \frac{f}{f_*} \, W \left( f \right) \right\vert^2 \, {\tilde R}_{ij}^{\ell m} \left( f \right) \, {\tilde I}_{\ell m} \left( f \right) \nonumber\\ 
&\equiv& \sum_{\ell m} \int_0^\infty d f \, R_{ij}^{\ell m} \left( f \right) \, {\tilde I}_{\ell m} \left( f \right) \;, 
\label{<DFDF>}
\end{eqnarray} 
where the frequency $f_*$ is related to the LISA arm length $L$ by 
\begin{equation}
f_* \equiv \frac{1}{2 \pi L} \simeq 0.019 \, {\rm Hz} \times \frac{2.5 \cdot 10^6 \, {\rm km}}{L} \;, 
\label{fstar} 
\end{equation} 
where 
\begin{equation}
\left\vert W \left( f \right) \right\vert^2 = 
\left\{ \begin{array}{l} 
1 \;\;,\;\;\;\; \quad\quad\quad\quad  {\rm for \; TDI \; 1.0} \\ 
4 \, \sin^2 \left( \frac{f}{f_*} \right)  \;\;,\;\;\;\; {\rm for \; TDI \; 1.5}\,.
\end{array} \right.
\end{equation} 
We introduced the anisotropic LISA response function 
\begin{eqnarray}
&& \!\!\!\!\!\!\!\!  \!\!\!\!\!\!\!\!  \!\!\!\!\!\!\!\! 
{\tilde R}_{ij}^{\ell m} \left( f \right) \equiv  
\frac{1}{8 \pi} \int d^2 {\hat k} \, {\rm e}^{-2 \pi i f \, {\hat k} \cdot \left( \vec{x}_i - \vec{x}_j \right) } \,  
{\tilde Y}_{\ell m} \left( {\hat k} \right) \nonumber\\
&& \times \sum_A 
R^A \left( f \, {\hat k} ,\, {\hat l}_{i,i+1} ,\, {\hat l}_{i,i+2} \right) 
\, R^{A*} \left( f {\hat k} ,\, {\hat l}_{j,j+1} ,\, {\hat l}_{j,j+2} \right) \,,  
\label{response}
\end{eqnarray} 
with the functions $R^A$ are given in Eq.~\eqref{RA}. In the isotropic case, the response function in Eq.~\eqref{response} agrees with Eq.~(A.21) of~\cite{Flauger:2020qyi}.

As we show in Appendix~\ref{Appendix_TDI}, under a rigid rotation of the instrument the response function transforms as a spherical harmonic. Specifically, if $R$ is a rotation under which the position of the three satellites changes according to $\vec{x}_i \to R \, \vec{x}_i$, we have 
\begin{equation}
{\tilde R}_{Ri Rj}^{\ell m} \left( f \right) = \sum_{m' = - \ell}^\ell \left[ D_{mm'}^{(\ell)} \left( R \right) \right]^* \, 
{\tilde R}_{i j}^{\ell m'} \left( f \right)  \;, 
\label{rot-R}
\end{equation} 
where $D_{mm'}^{(\ell)}$ are the elements of the Wigner $D$-matrix. For a rotation of an angle $\alpha$ about the $z-$axis we then have 
\begin{equation}
{\tilde R}_{R_z \left( \alpha \right) i ,\, R_z \left( \alpha \right) j }^{\ell m} \left( f \right) = 
{\rm e}^{i m \alpha} \, {\tilde R}_{i j}^{\ell m} \left( f \right) \,. 
\label{rotz-R}
\end{equation}  

Using this fact, and the property 
\begin{eqnarray} 
{\tilde R}_{ji}^{\ell m} \left( f \right) = \left( - 1 \right)^\ell \, {\tilde R}_{ij}^{\ell m} \left( f \right) \;,  
\label{transpose-R}
\end{eqnarray} 
(that we also prove in Appendix~\ref{Appendix_TDI}), we then learn that, if we place the three satellites in the $xy$ plane, the various components of the response function satisfy 
\begin{eqnarray} 
\left( \begin{array}{ccc} 
{\tilde R}_{11}^{\ell m}  & {\tilde R}_{12}^{\ell m}  & {\tilde R}_{13}^{\ell m}  \\ 
{\tilde R}_{21}^{\ell m}  & {\tilde R}_{22}^{\ell m}  & {\tilde R}_{23}^{\ell m}  \\
{\tilde R}_{31}^{\ell m}  & {\tilde R}_{32}^{\ell m}  & {\tilde R}_{33}^{\ell m} 
\end{array} \right) 
= \left( \begin{array}{ccc} 
{\tilde R}_{11}^{\ell m} & {\tilde R}_{12}^{\ell m} & 
\left( - 1 \right)^\ell \, {\rm e}^{\frac{4 \pi i m}{3}} \,  {\tilde R}_{12}^{\ell m}   \\ 
\left( - 1 \right)^\ell \, {\tilde R}_{12}^{\ell m}   &  {\rm e}^{\frac{2 \pi i m}{3}} \,  {\tilde R}_{11}^{\ell m} & {\rm e}^{\frac{2 \pi i m}{3}} \,  {\tilde R}_{12}^{\ell m} \\ 
{\rm e}^{\frac{4 \pi i m}{3}} \,  {\tilde R}_{12}^{\ell m}  & 
\left( - 1 \right)^\ell {\rm e}^{\frac{2 \pi i m}{3}} \,  {\tilde R}_{12}^{\ell m} 
&  {\rm e}^{\frac{4 \pi i m}{3}} \,  {\tilde R}_{11}^{\ell m} 
\end{array} \right) \;, 
\label{R-prop-ij}
\end{eqnarray} 

In Appendix~\ref{Appendix_TDI} we also show that the response function satisfies 
\begin{equation}
{\tilde R}_{ij}^{\ell, -m}   = {\tilde R}_{ij}^{\ell m \,*} \;, 
\label{R-prop-menom}
\end{equation} 
as well as 
\begin{equation}
\ell + m = {\rm odd } \;\; \Rightarrow \;\; {\tilde R}_{ij}^{\ell m} \left( f \right) = 0  \;. 
\label{R-prop-lm}
\end{equation}
Moreover, from Eq.~\eqref{transpose-R}, we notice that 
\begin{equation}
\ell \;\; {\rm odd} \;\; \Rightarrow \;\; 
{\tilde R}_{ii}^{\ell m} \left( f  \right) = 0  \;\;\; \left( {\rm no \; sum \; over \;} i \right) \;. 
\label{R-prop-m}
\end{equation}

\subsection{$\ell$-dependent response functions in the A, E, T channels}


As shown by Eq.~\eqref{rot-R}, the anisotropic LISA response functions transform as spherical harmonics under rotations. One can therefore consider the $\ell-$dependent response function 
\begin{equation}
{\tilde R}_{ij}^{\ell} \left( f \right) \equiv \left( \sum_{m = -\ell}^\ell \, \left\vert {\tilde R}_{ij}^{\ell m} \left( f \right) \right\vert^2 \right)^{1/2} \;, 
\label{Rl}
\end{equation} 
that is invariant under rotations, and therefore constant in time (it does not depend on the orientation of the LISA triangle). As we show in the next subsection, it provides an estimate for the response of LISA to a statistically isotropic SGWB, see Subsection \ref{subs:sensitivity-l}. From the properties in Eq.~\eqref{R-prop-ij} we learn that 
\begin{eqnarray} 
&& {\tilde R}_{11}^\ell = {\tilde R}_{22}^\ell  ={\tilde R}_{33}^\ell \;, \nonumber\\ 
&& {\tilde R}_{12}^\ell = {\tilde R}_{21}^\ell = {\tilde R}_{13}^\ell = {\tilde R}_{31}^\ell = {\tilde R}_{23}^\ell  = {\tilde R}_{32}^\ell \;.  
\end{eqnarray} 

It is customary to consider linear combinations of the $\Delta F_i$ measurements considered so far
\begin{equation}
\Delta F_A \equiv \frac{\Delta F_3-\Delta F_1}{\sqrt{2}} \;,\;\; 
\Delta F_E \equiv \frac{\Delta F_1-2\Delta F_2+\Delta F_3}{\sqrt{6}} \;,\;\; 
\Delta F_T \equiv \frac{\Delta F_1+\Delta F_2+\Delta F_3}{\sqrt{3}} \;, 
\end{equation}
which we write more compactly as 
\begin{equation}
\Delta F_O \equiv c_{Oi} \, \Delta F_i \;,\;\; 
O \in \left\{ A,\, E ,\, T \right\} \;,\;\; 
i \in \left\{ 1,\, 2 ,\, 3 \right\} \;. 
\label{DFO}
\end{equation} 
These combinations (that we have normalized as in ref.~\cite{Flauger:2020qyi}, so that the rotation matrix associated with these transformations is orthogonal) diagonalize the noise variance, in the hypothesis that LISA is an equilateral triangle, with identical instruments at the vertices. In terms of the $A,E,T$ channels the response function formally reads 
\begin{equation}
{\tilde R}_{OO'}^{\ell m} \left( f \right) = c_{Oi}  \, c_{O'j} \, {\tilde R}_{ij}^{\ell m} \left( f \right) \;. 
\label{Rlm-OOp}
\end{equation} 
We evaluate these linear combinations, accounting for the identities in Eq.~\eqref{R-prop-ij}, namely 
\begin{equation}
{\tilde R}_{OO'}^{\ell} \left( f \right) \equiv \left( \sum_{m = -\ell}^\ell \, \left\vert c_{Oi} \, c_{O'j} \, {\tilde R}_{ij}^{\ell m} \left( f \right) \right\vert^2 \right)^{1/2} \;. 
\label{Rl-OOp}
\end{equation} 
The resulting expressions acquire different forms for even and odd multipoles. Specifically, for odd $\ell$ we find 
%
\begin{eqnarray} 
&& {\tilde R}_{AA}^\ell \left( f  \right) = {\tilde R}_{EE}^\ell \left( f  \right) = {\tilde R}_{TT}^\ell \left( f  \right) = 0 \;, \nonumber\\ 
&& {\tilde R}_{AE}^\ell \left( f  \right) =   \left\{ \frac{1}{3} \sum_{m=-\ell}^\ell  \left[ 1 + 2 \cos \left( \frac{2 m \pi}{3} \right) \right]^2 \left\vert   {\tilde R}_{12}^{\ell m} \left( f \right) \right\vert^2 \right\}^{1/2} \;, \nonumber\\ 
&& {\tilde R}_{AT}^\ell \left( f  \right) = 
 {\tilde R}_{ET}^\ell \left( f  \right) = 
\left\{ 2  \sum_{m=-\ell}^\ell  \sin^2  \left( \frac{ m \pi}{3} \right) \, \left\vert  {\tilde R}_{12}^{\ell m} \left( f \right) \right\vert^2 \right\}^{1/2} \;. \nonumber\\ 
\label{R-AET-lodd} 
\end{eqnarray} 
and for even $\ell$
\begin{eqnarray} 
&& {\tilde R}_{AA}^\ell \left( f  \right) = {\tilde R}_{EE}^\ell \left( f  \right) =  \left\{ \frac{1}{4} \sum_{m=-\ell}^\ell \left\vert \left( 1 + {\rm e}^{-\frac{4}{3} i m \pi} \right) {\tilde R}_{11}^{\ell m} \left( f \right) 
- 2 \, {\tilde R}_{12}^{\ell m} \left( f \right) \right\vert^2 \right\}^{1/2} \;, \nonumber\\ 
&& {\tilde R}_{TT}^\ell \left( f  \right) =  \left\{ \frac{1}{9} \sum_{m=-\ell}^\ell \left[ 1 + 2 \, \cos \left( \frac{2 m \pi}{3} \right) \right]^2  \left\vert {\tilde R}_{11}^{\ell m} \left( f \right) + 2 {\tilde R}_{12}^{\ell m} \left( f \right) \right\vert^2 \right\}^{1/2} \;, \nonumber\\ 
&& {\tilde R}_{AE}^\ell \left( f  \right) =   \left\{ \frac{1}{3}  \sum_{m=-\ell}^\ell  \sin^2  \left( \frac{ m \pi}{3} \right) \, \left\vert \left( 1 + {\rm e}^{\frac{2 i m \pi}{3}} \right)  {\tilde R}_{11}^{\ell m} \left( f   \right) - 2 {\tilde R}_{12}^{\ell m} \left( f   \right)  \right\vert^2 \right\}^{1/2} \;, \nonumber\\  \nonumber\\ 
&& {\tilde R}_{AT}^\ell \left( f  \right) = 
 {\tilde R}_{ET}^\ell \left( f  \right) 
= \left\{ \frac{2}{3}  \sum_{m=-\ell}^\ell  \sin^2  \left( \frac{ m \pi}{3} \right) \, \left\vert 
\left( 1 + {\rm e}^{\frac{2 i m \pi}{3}} \right)  {\tilde R}_{11}^{\ell m} \left( f   \right) + {\tilde R}_{12}^{\ell m} \left( f   \right)  \right\vert^2 \right\}^{1/2} \;. \nonumber\\ 
\label{R-AET-leven} 
\end{eqnarray} 
We also note that the property in Eq.~\eqref{transpose-R} implies that the response fuction is symmetric in the channels, ${\tilde R}_{O'O}^\ell = {\tilde R}_{OO'}^\ell$. 

\begin{figure}[ht!]
\centerline{
\includegraphics[width=\textwidth,angle=0]{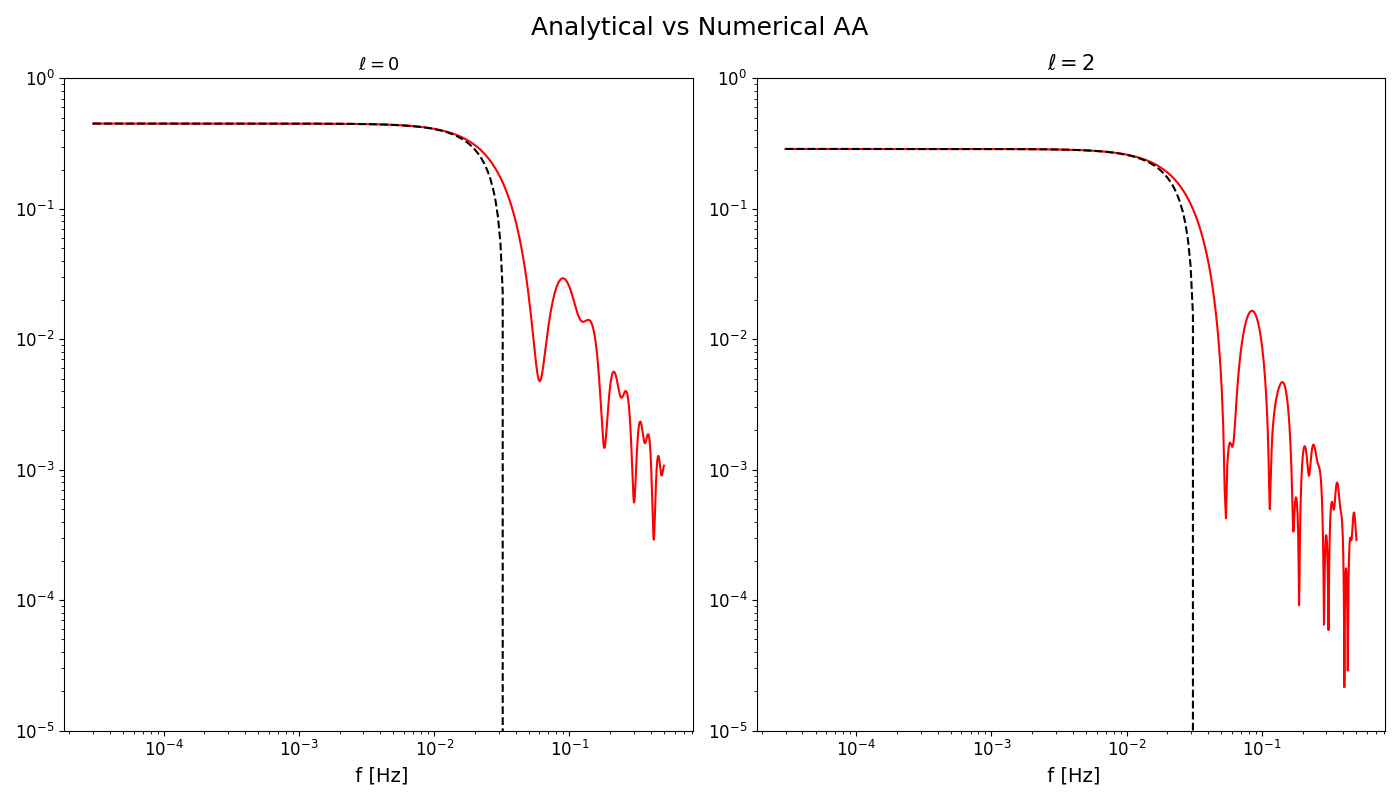}
}
\centerline{
\includegraphics[width=\textwidth,angle=0]{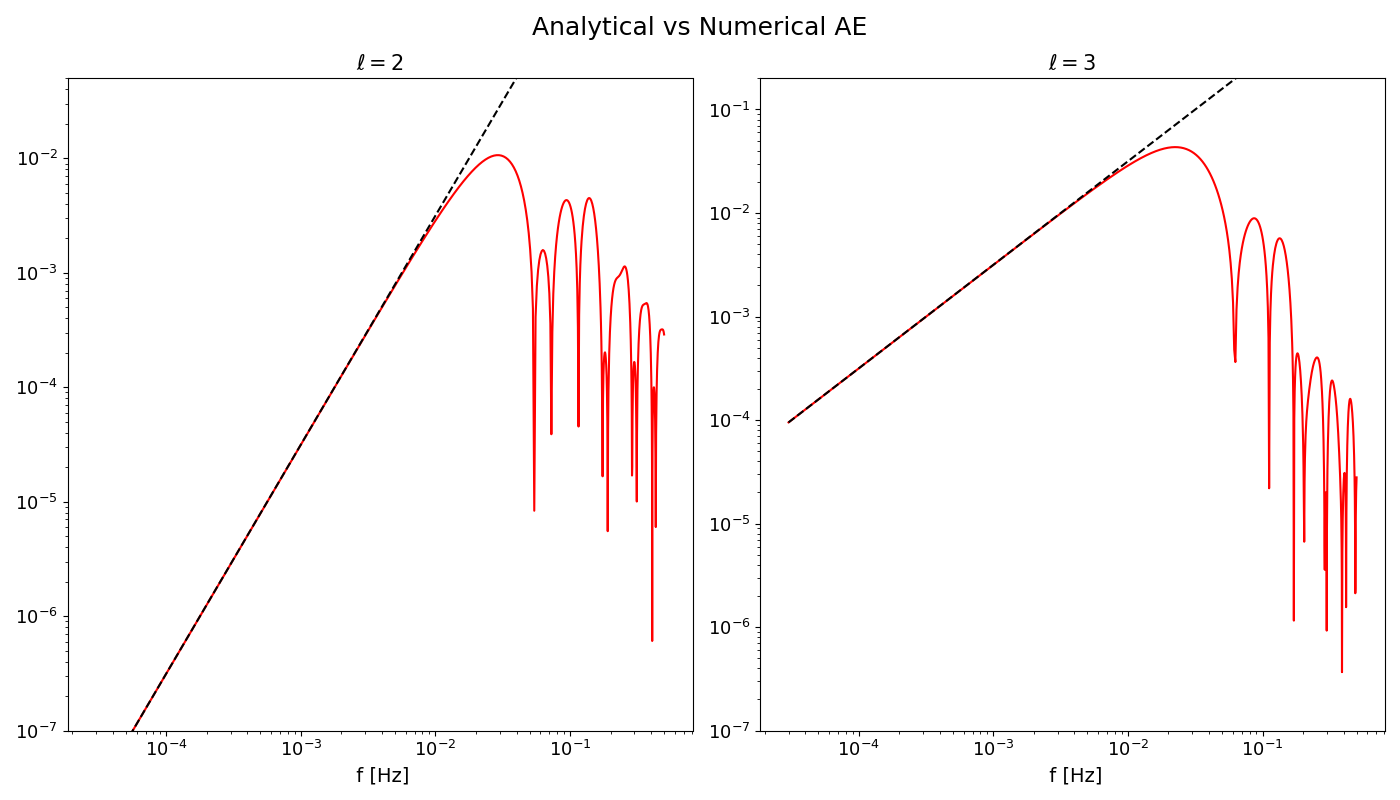}
}
\caption{\it Angular overlap functions, defined through Eqs.~\eqref{response} and~\eqref{Rl}, up to $\ell =6$ for the $AA=EE$ correlation (fist row) 
and for the $AE$ correlation (second row). The solid red line is from an exact evaluation. 
The dashed black line is the small frequency approximation in Table~\ref{table-small-f}. 
}
\label{fig:AA-AE}
\end{figure}

\begin{figure}[ht!]
\centerline{
\includegraphics[width=\textwidth,angle=0]{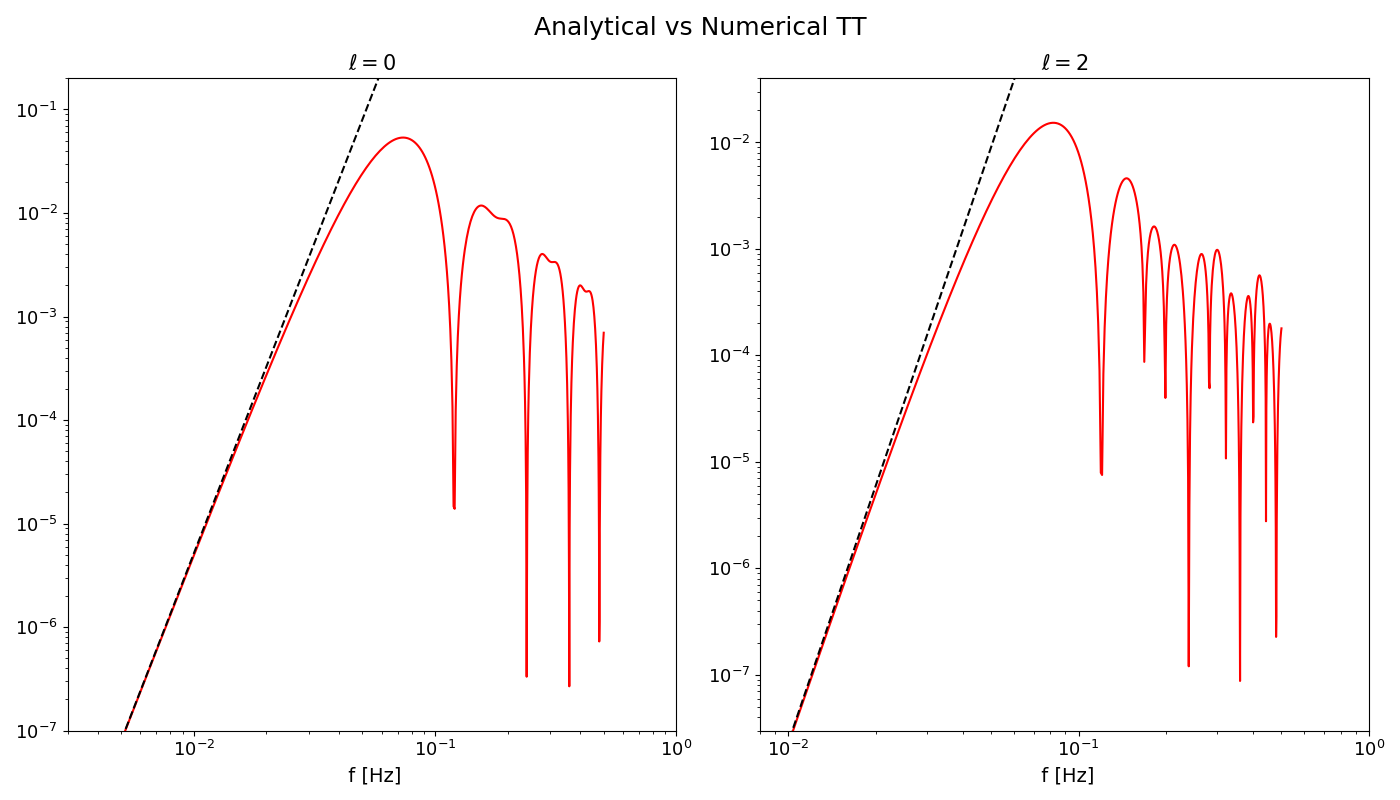}
}
\centerline{
\includegraphics[width=\textwidth,angle=0]{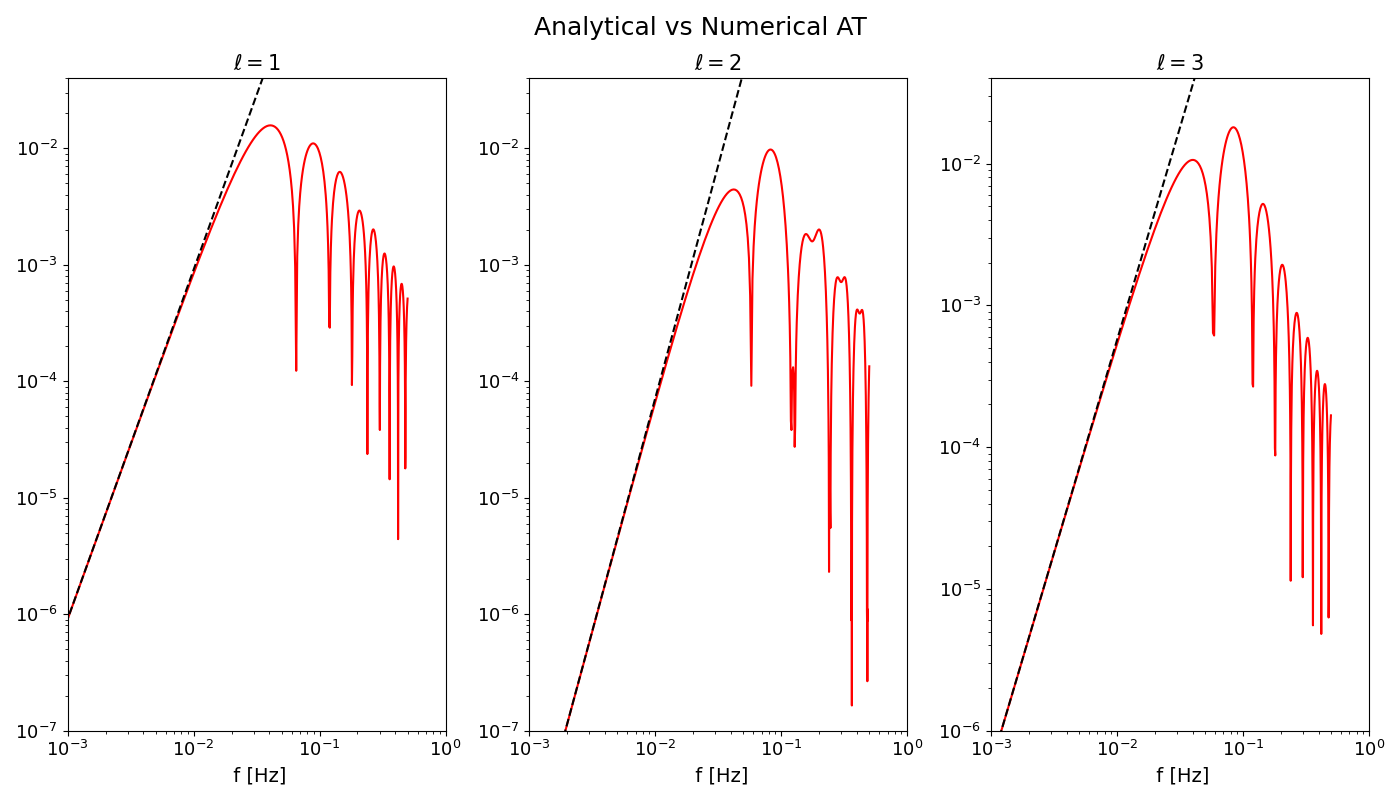}
}
\caption{\it Angular overlap functions, defined through Eqs.~\eqref{response} and~\eqref{Rl}, up to $\ell =6$ for the $TT$ correlation (fist row) 
and for the $AT=ET$ correlation (second row). The solid red line is from an exact evaluation. 
The dashed black line is the small frequency approximation in Table~\ref{table-small-f}. 
}
\label{fig:TT-AT}
\end{figure}

These expressions can be evaluated numerically, for arbitary frequency, or evaluated analytically in the small frequency regime.  For the first few multipoles, we obtain the values in the Table~\ref{table-small-f}. In Figures~\ref{fig:AA-AE} and~\ref{fig:TT-AT} we show instead a comparison between the full shape of the response functions and the small frequency expressions for these first multipoles.


\begin{table}
\centering
\begin{tabular}{|c|c|c|c|c|}
\hline
{ $\ell$} & $ {\tilde R}_{AA}^\ell $ & $ {\tilde R}_{AE}^\ell $ & ${\tilde R}_{TT}^\ell$ & ${\tilde R}_{AT}^\ell$ 
\\\hline \hline 
0   & $\frac{9}{20} - \frac{169 \, x^2}{1120 }$   & 0   & $\frac{x^6}{4032}$   & 0    \\\hline 
1   & 0   & 0   & 0   & $\frac{x^3}{112 \sqrt{2}}$   \\\hline 
2   & $\frac{9}{14 \sqrt{5}} - \frac{13 \, x^2}{56 \sqrt{5}} $   & $\sqrt{\frac{5}{3}} \, \frac{x^2}{112} $    & $\frac{73 \, x^8}{7983360  \, \sqrt{5} }$   & $\frac{x^4}{192 \sqrt{30}}$  \\\hline 
3   & 0   & $\sqrt{\frac{7}{30}} \, \frac{x}{8}  $   & 0   & $\frac{x^3}{96 \sqrt{7}}$  \\\hline 
4   & $\frac{9}{140} - \frac{3719 \, x^2}{147840} $    & 
$\frac{3}{8 \sqrt{35}} - \frac{27 \, x^2}{176 \sqrt{35}}  $  & $\frac{x^6}{12672}$   & $\sqrt{\frac{37}{35}}  \, \frac{x^4}{1056} $  \\\hline 
5   & 0   & $\frac{x}{8 \sqrt{2310}}$  & 0   & $\sqrt{\frac{211}{110}} \frac{x^3}{672}$  \\\hline 
6   & $\sqrt{\frac{1829}{195}} \, \frac{x^2}{4928}$   & $ \frac{x^2}{32 \sqrt{2730}} $ & 
$\sqrt{\frac{463}{13}} \, \frac{x^6}{88704}$  & $\sqrt{\frac{17}{2730}} \frac{x^4}{2112} $   \\\hline 
\end{tabular}
\caption{{\it Leading terms in a small frequency expansion of ${\tilde R}_{OO'}^\ell \left( x \right)$ 
where $x= f/f_*$, and $f_*$ is given in Eq.~\eqref{fstar}. In each term, we have kept up to the leading $f-$dependent term. We recall that the response functions are symmetric in the channels, that ${\tilde R}_{EE}^\ell ={\tilde R}_{AA}^\ell$, and that ${\tilde R}_{ET}^\ell ={\tilde R}_{AT}^\ell$.}
\label{table-small-f} 
}
\end{table}

\subsection{Signal-to-Noise Ratio for anisotropic signals}
We consider the Fourier transform of the signal in Eq.~\eqref{DFO}, performed with an integration time $\tau$ 
\begin{equation}
{\tilde \Delta F}_O \left( f ,\, t \right) \equiv \int_{t-\tau/2}^{t+\tau/2} d t' \, \Delta F_O \left( t' \right) e^{-2 i \pi f t'} \;. 
\label{Fourier-signal}
\end{equation}
This signal, if present, adds up with the instrumental noise in the measurement 
\begin{equation}
{\tilde m}_O\left( f ,\, t \right) \equiv {\tilde \Delta F}_O\left( f ,\, t \right) + {\tilde n}_O\left( f ,\, t \right) \;. 
\label{eq:data_noise_model}
\end{equation}
We assume that the noise is Gaussian and we recall that it is diagonal in the A,E,T basis, namely 
\begin{equation}
\left\langle n_O \left( f \right) n_{O'} \left( f \right) \right\rangle \equiv \frac{1}{2} \delta \left( f - f' \right) \delta_{OO'} \, N_O \left( f \right) \;,
\label{NO-def}
\end{equation} 
where the explicit expressions for $N_O \left( f \right)$ are given in Appendix \ref{app:C-SNR}. Then we define the estimator as
\begin{equation}
{\cal C} \equiv \sum_{O,O'} \int_0^T d t \int_{-\infty}^{+\infty} d f \left[ 
{\tilde m}_O\left( f ,\, t \right) \,  {\tilde m}_{O'}^* \left( f ,\, t \right) - 
\left\langle  {\tilde n}_O \left( f ,\, t \right) \,  {\tilde n}_{O'}^* \left( f ,\, t \right) \right\rangle \right] 
{\tilde Q}_{OO'} \left( t ,\, f \right)  \;, 
\label{C-def}
\end{equation} 
where the functions ${\tilde Q}_{OO'} \left( t ,\, f \right) $ are weights to be chosen in order to maximize the Signal-to-Noise Ratio (SNR) for this measurement \cite{Smith:2019wny}. 
The measurement time is denoted by $T$. For simplicity, we are integrating over equal times, disregarding correlations between measurements done at different times. In the estimator, we subtracted the expectation value of the instrumental noise ${\tilde n}_O$ associated with the measurement ${\tilde \Delta F}_O$, so to obtain an unbiased characterization of the SGWB. From the estimator, we get the SNR 
\begin{equation}
{\rm SNR} = \frac{\left\langle {\cal C} \right\rangle}{\sqrt{\left\langle \left \vert {\cal C} \right\vert^2 \right\rangle}} \;, 
\end{equation} 
that, as we will see, can be made real by an appropriate choice of the weights $Q$. 

As we show in Appendix \ref{app:C-SNR}, the expectation value of the estimator is
\begin{equation}
\left\langle {\cal C} \right\rangle =  \sum_{OO'}   \frac{\tau}{2} \, \int_0^T d t \int_0^{+\infty}  d f  
\sum_{\ell,m}  {\tilde I}_{\ell m} \left( f \right)  \, R_{OO'}^{\ell m} \left( f \right) \left[ {\tilde Q}_{OO'} \left( t ,\, f \right) + {\tilde Q}_{O'O} \left( t ,\, -f \right) \right] \;, 
\label{<C>}
\end{equation} 
where we recall that the intensity multipoles coefficients have been defined in Eq. (\ref{<hh>}), while, $R_{OO'}^{\ell m} = c_{Oi} c_{O'j} \, R_{ij}^{\ell m}$.

In Appendix  \ref{app:C-SNR} we also show that, under the hypothesis that the noise dominates over the signal, 
\begin{eqnarray}
\left\langle \left \vert {\cal C}  \right\vert^2 \right\rangle &=& \sum_{OO'} 
\frac{\tau^2}{4}  \, \int_0^T d t 
\int_0^{+\infty}  d f   N_O \left(  f  \right) N_{O'} \left(  f  \right)  \, 
\left\vert {\tilde Q}_{OO'} \left( t  ,\, f \right) +  {\tilde Q}_{O'O} \left( t  ,\, - f \right) \right\vert^2  \;. 
\label{<C2>}
\end{eqnarray}   

Choosing the weigth function as discussed in Appendix \ref{app:C-SNR}, see Eq. (\ref{SNR-app}) and the following discussion, leads to the optimal SNR 
\begin{eqnarray} 
{\rm SNR}  &=& 
\frac{3 H_0^2}{4 \pi^2 \sqrt{4 \pi}} \, \sqrt{ \sum_{O,O'} \int_0^\infty d f \int_0^T d t \, 
\frac{\Omega_{\rm GW}^2 \left( f \right) }{f^6 \, N_{OO} \left( f \right) \, N_{O'O'} \left( f \right)} 
\left\vert \sum_{\ell,m} \, 
\delta_{\rm GW,\ell m} \left( f \right) \, 
R_{OO'}^{\ell m} \left( f  \right) \right\vert^2 
} \;. \nonumber\\ 
\label{SNR-ell-m}
\end{eqnarray} 
where $\delta_{\rm GW,\ell m}$ has been defined in \eqref{dec}.

\subsection{Sensitivity to $\ell-$multipoles} 
\label{subs:sensitivity-l}

Eq. (\ref{SNR-ell-m}) provides the SNR for the detection of a SGWB which is the sum of all possible multipoles contributions. Although we have not explicitly written it, the response functions ${\cal R}_{OO'}^{\ell m} \left( f  \right)$ also depend on time, as they are functions of the positions of the satellites. A full analysis of the separate contributions of the various multipoles would then require a component separation, which is in practice the inversion of the time-dependent streams measured by the satellite to the multipole amplitudes $p_{\ell m}$. We leave this discussion to section \ref{sec:mapmaking}. Here we estimate the relative sensitivity of LISA to different $\ell-$multipoles by assuming that only one multipole dominates the SGWB and that multipoles with the same $\ell$ but different $m$ are obtained from the same Gaussian statistics. This amounts to assuming a statistically isotropic SGWB, with correlators given by Eq. (\ref{Cell-def}). 

Taking this into account, the expected SNR (\ref{SNR-ell-m}) can be written as a sum over the various multipoles, 
\begin{equation}
\left\langle {\rm SNR} \right\rangle \equiv \sqrt{\sum_\ell \left\langle {\rm SNR} \right\rangle_\ell^2 } \;, 
\end{equation}  
where, for each multipole, 
\begin{equation} 
\left\langle {\rm SNR} \right\rangle_\ell = 
\frac{3 \, H_0^2}{4 \pi^2 \sqrt{4 \pi}} \, \sqrt{ \sum_{O,O'} \int_0^\infty d f \int_0^T d t \, 
\frac{\Omega_{\rm GW}^2 \left( f \right)}{f^6 \, 
N_O \left( f \right) \, N_{O'} \left( f \right)} 
C_\ell^{\rm GW} \, \left[ R_{OO'}^\ell \left( f \right) \right]^2  } \;, 
\label{SNR-l}
\end{equation} 
where we recall that the response function $R_{OO'}^\ell \left( f \right)$ is the quantity 
defined in Eq. (\ref{Rl-OOp}) and rescaled as in Eq. (\ref{<DFDF>}). In the following, we can work directly in terms of ${\tilde R}_{OO'}^\ell \left( f \right)$ by rescaling the noise functions accordingly, see Eqs. (\ref{NA,E}) and (\ref{NT}). Moreover, ad discussed above, the response function ${\tilde R}_{OO'}^\ell$ to a statistically isotropic signal is time-independent, so that the integral over time in Eq. (\ref{SNR-l}) simply results in the usual property that the SNR grows with the square root of the observation time. Finally, we factor out the uncertainty in the Hubble rate by dividing it by its rescaled value $h$ and by considering the $\Omega \, h^2$ combination, as it is standard. This leads to 
\begin{equation} 
\left\langle {\rm SNR} \right\rangle_\ell = 
\frac{3   \left( H_0 /h \right)^2}{4 \pi^2 \sqrt{4 \pi}} \,  \sqrt{ T \, \sum_{O,O'} \int_0^\infty d f \,  
\frac{\Omega_{\rm GW}^2 \left( f \right) h^4 }{f^6 \, {\tilde N}_O \left( f \right) \, {\tilde N}_{O'} \left( f \right)} 
C_\ell^{\rm GW} \, \left[ {\tilde R}_{OO'}^\ell \left( f \right) \right]^2  } \;. 
\label{SNR-l-2}
\end{equation}    
From this expression we define the ``channel-channel'' sensitivity 
\begin{equation} 
\Omega_{\rm GW,OO',n}^\ell \left( f \right) \, h^2 \equiv \, \frac{4 \pi^2 \sqrt{4 \pi}}{3 \left( H_0 / h \right)^2} \, \frac{f^3 \, \sqrt{{\tilde N}_O \left( f \right) \, {\tilde N}_{O'} \left( f \right) } }{ {\tilde R}_{OO'}^\ell \left( f \right)} \;, 
\label{sensitivity-l-OOp}
\end{equation}  
as well as the optimally weighted sum over the three channels 
\begin{equation}
\Omega_{\rm GW,n}^\ell \left( f \right) \, h^2 \equiv
\left\{ \sum_{O,O'} \left[ \frac{1}{\Omega_{\rm GW,OO',n}^\ell \left( f \right) \, h^2 } \right]^2 \right\}^{-1/2} \;. 
\label{sensitivity-l}
\end{equation} 

\begin{figure}[t!]
\centerline{
\includegraphics[width=\textwidth,angle=0]{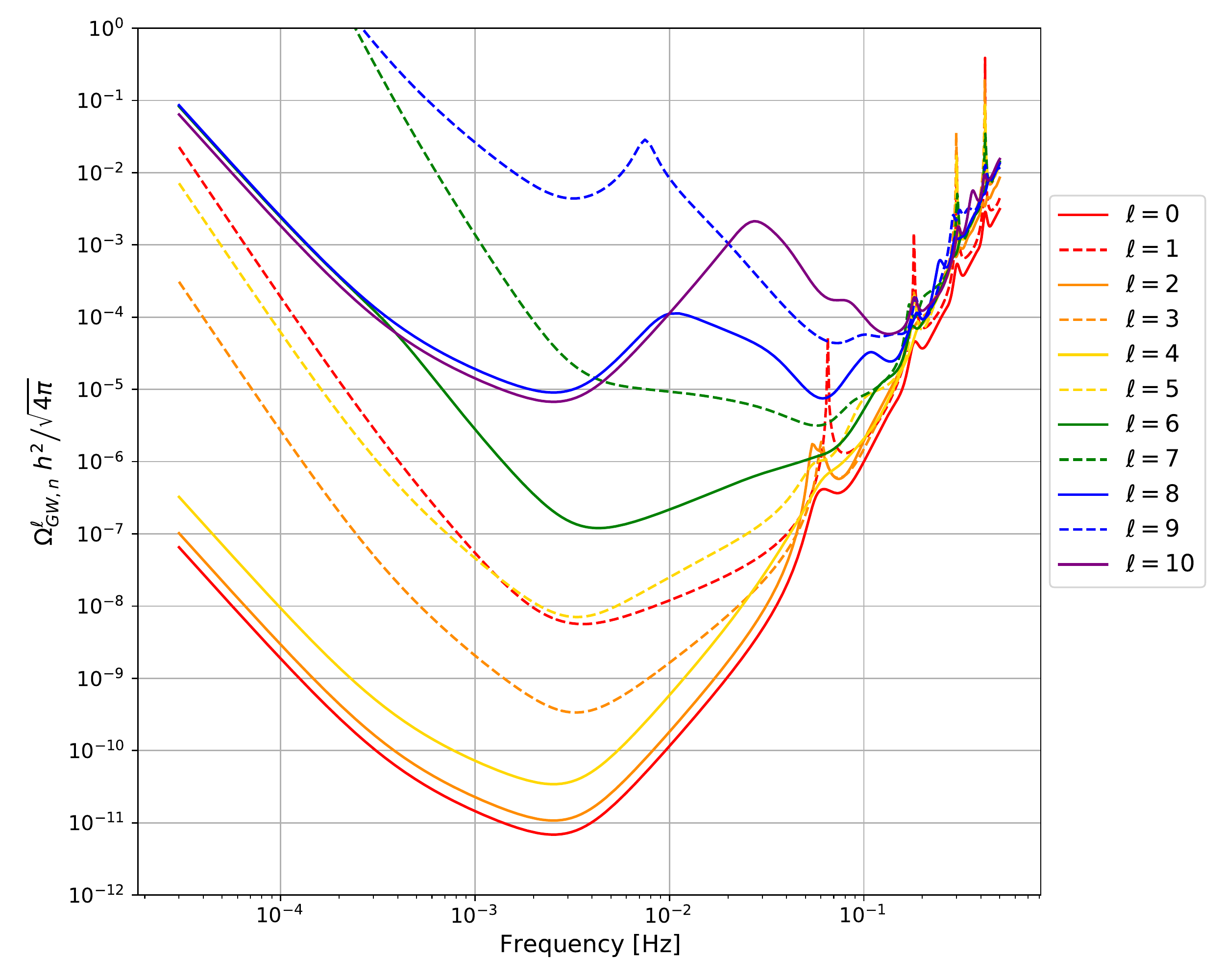}
}
\caption{\it Estimated LISA sensitivity to a given multipole $\ell$ of the SGWB, for multipoles up to $\ell=10$. Even (odd) multipoles are shown with solid (dashed) lines. The sensitivity is obtained by optimally summing over the LISA channels, see Eqs. (\ref{sensitivity-l-OOp}) and (\ref{sensitivity-l}).
}
\label{fig:LISA-sensitivity-l}
\end{figure}

The total sensitivity to the $\ell-$multiple is shown in Figure \ref{fig:LISA-sensitivity-l} for multipoles up to $\ell =10$. From this quantity, we can immediately obtain 
\begin{equation}\label{eq:ang_snr}
\left\langle {\rm SNR} \right\rangle_\ell^2 = T \, 
\int_0^\infty d f \left[ \frac{\sqrt{C_\ell^{\rm GW}} \, \Omega_{\rm GW} \left( f \right) \, h^2}{\Omega_{\rm GW,n}^\ell \left( f \right) \, h^2} \right]^2 \;. 
\end{equation} 
We note that the curves shown in Figure 
\ref{fig:LISA-sensitivity-l} are rescaled by $Y_{00} = 1/\sqrt{4 \pi}$, in such a way that the curve shown for $\ell =0$ coincides with the SciRD (Science Requirement Document) sensitivity curve for a homogeneous signal \cite{Babak:2021mhe} obtained from summing over the $A,E,T$ channels.

\subsection{Sensitivity to kinematic anisotropies}

Doppler anisotropies induced by the motion of the detector with respect to the SGWB rest frame count 
among the guaranteed features of the SGWB. In fact, already the early work \cite{Allen:1996gp}, which  sets the
basis for the analysis of SGWB anisotropies with ground-based GW interferometers, estimated the prospects for ground-based detectors to measure the kinematic
dipole of the SGWB. In this subsection we briefly consider the same question in the context of LISA. 

\begin{figure}[t!]
\centerline{
\includegraphics[width=.51 \textwidth,angle=0]{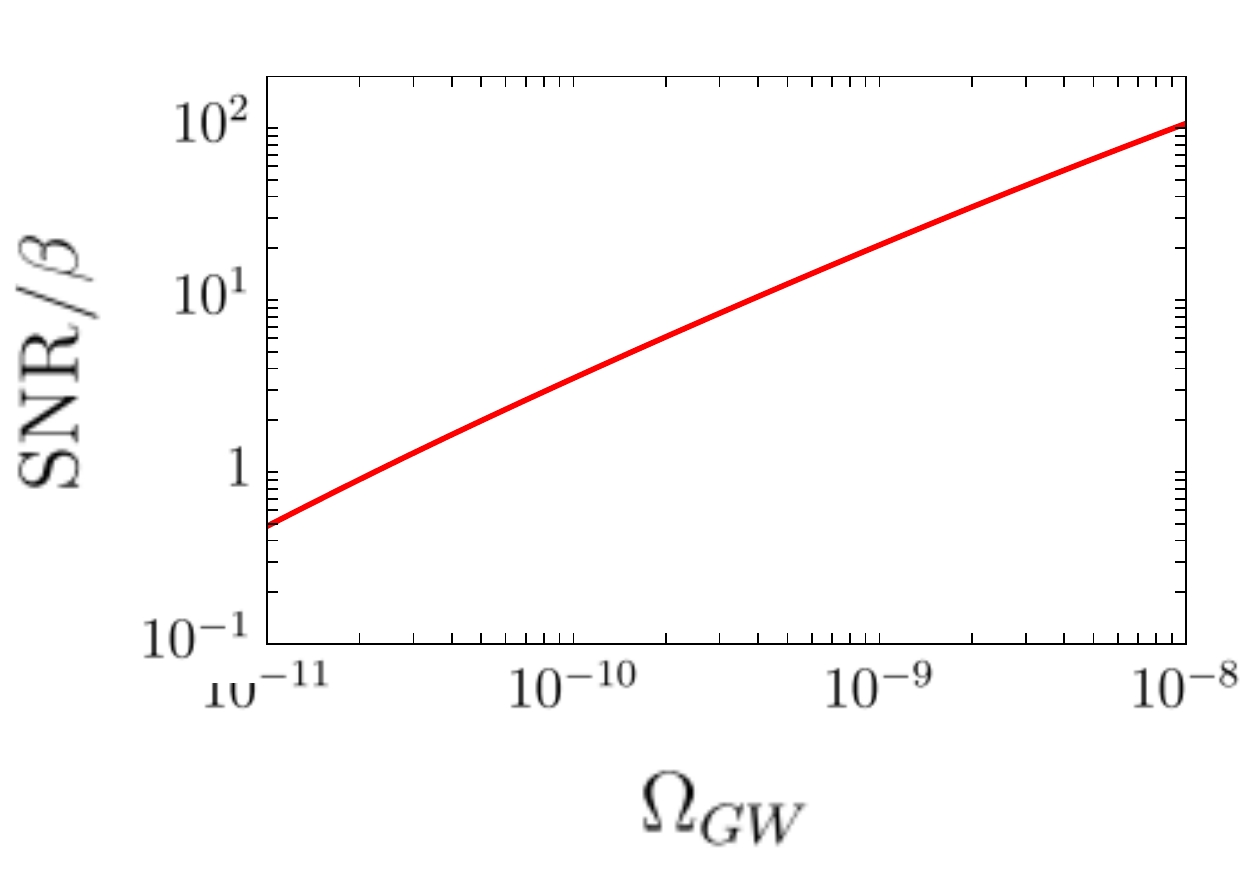}
\hspace{.5cm}
\includegraphics[width=.5 \textwidth,angle=0]{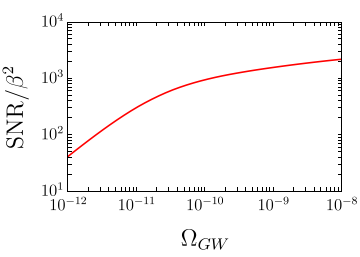}
}
\caption{\it The SNR for the dipole (left) and the quadrupole (right) induced by boosting an isotropic SGWB with fractional energy density $\Omega_{GW}$, assumed to be scale free across the LISA band. An observation time of $T=1$ year is assumed.}
\label{fig:boostdiquadrupole}
\end{figure}

The size and properties  of kinematic anisotropies  depend
on the frequency profile of the rest-frame SGWB energy density $\Omega_{\GW} (f)$. This fact can be important
for enhancing the amplitude of kinematic anistropies in certain early-universe scenarios where the SGWB has
rich features, as the ones discussed in Section \ref{sec:cosmo}. 

We 
consider two cosmological frames: the first,
denoted with  $\mathcal{S}'$,  is comoving with the SGWB rest frame; the second,  denoted with  $\mathcal{S}$,  moves with constant velocity  with respect to
  the rest frame  $\mathcal{S}'$. We assume that the SGWB density parameter
   in the rest frame,   $\Omega_{\rm GW}'(f)$, is perfectly isotropic and depends only on frequency $f$. 
  A boost transformation relates the SGWB density parameter in the rest frame
$\mathcal{S}'$ to the one in the moving one  $\mathcal{S}$. We indicate with 
 $\bv=\beta \hat{\bv}$ (where $\beta=v$ in units with $c=1$) the velocity of the frame $\mathcal{S}$ with respect to the rest frame $\mathcal{S}'$.

In the technical appendix \ref{app_boost} we derive the resulting expression of an anisotropic SGWB energy density $\Omega_{\GW}(f, \hat{\bf n})$ as a function of the rest-frame density $\Omega_{\rm GW}'(f)$. Assuming that the parameter $\beta$ is small, we can Taylor expand up to second order in $\beta$ and write
\bea
\Omega_{\rm GW}(f, \hat \bn)&=&\Omega'_{\rm GW}(f)
\left\{ \left[ 1+M \left(f \right)\right]+
 \hat \bn \, \hat \bv\, D \left( f \right)+
 \left[ 
 \left(\hat \bn \, \hat \bv \right)^2-\frac13
  \right]
 \,Q(f)
 \right\}\,,
  \label{genexpom4ab}
\eea
The functions of frequency $M$, $Q$, $D$,  control respectively the contributions of kinematic
effects to the monopole, dipole, and quadrupole of GW energy density in the detector frame. They read
\bea
\label{monanis1}
M(f)&=&\frac{\beta^2}{6} \left( 8+n_\Omega \left( n_\Omega-6\right)
+\alpha_\Omega
\right)\,,
\\
\label{dipanis1}
D(f)&=& \beta \left(4-n_\Omega\right)\,,
\\
\label{quapanis1}
Q(f)&=&\beta^2\left(10-\frac{9 n_\Omega}{2} +\frac{n_\Omega^2}{2}+\frac{\alpha_\Omega}{2}\right)\,.
\eea
In analogy with CMB literature, we introduce the SGWB spectral tilts
 \bea
\label{defno1a}
n_{\Omega}(f)&=&\frac{d\,\ln \Omega'_{\rm GW}(  f)}{d\,\ln f}\,,
\\
\label{defao1a}
\alpha_{\Omega}(f)&=&
\frac{d\,n_{\Omega}(f)}{d\,\ln f}\,.
\eea   
The expressions \eqref{monanis1}, \eqref{dipanis1}, \eqref{quapanis1} quantitatively demonstrate that 
enhanced spectral tilts can amplify kinematic anisotropies in certain scenarios.

We plot in Figure \ref{fig:boostdiquadrupole} the SNR for LISA  to  detect 
the kinematic dipole and quadrupole induced by a scale-invariant profile of $\Omega'_{\rm GW}(  f)\,=\,$ constant in
the SGWB rest frame. Notice the different vertical scale in the two plots, due to the fact that LISA sensitivity to the quadrupole is a factor $\sim 10^3$ better than that to the dipole, as discussed in the previous sections.

\begin{figure}[t!]
\centerline{
\includegraphics[width=.6 \textwidth,angle=0]{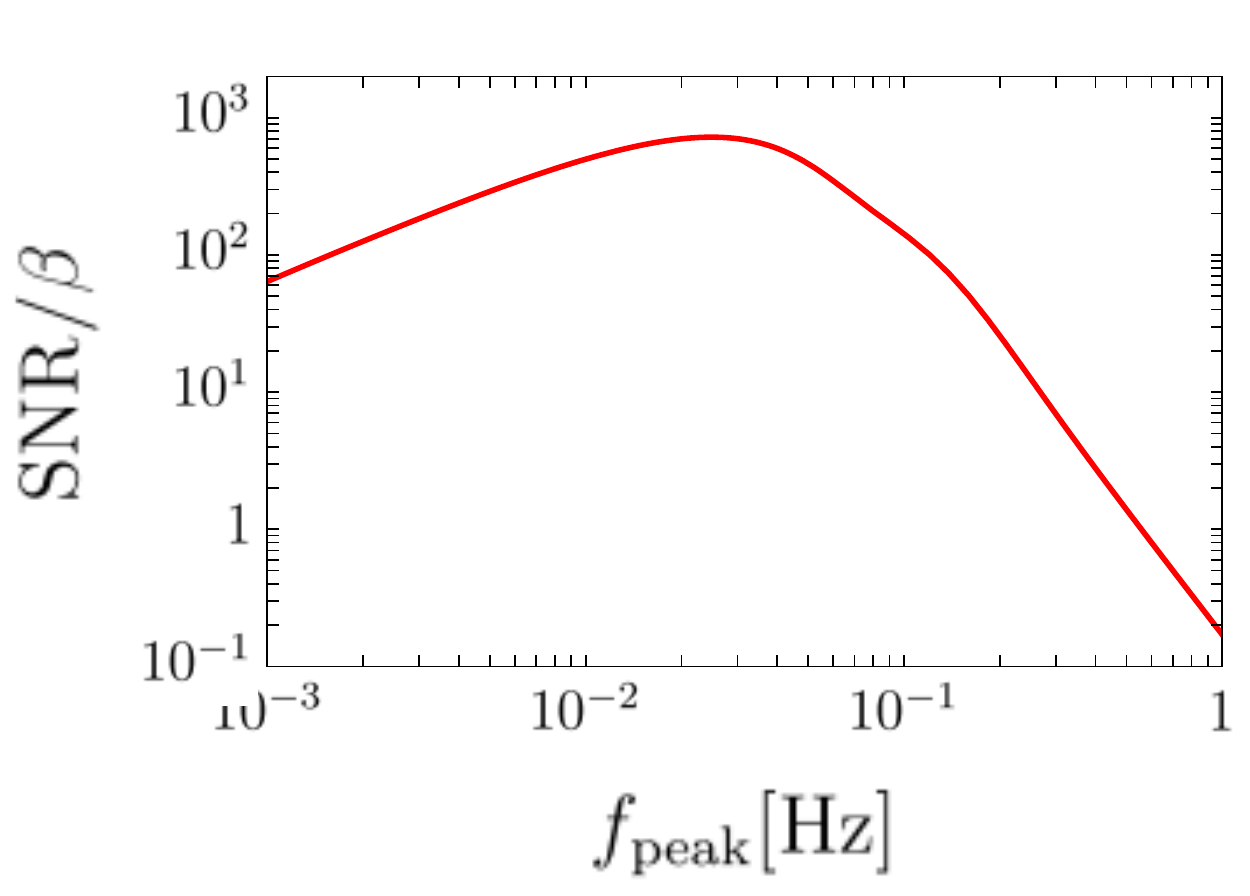}
}
\caption{\it The SNR for a broken power law, inspired by models of strongly first-order phase transitions, versus the break frequency. For these models, the total energy density contributes $0.1\%$ of the total energy density during the radiation era. An observation time of $T=1$ year is assumed.}
\label{fig:LISA-sensitivity-PT}
\end{figure}

We also show in Figure \ref{fig:LISA-sensitivity-PT} the sensitivity to the dipole induced by a boost with velocity $\beta$ on the SGWB spectrum generated by a strongly first order phase transition. We model the spectral density as a broken power law, using Eq.~(8) of Ref.~\cite{Caprini:2019egz}, and illustrated in Fig. 3 therein. We allow the location of the break to vary, but fix the amplitude so that the total energy density integrated over all frequencies contributes $0.1\%$ of the critical energy density during the radiation era. The SNR scales linearly with the amplitude, so boosting to $1\%$ raises the SNR by 10. In this case, the rich frequency profile of the SGWB energy density in the rest frame leads to a pronounced frequency-dependence of the amplitude of the SNR in the LISA band. 



\section{\sc Fisher Forecast}
\label{sec:fisher}

The next step of our analysis is to estimate, for the LISA strain and angular resolution sensitivity, statistical forecasts on the detectability of the lowest multipoles of the SGWB angular power spectrum, using a Fisher matrix method.   

We consider a total observation time of $t_{\rm obs} = 3$ years (corresponding to the total 4 years nominal mission assuming 75\% efficiency), and a frequency resolution $\Delta f = 10^{-6}$ Hz, which corresponds to segmenting the TDI data stream into chunks of 11.5 days (i.e.\ the inverse of the frequency resolution), and using as the final spectrum the average over the spectra of the chunks.

We work under the assumption of statistical isotropy, where different multipoles $\ell$ are uncorrelated and all orders $m$ are drawn from the same distribution for each multipole (only under this assumption it is justified to average over different parts of the sky, or in practice different time segments). We consider each multipole separately, in order to obtain a measure of the information contained in each of them. In practice, when trying to recover the angular power spectrum from a sky map, different multipoles become correlated, which would degrade the results obtained here.

Following the result obtained in Eq.\ \eqref{eq:ang_snr}, we define the SGWB power spectrum at multipole $\ell$ as
\begin{equation}
\Omega^\ell_{\mathrm{GW}} (f) h^2 =
\sqrt{C_\ell^{\rm GW}} \, \Omega_{\rm GW}(f) h^2 \,,
\end{equation}
where $C_\ell^{\rm GW}$ is the angular power spectrum of the GW density contrast as defined in Eq.\ \eqref{Cell-def}.

For the sake of generality, we consider a power-law SGWB spectrum peaking at a fiducial  multiple $L$ only, parameterized by the logarithmic amplitude $\log_{10}A_{\rm c}$ at a pivot frequency $f_{\rm c} = 2.5\cdot10^{-3}$ Hz, that is chosen close to the frequency where LISA has the best sensitivity, and by a spectral index $\alpha$,
\begin{equation} \label{eq:fisher:signal}
\Omega^\ell_{\mathrm{GW}} (f) h^2 = \delta_{\ell,L} 10^{\log_{10} A_\mathrm{c}}  \left(\frac{f}{f_{\rm c}}\right)^\alpha\,.
\end{equation}
For each multipole $\ell$ and channel combination $OO'$, we assume a Gaussian likelihood over the averaged data $\mathcal L_\ell$ given by
\begin{equation} \label{eq:fisher:logl}
\ln \mathcal{L}_\ell = - \frac{N_\text{c}}{2} \sum_{OO'} \sum_k \frac{\left(\mathcal{D}_{OO',\ell}^{(k)} - \mathcal{D}_{OO',\ell}^{(k),{\rm th}}\right)^2}{\sigma_{OO',\ell}^{(k) 2}}\,,
\end{equation}
where $N_{\rm c}$ is the number of data segments in the analysis; the sum runs over frequencies (or frequency bins) $f_{k}$, $\mathcal D_{OO',\ell}$ denotes the averaged signal over the data segments in the channel combination $OO'$, and $\mathcal D_{OO',\ell}^{\rm th}$ is the theoretical ansatz for the data,
\begin{equation}
\mathcal{D}^{(k),\text{\rm th}}_{OO',\ell} = \tilde{R}_{OO',\ell}(f_k)\Omega^\ell_{\mathrm{GW}} (f_k) h^2 + \tilde{N}_{OO'}^\Omega(f_k)\,.
\end{equation}
where $\tilde{R}_{OO',\ell}$ is the frequency response of the detector and $\tilde{N}_{OO'}^\Omega$ is the noise as defined in the previous section expressed in Omega units. The variance can be expressed in terms of the theoretical ansatz as $\sigma_{OO',\ell}^{(k) 2} = \left(\mathcal{D}^{(k),\text{th}}_{OO',\ell}\right)^2$. In practice, instead of summing over channels in the likelihood, we consider a single data vector and compare it with the effective noise combination defined by Eq. \eqref{sensitivity-l} and shown in figure \ref{fig:LISA-sensitivity-l}, and drop the $OO'$ channel indices in what follows.

Assuming a fixed noise model, the Fisher information matrix for the likelihood defined in Eq.\ \eqref{eq:fisher:logl} is simply
\begin{equation}
 \mathcal{C}_{\theta \rho}^{-1} \equiv \mathcal{F}_{\theta \rho} =  \left. - \partial_\theta \partial_\rho \ln \mathcal{L} \right|_{\rm best fit}=N_{\rm c} \sum_k \left(\partial_\theta \Omega^\ell_{\mathrm{GW}} (f_k) h^2\right) \left(\partial_\rho \Omega^\ell_{\mathrm{GW}} (f_k) h^2\right)  \frac{1}{\sigma_\ell^{(k) 2}}~,
\end{equation}
where $\theta, \rho$ are a combination of the signal model parameters $\log_{10}A_{\rm c}$ and $\alpha$, and the corresponding partial derivatives are
\begin{equation}
\partial_{\log_{10} A_\mathrm{c}} \Omega^\ell_{\mathrm{GW}} h^2 = \log(10) \Omega^\ell_{\mathrm{GW}} h^2
\qquad {\rm and}\qquad
\partial_{\alpha} \Omega^\ell_{\mathrm{GW}} h^2 = \log\left(\frac{f}{f_c}\right) \Omega^\ell_{\mathrm{GW}} h^2~.
\end{equation}

\begin{figure}
    \centering
    \includegraphics[width=0.48\textwidth]{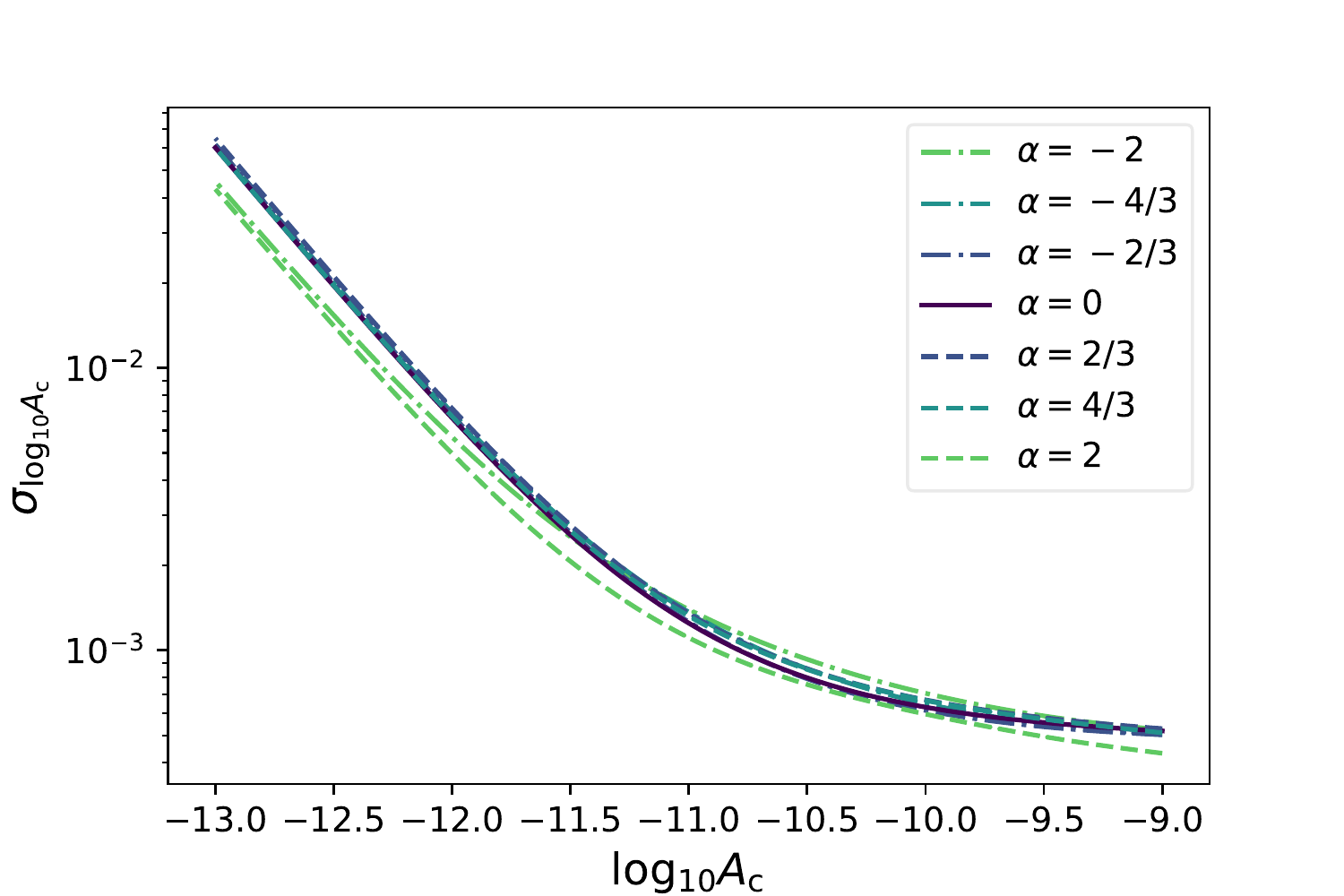} 
    \includegraphics[width=0.48\textwidth]{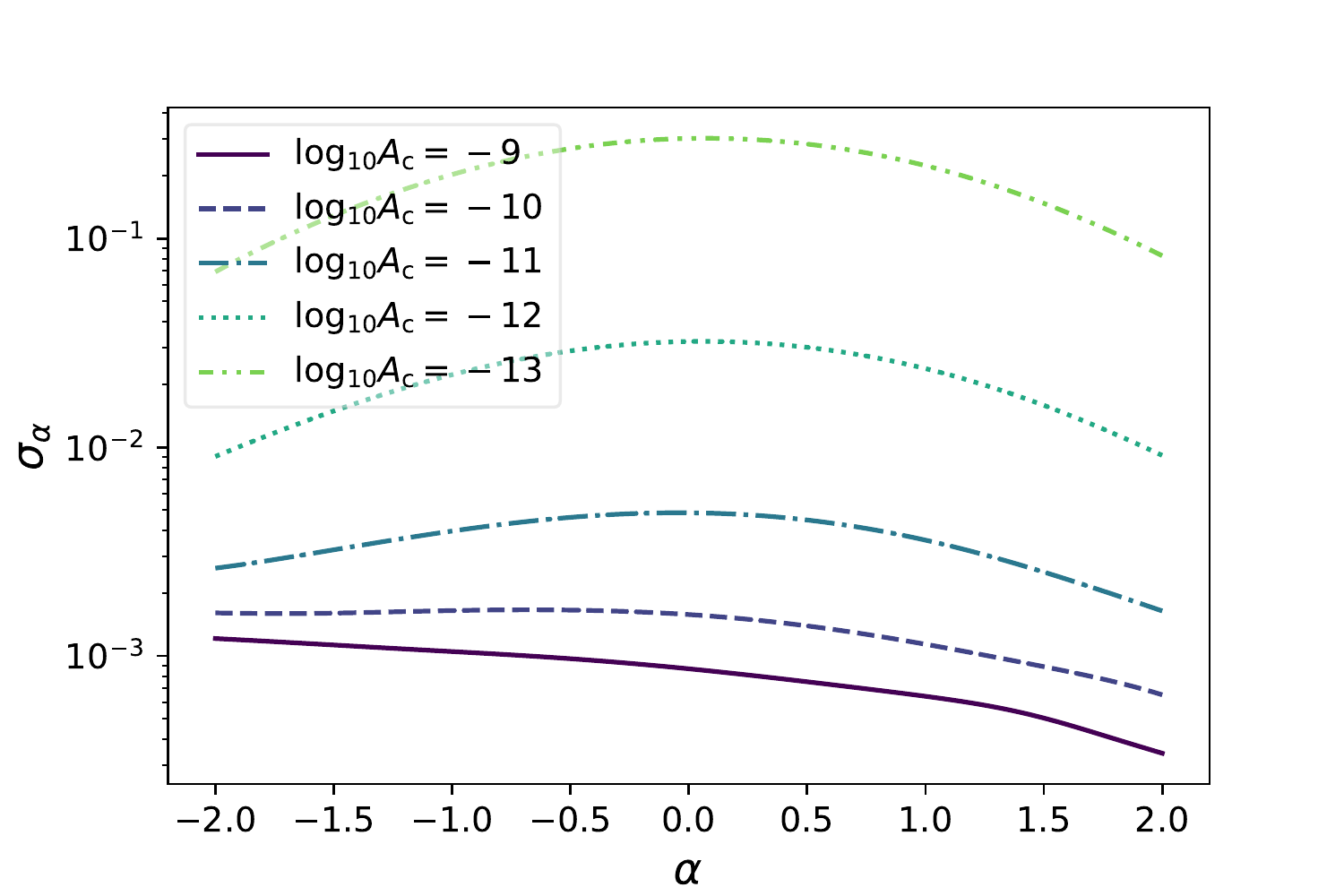} \\
    \includegraphics[width=0.48\textwidth]{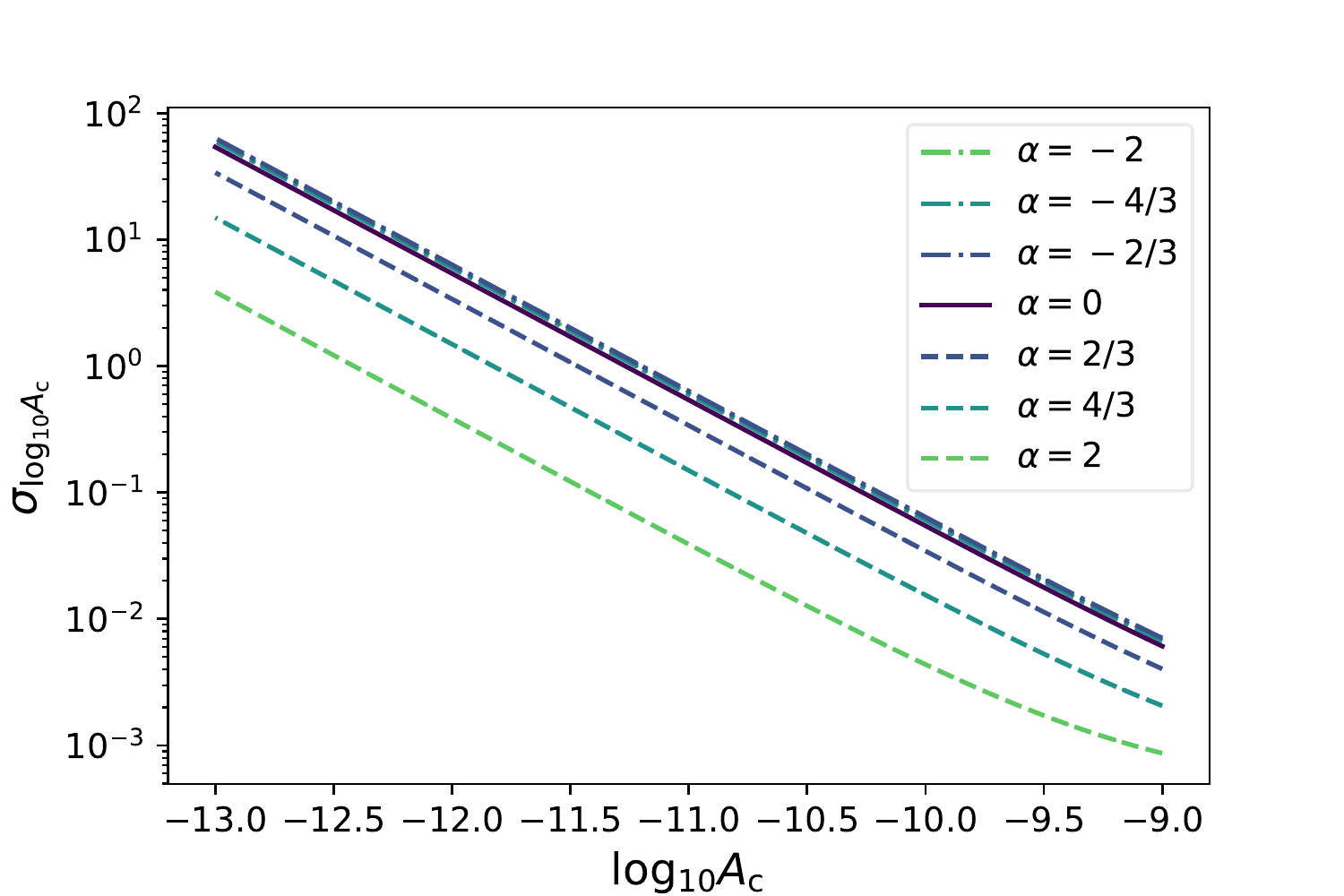}  \includegraphics[width=0.48\textwidth]{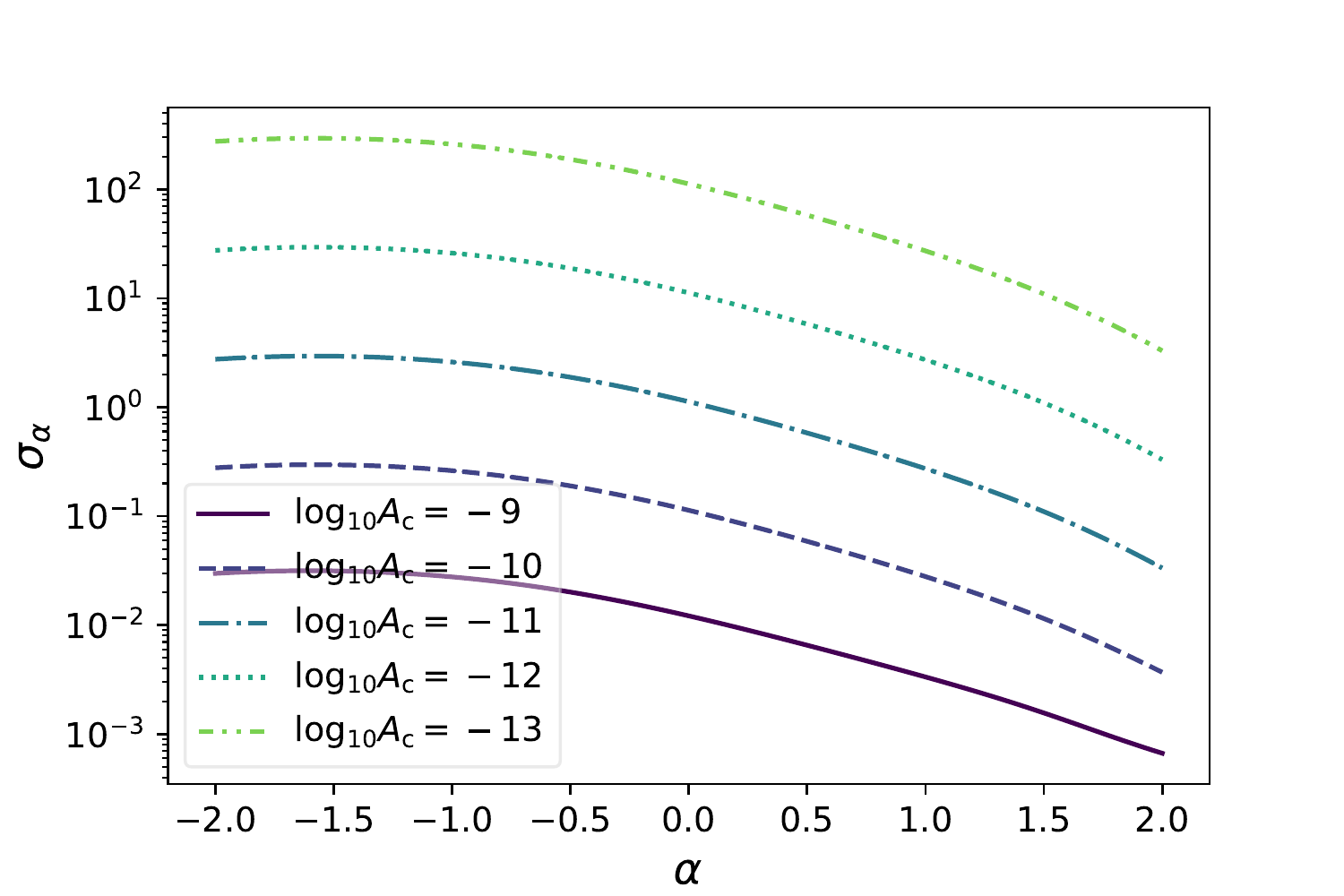} \\
    \includegraphics[width=0.48\textwidth]{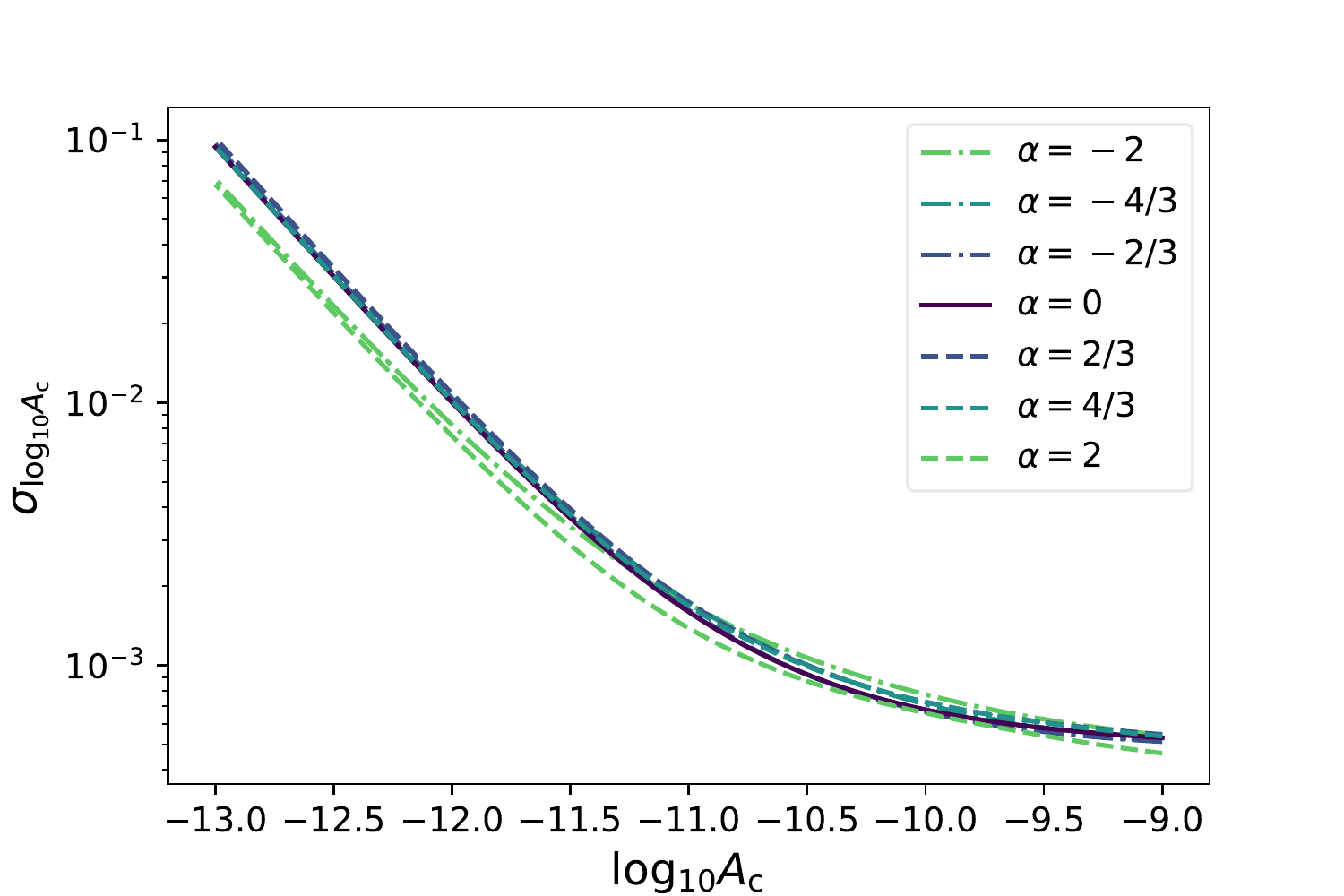}  \includegraphics[width=0.48\textwidth]{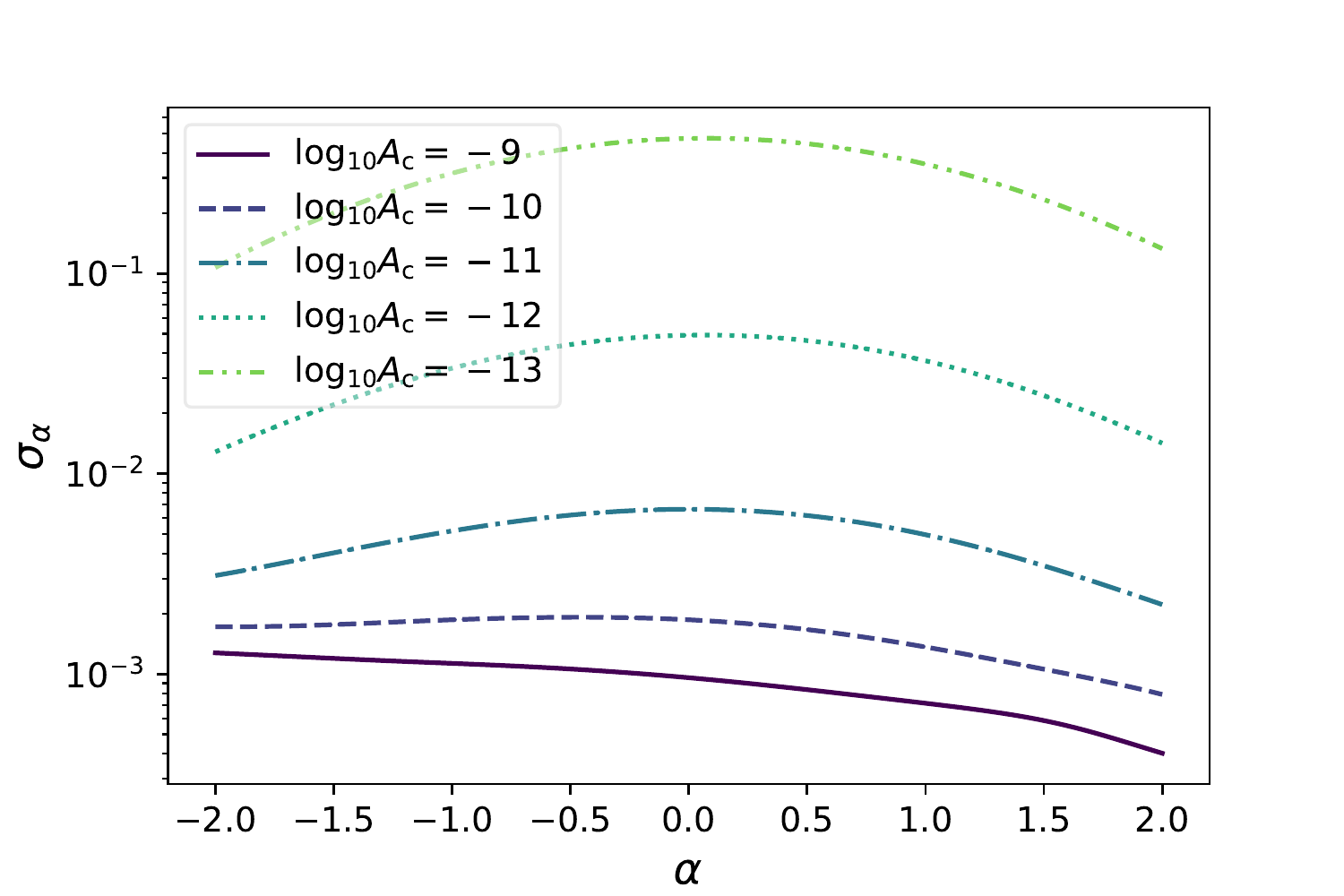} 
    \caption{\it The LISA marginalized $1\sigma$ forecasted limits on $\log_{10} A_\mathrm{c}$ (left panels) and $\alpha$ (right panels) at the multipole $l=0,1$ and $2$ (top, center and bottom panels), for a series of fiducial values the of SGWB amplitude and spectral index. See comments in main text.}
    \label{fig:forecasted_sigma_l}
\end{figure}

The estimated LISA sensitivity to a single-monopole power-law SGWB defined in Eq.\ \eqref{eq:fisher:signal} has been represented on Figure~\ref{fig:forecasted_sigma_l} for the monopole ($\ell=0$), dipole ($\ell=1$) and quadrupole ($\ell=2$), for a series of fiducial values of the SGWB amplitude $\log_{10}A_\mathrm{c}$ and spectral index $\alpha$. In all cases, the standard deviation for each parameter is considered marginalised over the other one (i.e.\ taken from the diagonal elements of the covariance matrix $\mathcal{C}_{\theta \rho}$, defined as the inverse of the Fisher information matrix).

As one can see in Figure~\ref{fig:forecasted_sigma_l}, for $\ell=0, 2$, sufficiently high log-amplitudes are recovered independently of the sign of the spectral index (but enhanced by stronger indices), due to the pivot frequency being chosen to approximately coincide with the peak in sensitivity at both multipoles. In contrast, for $\ell = 1$ positive spectral indices enhance the recovery of the amplitude. This is on the one hand because the corresponding sensitivity peaks at slightly larger frequency with respect to $f_c$; and on the other hand because of the milder slope of the sensitivity with respect to $\ell=0, 2$ towards high frequencies, so that the power law is closer to the high-frequency noise spectrum for lower $|\alpha|$ in $\ell=1$ than in $\ell=0, 2$ (see figure \ref{fig:LISA-sensitivity-l}).

For all multipoles, the spectral index is obviously recovered more effectively for higher log-amplitudes. In the optimal case of a signal amplitude of order $\Omega_{\rm GW} (f=f_\mathrm{c}) h^2 = 10^{-9}$, a null spectral index could be reconstructed with an uncertainty of order $10^{-3}, 10^{-2}, 10^{-3}$ for the $\ell=0,1,2$ multipoles respectively. In the more pessimistic case of $\Omega_{\rm GW} (f=f_\mathrm{c}) h^2= 10^{-13}$, for $\ell=0,2$ the spectral index could be reconstructed with an uncertainty of order $0.1$ or greater for largely positive or negative values of it, but this uncertainty approaches order one for SGWB spectra with a spectral index between $-1$ and $1$. In such a low-amplitude scenario, LISA will thus be more sensitive to models with a strongly varying SGWB spectrum. Notice how, for the same reasons described in the previous paragraph, the dipole $\ell = 1$ is more sensitive towards positive spectral indices,
whereas for $\ell = 0, 2$ the accuracy is almost symmetric with respect to the sign.

\section{\sc Map-making Method}
\label{sec:mapmaking}
In this section we briefly describe the maximum likelihood map-making method for stochastic backgrounds proposed in~\cite{Contaldi:2020rht} and provide estimates for the noise power spectrum $\cal N_\ell$ obtained by simulating and mapping the noise directly in the sky domain. Recently, another method to map the gravitational-wave sky with LISA has been developed and it is based on a Bayesian algorithm to map the power of the SGWB using a spherical harmonic approach~\cite{Banagiri:2021ovv}.

The maximum likelihood map-making with GW detectors relies on the specific {\it scan strategy} of the interferometer array, which describes how the sky signal is sampled as a function of time. The reconstruction of the GW sky and the angular resolution at which it may be achieved then depend on the amount of modes sampled throughout the whole duration of the observation.

To simplify the mapping procedure we assume that the anisotropic SGWB signal intensity $I$ has a simple power-law spectral shape which may be factored out, such that
\begin{equation}
    I(f,\hat{n}) = E(f)\,I(f_0,\hat{n})\,,
    \label{eq:specshape}
\end{equation}
where $E(f) = (f/f_0)^\gamma$, and $f_0$ is a specific reference frequency.

As for the scan strategy, we assume the spacecrafts follow three heliocentric, quasi-circular orbits remaining at a constant arm-length distance from each other, and that the noise in the detector is well understood. Specifically, as in Section~\ref{sec:response}, the noise is modelled by two contributions: acceleration noise and interferometer noise. For more details, see Equations (30) and (31) in~\cite{Contaldi:2020rht} and the description of the noise parameters in the official mock data release~\cite{Sangria}.

The data vector is defined (similarly to Equation~(\ref{eq:data_noise_model})) as $\textbf{d} = \textbf{R}h +\textbf{n}$, where the first term specifies the pure signal component, made up of the contraction between the linear detector response $\textbf{R}$ (see Equation~\ref{RA} in the appendix) and the SGWB strain $h$, and $\textbf{n}$ is the noise component. We keep the formalism general here for the sake of conciseness; note that for multiple LISA TDI channels $\textbf{d}$ is a vector in TDI space. To estimate the intensity $I(f_0,\hat{n})$ directly, the data are considered in frequency space: $\textbf{d}(f)$ with $f$ belonging to the appropriate frequency interval observed by LISA. We assume the noise is zero–mean and Gaussian with covariance $\textbf{N} = \textbf{n} \otimes \textbf{n}$ and the signal component is also Gaussian such that the total, signal plus noise, covariance of the data is $\textbf C = \textbf A \tilde{I} + \textbf N$. Here $\textbf A$ is the operator that describes the response of the detector to the strain intensity. This can be integrated in time and projected onto pixel or spherical harmonic space.  In the case of multiple TDI channels it represents the full correlated response matrix. Also note that in the case of multiple correlated TDI channels, e.g. $X$, $Y$, and $Z$ as considered in~\cite{Contaldi:2020rht}, the noise covariance is the full correlated covariance matrix with the auto-correlated noise model for the diagonal and cross-correlated model for the off-diagonal terms. $\tilde{I}$ is the observed realisation of GWB intensity on the sky. The likelihood $\mathcal{L}$ of the data is then 
\begin{equation}
    \mathcal{L} \propto \frac{1}{|\textbf C|^{1/2}} e^{-\frac{1}{2}{\textbf d}^\dagger \, \textbf C^{-1} \, {\textbf d} } \,, 
\label{Like}
\end{equation}
as described in \cite{Bond:1998zw} for the case of CMB mapping, it is possible to find the iterative solution which maximises $\mathcal{L}$,
\begin{align}
\tilde{I}_\alpha &=\frac{1}{2} \mathcal{F}_{\alpha\alpha'}^{-1} \cdot \text{Tr}\left[ \textbf C^{-1} \, \frac{\partial \textbf C}{\partial I_{\alpha'}} \, \textbf C^{-1} \, (\textbf D - \textbf N) \right]\,, \label{eq:itersol}\\
 \mathcal{F}_{\alpha\alpha'} &=\frac{1}{2} \text{Tr} \left[ \textbf C^{-1} \, \frac{\partial \textbf C}{\partial I_{\alpha}} \, \textbf C^{-1} \, \frac{\partial \textbf C}{\partial I_{\alpha'}} \right]\,, \label{eq:fisher}
\end{align}
where $\mathcal{F}$ is the Fisher information matrix and $\textbf D\equiv \textbf d^\dagger \otimes \textbf d$. In practical applications, given the constraints on scan strategies and response functions for gravitational wave observations, the Fisher matrix will need to be regularised in order the correct iterative solution to be found using Eq.~(\ref{eq:itersol}). Here, the intensity of the sky-map is indexed generically by $\alpha$ such that $I_\alpha$ are the set of ``parameters'', on which the signal component of $\textbf C$ depends, that have to be estimated. For any particular application, the indices $\alpha$ could stand for either map domain pixels ``$p$'' or spherical harmonic domain multipoles ``$\ell, m$''. The choice here is to work in the map domain, as it is not overly expensive in this case, and leads to a clearer understanding and regularisation of the Fisher matrix, as explained in~\cite{Renzini2018,Contaldi:2020rht}; in the latter, tests of the method and a regularisation technique are presented. In principle, the $\tilde{I}_\alpha$ could be estimated for a frequency band as narrow as the resolution permits, however this would result in a poorly regularised problem. To improve this the estimation must be done using wider frequency bands. Here we simply adopt the broad-band limit by assuming a spectral shape as in Eq.~(\ref{eq:specshape}). In practice this means the traces in Eqs.~(\ref{eq:itersol}, \ref{eq:fisher}) include a sum over the full frequency response, such that the final estimate is given with respect to a single reference frequency $\tilde{I}_\alpha(f_0)$.

Note that the term $\frac{\partial \textbf C}{\partial I_{\alpha'}} = \textbf A_\alpha$ represents the directional quadratic response of the detector, and is equal to $\textbf A = \textbf R \otimes \textbf R$ up to appropriate normalisation factors. $\textbf A_\alpha$ will, in general be time dependent, presenting a sky modulation with period of one year. Hence, to apply this mapping algorithm effectively, the data must be segmented into short observation time-frames, throughout which the sky response is assumed to be constant {\sl and} for which the noise can be estimated accurately. The duration of each segment $\Delta t$ also sets the lower bound on the observable frequency window, hence there is a trade-off between frequency and (potential) sky resolution: in principle a shorter time window $\Delta t$ will allow access to higher pixel resolution, but it will also fix the frequency resolution to $1/\Delta t$. Assuming statistical independence between time-frames, the two traces in Eqs.~(\ref{eq:itersol}, \ref{eq:fisher}) are then obtained by averaging\footnote{For actual data this average would be weighted by noise estimates but here we assume the noise is constant and uncorrelated between time-frames.} over all the frames available. 

\begin{figure}
    \centering
    \includegraphics[width=0.75\linewidth]{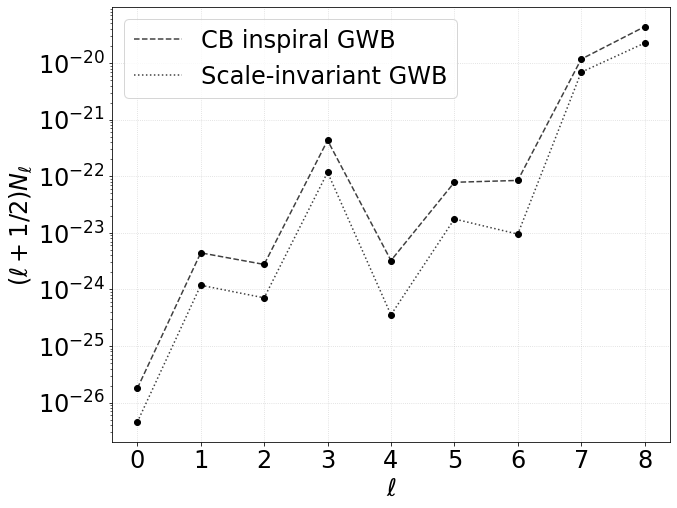}
    \caption{\it Noise power spectrum for LISA in $\Omega_{\rm GW}$ units at reference frequency $f_0 = 0.01$ Hz. Note the $\cal N_0$ here is in good agreement with the curve shown in~\cite{Alonso2020}, Figure 4, taking into account that the observation time considered here is one year, whereas in~\cite{Alonso2020} it is four.}
    \label{fig:Nell_plot}
\end{figure}

For the purpose of this paper, we use the method described above directly in pixel space to calculate the noise power spectrum $\cal N_\ell$, assuming the LISA noise curves as in~\cite{Contaldi:2020rht}, over an observation time of one year. This is simply done by running the iterative map-maker over a noise-dominated data set, such that the Fisher matrix -- setting now $\alpha \equiv p$ -- reduces to
\begin{align}
 \mathcal{F}_{pp'} &= \frac{1}{2}\text{Tr} \left[ \textbf N^{-1} \, \textbf A_p \, \textbf N^{-1} \, \textbf A_{p'} \right]\,. \label{eq:fisherpp}
\end{align}
The statistically isotropised, angular power spectrum of the noise $\cal N_\ell$ can then be estimated by expanding and inverting the Fisher matrix~\cite{Renzini2019b}:
\begin{equation}
    \tilde {\cal N}_\ell  = \frac{1}{2\ell + 1}\sum_m\,\Big|\left( {\cal Y}^{}_{\ell m, p} \, {\cal F}_{pp'} \,{\cal Y}^\dagger_{p',\ell m} \right)^{-1}\Big|\,,
    \label{eq:rotate2}
\end{equation}
where the linear operator ${\cal Y}_{\ell m, p}\equiv Y_{\ell m}(\hat{p})$ and similar for its adjoint. 
This is similar in spirit to what is presented in~\cite{Alonso2020}, however note that here the estimate is obtained by simulating and integrating the full scan strategy, without assuming the noise is isotropic to begin with. The $\cal N_\ell$ estimates are shown in Figure~\ref{fig:Nell_plot}, in units of $\Omega^2_{GW}$ at reference frequency $f_0 = 0.01$ Hz for a quicker comparison with signal models. This matches the convention chosen in~\cite{Alonso2020}, and the $\ell=0$ mode of the two estimated noise power spectra matches well. Note that in Figure~\ref{fig:Nell_plot} an observation time of four years is assumed.

The discrepancy at odd $\ell$s between the noise power spectrum obtained by the inverse Fisher matrix and the analytic computation presented in~\cite{Alonso2020} is in part due to the substantially different sampling of the $m$ modes. While these are marginalised over at a fixed time in Equation~(\ref{Rl-OOp}) to obtain the instantaneous $\ell$--mode response, in the map--making procedure these are kept into account, and through the scan strategy contribute to breaking the degeneracies in the odd $\ell$--modes of (instantaneous) $\tilde{R}^\ell_{OO}$. The higher-$\ell$ section of the curve is highly dominated by the conditioned inversion; in fact, the noise covariance matrix found in this case is highly singular, and $\sim$90\% of its eigenvalues has been discarded to produce the curve in Figure~\ref{fig:Nell_plot}. The conditioning has a stronger impact on the higher-$\ell$ end of the angular spectrum, as the response of the detector is weaker at higher angular scales. It is therefore difficult to compare numerical estimates of the effective noise at different angular scales to the result of analytical estimates. This is an active field of research, and it is clear that a robust regularisation scheme will be required when attempting to reconstruct the higher modes of the angular power spectrum with this configuration of the LISA instrument. Inevitably, the presence of complicating factors such as non-stationarity, noise uncertainties, and scanning systematics in real data will exacerbate this issue and mapping techniques will require significant developments in order to reconstruct anisotropies as optimally as possible.

\section{\sc Conclusions}
\label{ sec:conclusions}
The anisotropies of the SGWB represent a powerful tool to characterize and distinguish the different sources of GWs. We have seen in this paper how different GW sources are characterized by different angular spectra. Such anisotropies have mainly two contributions: one directly related to the production mechanism of each particular GW source, and one being an effect of the propagation of GWs on our perturbed Universe, which is common for all the GWs sources. We have made an overview of the main cosmological and astrophysical sources characterized by anisotropies, that are expected to be present in the LISA frequency band. We have presented the angular spectrum for different cosmological backgrounds (i.e., inflation, phase transition, PBH and cosmic strings) and an astrophysical one (Solar Mass Black Hole Binaries). We have then built a SNR estimator to quantify the sensitivity of LISA to different multipoles. To do this, we have computed the responses of LISA in harmonic space as functions of frequency for the AET TDI channels. We have also derived the analytic form of the responses in the low frequency limit. It is important to stress that, when anisotropic signals are considered, both the auto-correlation responses (i.e., AA, EE) and cross-correlation ones (i.e., AE, AT) are different from zero. We have shown how LISA will have a better sensitivity to detecting a quadrupole (i.e., $\ell=2$) than it will for the dipole (i.e., $\ell=1$). We have quantified the SGWB energy density required to observe the kinematic dipole and quadrupole induced by the motion of the LISA detector with respect to the SGWB rest frame. We found that an $\Omega_{\rm GW}\sim 2\times 10^{-11}$ is required to observe a dipolar signal, while the sensitivity to the quadrupole is a factor $\sim 10^3$ larger than that to the dipole. We have also performed a forecast of the detectability of the lowest multipoles of the SGWB angular power spectrum through a Fisher matrix analysis. We have shown that for $\ell=0, 2$ sufficiently high amplitudes are recovered independently of the sign of the spectral index (but enhanced by stronger indices). On the other hand for $\ell = 1$, positive spectral indices enhance the recovery of the amplitude; conversely the spectral index is recovered more effectively for higher log-amplitudes, for all multipoles. Finally, taking into account the LISA motion and the sky scan strategy, we have applied a maximum likelihood map-making technique to extract the noise angular power spectrum $\cal N_\ell$ as a function of the multipole $\ell$.

The LISA sensitivity and angular resolution will allow to detect the anisotropies of the SGWB, opening the possibility to use them in the process of characterization of the SGWB, and also to study their correlation to other cosmological tracers such as the Cosmic Microwave Background~\cite{Adshead:2020bji, Ricciardone:2021kel, Braglia:2021fxn, Dimastrogiovanni:2021mfs} and galaxies as tracers of the Large-Scale Structure~\cite{Canas-Herrera:2019npr, Mukherjee:2019oma, Mukherjee:2019wcg, Mukherjee:2020hyn, Mukherjee:2020mha}. This represents an exciting possibility to use LISA to explore our universe in a completely new perspective.

\acknowledgments
It is a pleasure to thank Valerie Domcke, Juan Garcia-Bellido and Sabino Matarrese for useful discussions. We acknowledge the LISA Publication and Presentation committee, in particular Sharan Banagiri for carefully reading and useful comments on the draft. N.B. and D.B. acknowledge partial financial support by ASI Grant No. 2016-24-H.0. R.C. is supported in part by U.S. Department of Energy Award No. DE-SC0010386.
CRC acknowledges support under a UKRI Consolidated Grant ST/T000791/1.
V.DL. and A.R. are supported by the Swiss National Science Foundation 
(SNSF), project {\sl The Non-Gaussian Universe and Cosmological Symmetries}, project number: 200020-178787. M.F.  would like to acknowledge support from the “Atracción de
Talento” CAM grant 2019-T1/TIC15784. DGF (ORCID 0000-0002-4005-8915) is supported by a Ram\'on y Cajal contract with Ref.~RYC-2017-23493, by the project PROMETEO/2021/083 from Generalitat Valenciana, and by the project PID2020-113644GB-I00 from Ministerio de Ciencia e Innovaci\'on.
G.F. acknowledges financial support provided under the European
Union's H2020 ERC, Starting Grant agreement no.~DarkGRA--757480 and under the MIUR PRIN programme, and support from the Amaldi Research Center funded by the MIUR program ``Dipartimento di Eccellenza" (CUP:~B81I18001170001).
M.Pe. is supported by Istituto Nazionale di Fisica Nucleare (INFN) through the Theoretical Astroparticle Physics (TAsP) and the 
Inflation, Dark Matter and the Large-Scale Structure of the Universe (InDark) project.
The work of M.Pi. was supported by STFC grants ST/P000762/1 and ST/T000791/1. M.Pi. acknowledges support by the European Union's Horizon 2020 Research Council grant 724659 MassiveCosmo ERC- 2016-COG.
A.R. acknowledges funding from Italian Ministry of Education, University and Research (MIUR) through the ``Dipartimenti di eccellenza'' project Science of the Universe.
M.S. is supported in part by the Science and Technology Facility Council (STFC), United Kingdom, under the research grant ST/P000258/1. 
The work of LS is partially supported by the US-NSF grants PHY-1520292 and PHY-1820675.
G.T. is partially supported by the STFC grant ST/T000813/1.
S.C. acknowledges support from the Belgian Francqui Foundation through a Francqui Start-up Grant. 

\newpage
\begin{appendix}
\section*{Appendixes}
\label{sec:appendix}
\section{Properties of the anisotropic response function}
\label{Appendix_TDI}

We insert the expression~\eqref{GW-deco} into~\eqref{deltaT} and perform the line of sight integration, obtaining
\begin{eqnarray} 
\Delta T_{12} \left( t \right)  &=&  L \, 
\int d^3 k {\rm e}^{-2 \pi i \vec{k} \cdot \vec{x}_2  }  \sum_A \Bigg[ 
{\rm e}^{2 \pi i k  \left( t - L  \right)} \, 
{\cal M} \left( \vec{k} ,\, {\hat l}_{12} \right)
{\tilde h}_A \left( \vec{k}  \right) 
{\cal G}^A \left( {\hat k} ,\, {\hat l}_{12} \right) 
\nonumber\\ 
&& \quad\quad\quad\quad +  
{\rm e}^{-2 \pi i k  \left( t - L  \right) }  
{\cal M}^* \left( - \vec{k} ,\, {\hat l}_{12} \right) 
{\tilde h}_A^* \left( - \vec{k}  \right) \, 
{\cal G}^{A*} \left( - {\hat k} ,\,  {\hat l}_{12} \right) \Bigg] \,,
\end{eqnarray} 
where we have defined ${\tilde h}_A \left( \vec{k}  \right) \equiv {\tilde h}_A \left( k ,\, {\hat k} \right) /\, k^2$ and 
\begin{equation}
{\cal M} \left( \vec{k} ,\, {\hat l}_{ij} \right) \equiv  
{\rm e}^{i \pi L k \left( 1 + {\hat k} \cdot {\hat l}_{ij} \right) } 
\frac{\sin \left( \pi L k \left( 1 + {\hat k} \cdot {\hat l}_{ij} \right) 
\right)}{\pi L k \left( 1 + {\hat k} \cdot {\hat l}_{ij} \right)} \;\;,\;\; {\cal G}^A \left( {\hat k} ,\, {\hat l}_{ij} \right) \equiv \frac{{\hat l}_{ij}^a \, {\hat l}_{ij}^b}{2} \, e_{ab}^A \left( {\hat k} \right) \,. 
\label{calM-G-def}
\end{equation} 

Lengthy but straightforward algebra then leads to the TDI combinations (defined in Eqs.~\eqref{TDI1.0} and~\eqref{TDI1.5}): 
\begin{eqnarray} 
\Delta F_{1(23)} \left( t \right) 
&=& - 
\int d^3 k {\rm e}^{-2 \pi i \vec{k} \cdot \vec{x}_1  } \,  \frac{i k}{f_*} \, \sum_A 
\Bigg[ 
{\rm e}^{2 \pi i k  \left( t -  L  \right)} \, W \left( k \right) \, {\tilde h}_A \left( \vec{k}  \right) \, 
R^A \left( \vec{k} ,\, {\hat l}_{12} ,\, {\hat l}_{13} \right) \nonumber\\ 
&& \quad\quad\quad\quad - 
{\rm e}^{-2 \pi i k  \left( t - L  \right) }  \, W^* \left( k \right) {\tilde h}_A^* \left( -\vec{k}  \right) \, 
R^{A*} \left( - \vec{k} ,\, {\hat l}_{12} ,\, {\hat l}_{13} \right) 
\Bigg] \;. 
\label{TDI-integral}
\end{eqnarray}  

In this expression, $f_*$ is the frequency defined in Eq.~\eqref{fstar}, and we have introduced the function 
\begin{eqnarray} 
R^A \left( \vec{k} ,\, {\hat l}_{ij} ,\, {\hat l}_{ik} \right) \equiv 
{\cal G}^A \left( {\hat k} ,\, {\hat l}_{ij} \right)  {\cal T} \left( \vec{k} ,\, {\hat l}_{ij} \right) -
{\cal G}^A \left( {\hat k} ,\, {\hat l}_{ik} \right)  {\cal T} \left( \vec{k} ,\, {\hat l}_{ik} \right) \;, 
\label{RA}
\end{eqnarray}  
with 
\begin{eqnarray} 
{\cal T} \left( \vec{k} ,\, {\hat l}_{12} \right) \equiv 
{\rm e}^{- i k / f_* } {\cal M} \left( \vec{k} ,\, {\hat l}_{21} \right) 
+ {\rm e}^{- i  \vec{k} \cdot {\hat l}_{12} / f_* } {\cal M} \left( \vec{k} ,\, {\hat l}_{12} \right) \;, 
\label{calT-def}
\end{eqnarray} 
as well as the the function $W$ which is different for the two TDI combinations: 
\begin{eqnarray}
W \left( k \right) = \left\{ \begin{array}{l} 
1 \;\;,\;\;\;\; \quad\quad\quad\quad  {\rm for \; TDI \; 1.0} \\ 
{\rm e}^{-2 i k / f_*} - 1  \;\;,\;\;\;\; {\rm for \; TDI \; 1.5} \end{array} \right. 
\end{eqnarray}

The correlation between the TDI measurements in Eq.~\eqref{TDI-integral} is expressed by Eq.~\eqref{<DFDF>}. As stated in the main text, the anisotropic LISA response function in Eq.~\eqref{response} satisfies the properties in Eqs.~\eqref{R-prop-ij},~\eqref{R-prop-menom},~\eqref{R-prop-lm}, and~\eqref{R-prop-m}, that we now prove.

To prove the first property, we consider a rigid rotation of the instrument, for which the position of the three satellites changes according to $\vec{x}_i \to R \vec{x}_i$. 

We perform an analogous rotation on the integration variable in Eq.~\eqref{response}, and, accounting for the fact that scalar products of two vectors are invariant under a rotation we arrive to  
\begin{eqnarray} 
{\tilde R}_{Ri Rj}^{\ell m} \left( f \right) &=& \frac{1}{8 \pi} \int d^2 {\hat k} \, {\rm e}^{-2 \pi i f \, {\hat k} \cdot \left( \vec{x}_i - \vec{x}_j \right) } \,  
{\tilde Y}_{\ell m} \left( R {\hat k} \right) \sum_A \nonumber\\ 
&& \!\!\!\!\!\!\!\!  \!\!\!\!\!\!\!\! 
\left[ {\cal G}^A \left( R {\hat k} ,\, R {\hat l}_{i,i+1} \right)  {\cal T} \left( f  {\hat k} ,\,  {\hat l}_{i,i+1} \right) - {\cal G}^A \left( R {\hat k} ,\, R {\hat l}_{i,i+2} \right)  {\cal T} \left( f  {\hat k} ,\,  {\hat l}_{i,i+2} \right) \right] 
\nonumber\\ 
&& \!\!\!\!\!\!\!\!  \!\!\!\!\!\!\!\! 
\left[ {\cal G}^{A*} \left( R {\hat k} ,\, R {\hat l}_{j,j+1} \right)  {\cal T}^* \left( f  {\hat k} ,\,  {\hat l}_{j,j+1} \right) - {\cal G}^{A*} \left( R {\hat k} ,\, R {\hat l}_{j,j+2} \right)  {\cal T}^* \left( f  {\hat k} ,\,  {\hat l}_{j,j+2} \right) \right] \,. \nonumber\\ 
\label{Rrot-patial}
\end{eqnarray}  

The behavior of the polarization operators under a rotation can be found in Eq.~(A.17) of ref.~\cite{Bartolo:2018qqn}. Using that result, we can see by direct computation that, for any two unit vectors ${\hat u},\, {\hat v}$, 
\begin{eqnarray} 
\sum_A {\cal G}^A \left( R {\hat k} ,\, R {\hat u} \right) {\cal G}^{A*} \left( R {\hat k} ,\, R {\hat v} \right)  =  \sum_\lambda {\cal G}^A \left( {\hat k} ,\,  {\hat u} \right) {\cal G}^{A*} \left(  {\hat k} ,\,  {\hat v} \right) \;. 
\end{eqnarray} 
As a consequence, the rotation matrix is eliminated from the last two lines of Eq. \eqref{Rrot-patial}, and one is left with the rotation of the spherical harmonic, from which Eq.~\eqref{rot-R} is obtained. 

Inserting the expression in Eq.~\eqref{calM-G-def} for ${\cal M}$ in Eq.~\eqref{calT-def}, we see that ${\cal T}^* \left( - \vec{k} ,\, {\hat l}_{ij} \right) =  {\cal T} \left( \vec{k} ,\, {\hat l}_{ij} \right)$. An identical property is shared by the GW polarization operators, and therefore by the functions ${\cal G}^A$. As a consequence, 
\begin{eqnarray} 
\sum_A R^{A*} \left( - \vec{k} ,\, {\hat l}_{ij} ,\, {\hat l}_{ik} \right) \, R^A \left( - \vec{k} ,\, {\hat l}_{lm} ,\, {\hat l}_{ln} \right) =  \sum_A R^A \left(  \vec{k} ,\, {\hat l}_{ij} ,\, {\hat l}_{ik} \right)  \, R^{A*} \left(  \vec{k} ,\, {\hat l}_{lm} ,\, {\hat l}_{ln} \right) \;. \nonumber\\ 
\label{RARA-prop}
\end{eqnarray} 
We start from Eq.~\eqref{response} for ${\tilde R}_{ji}^{\ell m}$. We send ${\hat k} \to - {\hat k}$ in the integrand, and we use the property that we have just proven. We arrive to an expression that is identical to the r.h.s. of Eq.~\eqref{response}, with the only difference that the argument of the spherical harmonic is $= {\hat k}$. From the transformation of the spherical harmonics under parity we then obtain the property in Eq.~\eqref{transpose-R}. 

Let us now prove the property in Eq.~\eqref{R-prop-lm}. We place the LISA satellites in the $xy$ plane, with the center of LISA at the origin, and we simultaneously send the positions of the satellites $\vec{x}_i \to \vec{x}_i$, and change sign to the integration variable ${\hat k}$ in Eq.~\eqref{response}. These two operations do not change the 
scalar products ${\hat k} \cdot {\hat l}$ entering in the integrand of  Eq.~\eqref{response}. Therefore, they do not modify the first facor nor the second line of the integrand of Eq.~\eqref{response}, but only affect the spherical harmonics. Next, we rotate the LISA triangle and the integration variable by $180^\circ$ around the $z-$axis. These two operations only affect the spherical harmonic in the integrand of~\eqref{response}. Under both sets of operations, the spherical harmonic changes to 
\begin{equation}
{\tilde Y}_{\ell m} \left( {\hat k} \right) \to {\tilde Y}_{\ell m} \left( - {\hat k} \right) = 
\left( - 1 \right)^\ell {\tilde Y}_{\ell m} \left( {\hat k} \right) \to 
\left( - 1 \right)^\ell {\tilde Y}_{\ell m} \left( R_{z,\pi} {\hat k} \right) = 
\left( - 1 \right)^{\ell + m} {\tilde Y}_{\ell m} \left( {\hat k} \right) \;. 
\end{equation} 
On the other hand, performing both sets of operations leaves the position of the LISA satellites unaffected, and therefore cannot change the response function. It follows that the response function must vanish whenever $\ell + m$ is odd, as stated in Eq.~\eqref{R-prop-lm}. 

Finally, let us prove the property in Eq.~\eqref{R-prop-menom}. We start from Eq.~\eqref{response} for ${\tilde R}_{ij}^{\ell,-m}$. We change integration variable ${\hat k} \to - {\hat k}$, we use the property $Y_{\ell,-m} \left( - {\hat k} \right) = \left( - 1 \right)^{\ell + m} \, Y_{\ell m}^* \left( {\hat k} \right)$, as well as Eq.~\eqref{RARA-prop}. We end up with the conjugate of the r.h.s. of Eq.~\eqref{response} times the factor $\left( - 1 \right)^{\ell + m}$. From the last property that we have proven, we know that the response function is non vanishing only if $\ell + m$ is even, namely only if this additional factor is one. This proves the property in Eq.~\eqref{R-prop-menom}. 

\section{Optimal Signal-to-Noise Ratio}
\label{app:C-SNR}

In this appendix we derive Eqs. (\ref{<C>}), (\ref{<C2>}), and (\ref{SNR-ell-m}) given in the main text. Moreover, we give the explicit expressions for the noise functions (\ref{NO-def}). 

We start from the evaluation of the expectation value $\left\langle C \right\rangle$ of the estimator (\ref{C-def}). Thanks to the subtraction of the noise expectation value, only the signal contributes to $\left\langle {\cal C} \right\rangle$. We insert the expression (\ref{TDI-integral}) into the Fourier transform (\ref{Fourier-signal}) of the signal. Lengthy but straightforward algebra then leads to the two-point function 
\begin{eqnarray} 
\left\langle {\tilde \Delta F}_O \left( f ,\, t \right) \, {\tilde \Delta F}_{O'}^* \left( f' ,\, t \right) \right\rangle 
& = & \int d k \, k^2  \frac{k^2}{f_*^2}  \sum_{\ell,m}  {\tilde I}_{\ell m} \left( k \right) \, 
\frac{2}{k^2}  \left\vert W \left( k L \right) \right\vert^2 {\tilde R}_{OO'}^{\ell m} \left( k \right)  \nonumber\\ 
&& 
\times \left[ \delta_\tau \left( f - k \right) \delta_\tau \left( f' - k \right) 
+ \delta_\tau \left( f + k \right)  \delta_\tau \left( f' + k \right)  \right] \;, \nonumber\\ 
\label{DFDF}
\end{eqnarray} 
where Eq. (\ref{<hh>}) has been used for the two-point function of the SGWB. In this expression we have denoted by $\delta_\tau$ the (rescaled) sinc function 
\begin{equation}
\delta_\tau \left( f \right) \equiv \frac{\sin \left( \pi \, \tau \, f \right)}{\pi \, f} \;, 
\end{equation}
that emerges from the integration over $dt'$ in eq. (\ref{Fourier-signal}). The notation is justified by the fact that 
$\delta_\tau \left( f \right)$ approaches the Dirac delta function $\delta_D \left( f \right)$ in the limit of infinite $\tau$, or, in practical terms, for $\tau \gg 1/f$. In this limit the above expression for the two-point function simplifies to 
\begin{equation}
\left\langle {\tilde \Delta F}_O \left( f' ,\, t \right) \, {\tilde \Delta F}_{O'}^* \left( f ,\, t \right) \right\rangle 
= \frac{ \delta \left( f - f' \right)}{2} \, \sum_{\ell,m}  {\tilde I}_{\ell m} \left( f \right) \, R_{OO'}^{\ell m} \left( f \right) \;, 
\end{equation}
while, in the case of equal frequencies, one of the time integration involved in the Fourier transform becomes trivial, leading to 
\begin{equation}
\left\langle {\tilde \Delta F}_O \left( f ,\, t \right) \, {\tilde \Delta F}_{O'}^* \left( f ,\, t \right) \right\rangle 
= \frac{\tau}{2} \, \sum_{\ell,m}  {\tilde I}_{\ell m} \left( f \right) \, R_{OO'}^{\ell m} \left( f \right) \;. 
\end{equation}
We insert this into Eq. (\ref{C-def}), split the integral in positive and negative frequencies, rename $f \to - f$ in the negative frequency range, and use the fact that both ${\tilde I}_{\ell m}$ and $R_{OO'}^{\ell m}$ are even functions of the frequency. This leads to Eq. (\ref{<C>}) for the expectation value of the estimator. 

In the computation of the variance of the estimator disregard the contribution of the signal, under the assumption that it is dominated by the noise. Analogously  to Tq. (\ref{Fourier-signal}), the Fourier transform of the noise reads 
\begin{eqnarray} 
{\tilde n}_O \left( f ,\, t \right) = \int_{t-\tau/2}^{t+\tau/2} d t' \, {\rm e}^{-2 \pi i f t'} \, n_O \left( t' \right) 
& = & \int_{t-\tau/2}^{t+\tau/2} d t' \, {\rm e}^{-2 \pi i f t'} \,   \int d k \, {\rm e}^{2 \pi i k t'} \, n_O \left( k \right) \nonumber\\ 
& = & \int d k \, {\rm e}^{-2 \pi i t \left( f - k \right)} \, \delta_\tau \left( f - k \right)  n_O \left( k \right) \;.
\end{eqnarray} 
The noise correlators then (\ref{NO-def}) lead to 
\begin{eqnarray} 
&& \!\!\!\!\!\!\!\!  \!\!\!\!\!\!\!\! 
\left\langle {\tilde n}_O \left( f ,\, t \right)  {\tilde n}_{O'} \left( f' ,\, t' \right) \right\rangle =  \frac{\delta_{OO'}}{2} \,  \int d k \, {\rm e}^{-2 \pi i t \left( f - k \right)}  {\rm e}^{2 \pi i t' \left( - f' - k \right)} 
\, \delta_\tau \left( f - k \right)  \, \delta_\tau \left( f' + k \right) \,  N_O \left( k \right)\,. \nonumber\\ 
\end{eqnarray} 

We use this in the evaluation of $\left\langle {\cal C}^2 \right\rangle$, that we evaluate under the assumption that the noise is Gaussian, obtaining 
\begin{eqnarray} 
&&
\left\langle \left \vert {\cal C}  \right\vert^2 \right\rangle  =\\
&&\frac{1}{2} \sum_{OO'O''O'''} 
\left( \int_0^{T/2} d t_{\rm av} \int_{-t_{\rm av} }^{t_{\rm av} } d t_d + 
\int_{T/2}^T d t_{\rm av} \int_{t_{\rm av}-T}^{T-t_{\rm av} } d t_d  \right) 
\int_{-\infty}^{+\infty} d f \int_{-\infty}^{+\infty} d f'  \int d k \int d k' \nonumber\\ 
&& 
{\tilde Q}_{OO'}^* \left( t_{\rm av} + t_d ,\, f \right) {\tilde Q}_{O''O'''} \left( t_{\rm av} - t_d ,\, f' \right) N_O \left( k \right) N_{O'} \left( k' \right)
\delta_\tau \left( f - k \right)  \delta_\tau \left( f - k' \right)  {\rm e}^{4 \pi i t_d \left( k' - k \right)} \nonumber\\ 
&& 
\left[ \delta_{OO''} \,   \delta_{O'O'''} \, 
\, \delta_\tau \left( f' - k \right)  \, \delta_\tau \left( f' - k' \right) 
+  \delta_{OO'''} \,  \delta_{O'O''} \, \delta_\tau \left( f' + k \right)  \, \delta_\tau \left( f' + k' \right) 
\right] \;, 
\label{C2-app-int}
\end{eqnarray} 
where $t=t_{\rm av} + t_d$ and $t'=t_{\rm av} - t_d$, and $t$ (respectively, $t'$) is the time integration variable in the first (respectively, second) ${\cal C}$ entering in the variance. 

We assume that the weight $Q$ changes slowly over timescales comparable with the measured inverse frequencies, so that we can assume that it depends only on the combination $t_{\rm av}$. We can then integrate over $t_d$. In doing so, the only quantity depending on $t_d$ in Eq. (\ref{C2-app-int}) is the last phase of the second line, and the two integrals of this quantity expressed by the parenthesis in the first line give, respectively, $2 \delta_{t_{\rm av}} \left( 4 \left( k - k' \right) \right)$ and $2 \delta_{T-t_{\rm av}} \left( 4 \left( k - k' \right) \right)$. The measurement times are much grater than the inverse of the frequencies, so that both these quantities can be approximated by $2 \delta_D \left( k - k' \right)$. The two integrals then provide the same result and we can simply add up to the intervals of the integral over $t_{\rm av}$. Performing the  $k'$ integration, we then obtain 
\begin{eqnarray} 
&& \!\!\!\!\!\!\!\!  \!\!\!\!\!\!\!\! 
\left\langle \left \vert {\cal C}  \right\vert^2 \right\rangle  = \frac{1}{4} \sum_{OO'O''O'''} 
\int_0^T d t_{\rm av} 
\int_{-\infty}^{+\infty} d f \int_{-\infty}^{+\infty} d f'  \int d k 
\, {\tilde Q}_{OO'}^* \left( t_{\rm av}  ,\, f \right) {\tilde Q}_{O''O'''} \left( t_{\rm av}  ,\, f' \right) \nonumber\\ 
&& \quad\quad \times \, N_O \left( k \right) N_{O'} \left( k \right) \, 
\delta_\tau \left( f - k \right)  \delta_\tau \left( f - k \right) 
\Bigg[ \delta_{OO''} \,   \delta_{O'O'''} \, 
\, \delta_\tau \left( f' - k \right)  \, \delta_\tau \left( f' - k \right) \nonumber\\ 
&& \quad\quad \quad\quad \quad\quad \quad\quad \quad\quad \quad\quad \quad\quad \quad\quad \quad\quad + \delta_{OO'''} \,  \delta_{O'O''} \, \delta_\tau \left( f' + k \right)  \, \delta_\tau \left( f' + k \right) 
\Bigg] \;. \nonumber\\ 
\end{eqnarray} 

As we did for the expectation value, we can then substitute the functions $\delta_\tau$ with the Dirac delta-function, since the time $\tau$ is much greater than the inverse frequencies. We then perform the integrals overt $f$ and $f'$, the sums over $O''$ and $O'''$, and we relabel $k \to f$ and $t_{\rm av} \to f$ in the resulting expression 
\begin{equation}
\!\!\!\!\!\!\!\! \!\!\!\!\!\!\!\! 
\left\langle \left \vert {\cal C}  \right\vert^2 \right\rangle  = \frac{\tau^2}{4} \sum_{OO'} 
\int_0^T d t  \int_{-\infty}^{+\infty} d k \, 
{\tilde Q}_{OO'}^* \left( t  ,\, f \right) 
\left[  {\tilde Q}_{OO'} \left( t  ,\, f \right) + {\tilde Q}_{O'O} \left( t  ,\, -f \right) \right] 
N_O \left( f \right) N_{O'} \left( f \right) \;. 
\end{equation} 
Using the fact the the noise is an even function of $f$, this expression can be finally written as Eq. (\ref{<C2>}) of the main text. 

Starting for the expressions (\ref{<C>}) and (\ref{<C2>}), for, respectively, the expectation value and the variance of the estimator ({C-def}), it is convenient to define 
\begin{equation}
{\cal Q}_{OO'} \left( t ,\, f \right) \equiv \frac{\tau}{2}  
\sqrt{N_O \left(  f  \right) \, N_{O'} \left(  f  \right)}  \, 
\left[  {\tilde Q}_{OO'} \left( t  ,\, f \right) +  {\tilde Q}_{O'O} \left( t  ,\, - f \right) \right] \,, 
\end{equation} 
in terms of which, 
\begin{equation}
{\rm SNR} = \frac{\left\langle C  \right\rangle}{\left\langle \left \vert {\cal C}  \right\vert^2 \right\rangle} = \frac{\sum_{OO'} \int_0^\infty d f \int_0^T d t \, \gamma_{OO'} \left( f ,\, t \right) \, {\cal Q}_{OO'} \left( t ,\, f \right) }{\sqrt{ \sum_{OO'} \int_0^T d t \int_0^{+\infty}  d f    \, \left\vert {\cal Q}_{OO'} \left( t  ,\, f \right)  \right\vert^2  }} \;, 
\label{SNR-app}
\end{equation} 
where, making use of Eqs. (\ref{<C>}) and (\ref{I-to-delta}), 
\begin{equation}
\gamma_{OO'} \left( f ,\, t \right) \equiv 
\frac{3  H_0^2}{4 \pi^2 \sqrt{4 \pi}} 
\frac{\Omega_{\rm GW} \left( f \right)}{f^3} \;  
\frac{\sum_{\ell,m} 
\delta_{\rm GW,\ell m} \left( f \right) 
{\cal R}_{OO'}^{\ell m} \left( f  \right) }{  \sqrt{N_O \left(  f  \right) \, N_{O'} \left(  f  \right) } } \;. 
\end{equation} 
We then see that the SNR is maximized by ${\cal Q}_{OO'} \left( t ,\, f \right) = c \times \gamma_{OO'}^* \left( f ,\, t \right) $, where $c$ is an arbitrary constant that we can set to one. This leads to Eq. (\ref{SNR-ell-m}) of the main text. 

We conclude this appendix by providing the LISA noise functions used in our computations, referring the interested reader to ref.~\cite{Flauger:2020qyi} for a detailed discussion of these quantities. For the $A$ and $E$ channels one has
\begin{eqnarray}
&& \!\!\!\!\!\!\!\!  \!\!\!\!\!\!\!\! 
{\tilde N}_{A,E} \equiv \frac{N_{A,E}}{ 4 \left( f / f_* \right)^2 \, \left\vert W \left( f \right) \right\vert^2 } \nonumber\\ 
&& 
= \frac{1}{2}
\left[ 2 +  \cos \left( \frac{f}{f_*} \right) \right]  \frac{P^2}{L^2} \, \frac{\rm pm^2}{\rm Hz} \left[ 1 + \left( \frac{2 \, {\rm mHz}}{f} \right)^4 \right]  \nonumber\\ 
&& 
+ 2 \left[ 1 + \cos \left( \frac{f}{f_*} \right) + \cos^2  \left( \frac{f}{f_*} \right) \right] \frac{A^2}{L^2} \, \frac{\rm fm^2}{{\rm s}^4 \, {\rm Hz} }  
\left[ 1 + \left( \frac{0.4 \, {\rm mHz}}{f} \right)^2 \right] 
\left[ 1 + \left( \frac{f}{8 \, {\rm mHz}} \right)^4 \right] 
\left( \frac{1}{2 \pi f} \right)^4 \;, \nonumber\\ 
\label{NA,E}
\end{eqnarray}
where the coefficients $P$ and $A$ provide, respectively the aplitude of the Interferometry Metrology System and the acceleration noise. We assume the central values for these coefficients from ESA mission specifications requirements, namely $P = 15$ and $A = 3$. For the $T$ channel one has instead 
\begin{eqnarray}
&& \!\!\!\!\!\!\!\!  \!\!\!\!\!\!\!\! 
{\tilde N}_T \equiv \frac{N_{TT}}{ 4 \left( f / f_* \right)^2 \, \left\vert W \left( f \right) \right\vert^2 } \nonumber\\ 
&& 
= \left[ 1 - \cos \left( \frac{f}{f_*} \right) \right]  \frac{P^2}{L^2} \, \frac{\rm pm^2}{\rm Hz} \left[ 1 + \left( \frac{2 \, {\rm mHz}}{f} \right)^4 \right]  \nonumber\\ 
&& 
+  2 \left[ 1 - \cos \left( \frac{f}{f_*} \right) \right]^2  \frac{A^2}{L^2} \, \frac{\rm fm^2}{{\rm s}^4 \, {\rm Hz} }  
\left[ 1 + \left( \frac{0.4 \, {\rm mHz}}{f} \right)^2 \right] 
\left[ 1 + \left( \frac{f}{8 \, {\rm mHz}} \right)^4 \right] 
\left( \frac{1}{2 \pi f} \right)^4 \;. \nonumber\\ 
\label{NT}
\end{eqnarray}

\section{Boost-induced anisotropies of the SGWB}
\label{app_boost}

We   derive the expressions for the anisotropies of the SGWB induced by a boost transformation. We use the same methods as in ref.~\cite{Landau:1987gn,Peebles:1968zz,mckinley,Kosowsky:2010jm}. We 
consider two frames: the first,
denoted with  $\mathcal{S}'$,  is comoving with the SGWB rest frame; the second,  denoted with  $\mathcal{S}$,  moves with constant velocity ${\bf v}$ with respect to
  the rest frame  $\mathcal{S}'$.

A boost transformation relates the SGWB density parameter in the rest frame
$\mathcal{S}'$ to the one in the moving one  $\mathcal{S}$. We denote with  $f'$ the frequency  of the GW in the  SGWB rest frame.
 and with $\hat \bn'$
the unit vector denoting 
its direction. The frequency $f$ in the frame in motion is associated with $f'$ by a Lorentz
transformation reading
\be\label{Eshift}
f\,=\,
{\cal D}\,
f'\,,
\ee
with
\be
\label{defcald}
{\cal D}\,=\,\frac{\sqrt{1-\beta^2}}{1- \beta \,\hat\bn \cdot \hat \bv}\,. 
\ee
where $\bv=\beta \hat{\bv}$ is the relative velocity of the two frames, and $\beta=v$ in units with $c=1$.

In order  to compute how the GW energy density changes under a Doppler boost, we work in terms of the GW distribution function, denoted with
 $\Delta'( f' )$. We assume for simplicity it only depends
 on the frequency $f'$
 in the SGWB rest frame (i.e. the SGWB is perfectly isotropic in the frame ${\cal S}'$). 
 We  express  the number of gravitons for unit of phase space  in the rest-frame ${\cal S}'$ as:
 \be
 \label{dNgrnumb}
d N' \,=\,\Delta'(f'   )\,f'^2\,d f'\,d^2 \hat \bn'\,d V'\,,
\ee
where  $dV'$ corresponds to the infinitesimal volume containing
 gravitons with propagation vector $\hat \bn'$ in the element of measure $ d f'\, d^2 \hat \bn'$.
  It is not difficult to prove that the combination  $f'^2\,d f'\,d^2 \hat \bn'\,d V'$ is invariant
  under boosts. In fact, we have the relations $f'\,=\,{\cal D}^{-1}\,f$, $d^2\hat  \bn'\,=\,{\cal D}^{2}\, d^2 \hat \bn$,
  $d V'\,=\,{\cal D}\,d V$ (see  ref.~\cite{mckinley,Kosowsky:2010jm}).  On the other hand, the number of gravitons~\eqref{dNgrnumb} is independent
of the frame, and $dN '\,=\,d N$. Hence~\cite{Landau:1987gn}
  \be
  \label{samdi1}
\Delta'( f'   )\,=\,\Delta(f   )\,.
\ee
The GW distribution function $\Delta$ can be used to define the energy density of GW in the rest frame as
energy per unit volume and unit solid angle:
\bea
d \rho'_{\rm GW}(f', \hat \bn'  )&=& \frac{f'\,d N' }{d^2 \hat \bn'\,d V'}
\,=\,\Delta'(f'  )\,f'^3\,d f'\,.
\eea
This definition allows us to express the GW density parameter $\Omega_{\rm GW}'(\omega', \hat \bn'  \hat \bv)$
in the  rest frame ${\cal S}'$ as
\begin{eqnarray}
\label{genexpom}
\Omega'_{\rm GW}(f', \hat \bn'  )\equiv\frac{1}{\rho_c}\,\frac{d  \rho'_{\rm GW}}{d \ln f'}&=&\frac{3\pi\,f'^4}{2\,H_0^2\,M_{\rm Pl}^2}\,\Delta'(f' )\,.
\end{eqnarray}
Using Eq.~\eqref{samdi1}, we have the equality
\be
\Omega_{\rm GW}(f  )\,=\,\left( \frac{f}{f'} \right)^4 \Omega'_{\rm GW}(f' )\,.
\ee
Hence,
we find that the GW density parameter in the moving frame ${\cal S}$ is related with the corresponding quantity in the frame  ${\cal S'}$ at rest through the
general formula  
\be
\label{genexpom4}
\Omega_{\rm GW}(f, \hat \bn )\,=\,{\cal D}^4\,\, \Omega'_{\rm GW}\left({\cal D}^{-1}\,f\right)
\ee
 with ${\cal D}$  given in Eq.~\eqref{defcald}. 
  Notice that in the moving frame ${\cal S}$ the expression of  $\Omega_{\rm GW}$
   is anisotropic, due to the dependence of ${\cal D}$ on $\hat \bn$. 
 The parameter $\beta$ is usually small: 
 for example, for cosmological backgrounds,  CMB suggests that 
  $\beta\simeq 1.23 \times 10^{-3}$. Under the assumption of small $\beta$, 
  we Taylor expand  Eq.~\eqref{genexpom4}.

\bigskip

We introduce the  tilts of the SGWB spectrum as
\bea
\label{defno}
n_{\Omega}(f)&=&\frac{d\,\ln \Omega'_{\rm GW}(  f)}{d\,\ln f}\,,
\\
\label{defao}
\alpha_{\Omega}(f)&=&
\frac{d\,n_{\Omega}(f)}{d\,\ln f}\,.
\eea
Expanding Eq.~\eqref{genexpom4}  in powers of 
 $\beta$,
 and limiting the expansion to order
 $\beta^2$  we find that 
 the GW density parameter in the moving frame ${\cal S}$ receives
   a kinematic modulation of the monopole, and the generation of a kinematic dipole  and 
 a kinematic quadrupole due to boost effects:
\bea
\Omega_{\rm GW}(f, \hat \bn)&=&\Omega'_{\rm GW}(f)
\left[ 1+M(f)+
 \hat \bn \cdot \hat \bv\,D(f)+
 \left( 
 (\hat \bn \cdot \hat \bv)^2-\frac13
  \right)
 \,Q(f)
 \right]\,,
 \nonumber\\
  \label{genexpom4a}
\eea
 The frequency-dependent coefficients (we
understand  the explicit frequency-dependence of the spectral tilts)
\bea
\label{monanis}
M(f)&=&\frac{\beta^2}{6} \left( 8+n_\Omega \left( n_\Omega-6\right)
+\alpha_\Omega
\right)\,,
\\
\label{dipanis}
D(f)&=& \beta \left(4-n_\Omega\right)\,,
\\
\label{quapanis}
Q(f)&=&\beta^2\left(10-\frac{9 n_\Omega}{2} +\frac{n_\Omega^2}{2}+\frac{\alpha_\Omega}{2}\right)\,,
\eea
indicate respectively the monopole, dipole, quadrupole  boost contributions.


\end{appendix}

\bibliographystyle{JHEP}
\bibliography{AnisGWB.bib}

\end{document}